\documentclass[12pt,preprint]{aastex}

\newfont{\Fr}{eufm10 scaled\magstep1}
\newfont{\Sc}{eusm10 scaled\magstep1}

\newcommand{\vct}[1]{\mbox{\boldmath${#1}$}}
\newcommand{\lsim}{\mbox{\raisebox{-1.ex}{$\stackrel{<}{\sim}$}}}
\newcommand{\gsim}{\mbox{\raisebox{-1.ex}{$\stackrel{>}{\sim}$}}}

\newcommand{\rD}{r_{\mbox{\tiny \hspace{-0.4mm}$D$}}}
\newcommand{\rQ}{r_{\mbox{\tiny \hspace{-0.4mm}$Q$}}}

\newcommand{\zD}{z_{\mbox{\tiny \hspace{-0.2mm}$D$}}}
\newcommand{\zQ}{z_{\mbox{\tiny \hspace{-0.2mm}$Q$}}}

\def \blankline{\vspace{0.3 cm}}

\shorttitle{Primary electron}
\shortauthors{Shibata and Sekiguchi}

\begin{document}


\title{A POSSIBLE APPROACH TO THREE-DIMENSIONAL\\ 
       COSMIC-RAY PROPAGATION IN THE GALAXY. IV.\\ 
       ELECTRONS and ELECTRON-INDUCED $\gamma$-RAYS\\ 
        (To appear in ApJ, December 2010)}


\author{T. Shibata, T. Ishikawa, and S. Sekiguchi}

\affil{}

\affil{\footnotesize 
Department of Physics and Mathematics, Aoyama-Gakuin University,
Kanagawa 229-8558, Japan}







\begin{abstract}

Based on the diffusion-halo model for cosmic-ray (CR) propagation,
including stochastic reacceleration due to collisions with 
hydromagnetic turbulence, we study the behavior of the electron
component and the diffuse $\gamma$-rays (D$\gamma$'s) induced by
them.  The galactic parameters appearing in these
studies are essentially the same as those appearing in the hadronic CR
components, while we additionally need information on the
interstellar radiation field, taking into account dependences on both
the photon energy, $E_{\scriptsize \mbox{ph}}$, and the position, $\vct{r}$. 
We compare our numerical results with the data on hadrons, 
electrons and  D$\gamma$'s, 
including the most recent results from FERMI, which gives two remarkable results;
1) the electron spectrum falls with energy as $E_e^{-3}$ up to 1\,TeV, and
   does not exhibit prominent spectral features around 500\,GeV, 
 in contrast to
 the dramatic excess appearing in both ATIC and PPB-BETS spectra, and
2) the EGRET GeV-excess in the D$\gamma$ spectrum is due 
 neither to an astronomical origin (much harder CR spectrum in
 the galactic center) nor a cosmological one (dark matter
 annihilation or decay), but due to an instrumental problem.
In the present paper, however, we focus our interest rather 
conservatively upon the 
internal relation between these three components,
 using  {\it common} galactic parameters. We find that
they are in reasonable harmony with each other within both 
 the theoretical and experimental uncertainties, apart from 
the electron-anomaly problem, while some enhancement of
 D$\gamma$'s appears in the high galactic latitude with
 $|b| > 60^\circ$ in the GeV region.
\end{abstract}

\keywords{cosmic rays ---  Galaxy: structure --- electrons: diffuse background}

\section{Introduction}

Although the electron component is only a small fraction of all
 cosmic-ray (CR) 
components, around 1\% of the proton intensity around
10\,GeV, it plays a key role in understanding the structure of
our Galaxy and the galactic phenomena occurring within it. This
is because electrons have electromagnetic interactions with
the interstellar radiation field, such as photons and magnetic
fields, resulting in drastic energy loss during propagation
through the Galaxy, in contrast to the hadronic component.

This peculiar nature yields valuable information for the study of
CR astrophysics, which can not be obtained by the hadronic components
alone.  Namely, due to the rapid energy-loss rate, proportional to
$E_e^2$ in the high energy region, from
the inverse Compton scattering off photons and 
synchrotron radiation in magnetic fields,
the life-time of TeV electrons is at most $10^5$\,yr,
indicating that detected electrons have originated in nearby sources,
less than 1\,kpc from the solar system (SS).
Therefore, accurate observations of TeV electrons will provide
a direct signature of nearby CR sources as well as the mechanism of the CR
acceleration, while depending on the release time from supernova
remnants and their distance from the SS.  

Qualitative studies of such
a possibility have been performed by many authors 
(Shen 1970; Nishimura et al.\ 1979; Cowsik \& Lee 1979; 
Berezinskii et al.\ 1990; Aharonian et al.\ 1995; 
Ptuskin \& Ormes 1995; Pohl \& Esposito 1998; Kobayashi et al.\ 2004;
 Delahaye et al.\ 2010), 
with Kobayashi et al.\ and Delahaye et al.\ 
presenting explicitly
several candidates for nearby sources of high energy CR electrons,
based on the most recent data for the age and distance of each 
 supernova remnant 
near the SS, although the statistics of high energy electron data are
currently too poor to identify sources definitely.
 
Particle identification and the energy determination of high energy
electrons is, however, quite difficult, while direct observation of
low energy electrons is relatively easy using, for instance, magnetic
spectrometers, and has been performed by several groups (Golden et
al.\ 1994; Boezio et al.\ 2000; DuVernois et al.\ 2001; Aguilar et
al.\ 2002).

  Although the statistics are not sufficient, the only group
that succeeded in observing {\it directly} TeV electrons is Nishimura
et al.\ (1980; see also Kobayashi et al.\ 1999) with the use of the
balloon-borne emulsion chamber. It should be noted that they
actually observe event by event the vertex point of the electron with
subsequent $e^{\pm}$-pair due to bremsstrahlung $\gamma$, with no
uncertainty from proton contamination. The precision in the
energy determination is approximately 10\% for electrons 
in the energy region larger than 50\,GeV, based on both the
three-dimensional cascade theory (Nishimura 1964) and
the simulations (Kasahara 1985; Okamoto \& Shibata 1987),
which have been well established
by the use of accelerator beams (Hotta et al.\ 1980; Sato \& Sugimoto 1979).

Recent development in high energy electron observations is indeed
remarkable, particularly those of ATIC (Chang et al.\ 2008) and
PPB-BETS (Torii et al.\ 2006), which
showed an anomaly in the electron spectrum with a significant bump
around 500\,GeV. Both groups point out that the excess 
indicates either a nearby source of energetic electrons, or
those coming from the annihilation of dark matter particles.

On the other hand, the most recent results obtained by the
FERMI Large Area Telescope (FERMI-LAT; Abdo et al.\ 2009) present no
prominent excess, with the electron spectrum falling with
energy as $E_e^{-3.04}$ up to 1\,TeV, which is not inconsistent
 with the emulsion chamber data (Kobayashi et al.\ 1999) within the statistical
 errors. The H.E.S.S.\ ground-based telescope 
(Aharonian et al.\ 2008, 2009) also 
shows no indication of structure in the electron spectrum,
but rather a power-law spectrum with 
$E_e^{-3.0\pm 0.1 \pm 0.3}$ (0.1: stat.\ error, 0.3: syst.\ error), 
albeit this being an indirect observation.

Nevertheless, looking carefully FERMI data
around the anomaly-energy, they still show systematically 
an enhancement as large as 30\%
compared to the numerical results (Abdo et al.\ 2009;
 Strong et al.\ 2004; 
 see also Figure 14 in this paper), 
so that we can not exclude
the possibility of an additional component such as local sources
and/or the dark matter scenario, while strength of the anomaly 
compared to the background {\it diffuse} electrons 
is not as dramatic as
presented by ATIC and PPB-BETS. 

 In any
case, both observational and theoretical studies for high energy
electrons are becoming increasingly important not only for
astrophysics, but also for particle physics and cosmology.
It is, therefore, desirable to find a reasonable model
for electron propagation in the Galaxy, which must explain
consistently and simultaneously all CR observables and not just
electrons, using common galactic parameters with the smallest number
of variables possible. In the sense, the recent review article by
Strong, Moskalenko, \& Ptuskin (2007) is a useful
survey of both the theory and relevant 
experimental data for the propagation of CRs, 
comprehensively summarizing the current landscape  
and open questions, although it was
published just before the anomaly problem mentioned above. 

 Under these situations,
 we have studied the three-dimensional CR propagation model analytically,  
 and found excellent agreement with
the experimental data for various
hadronic components, stable primaries, secondaries such as boron and
sub-iron elements ($Z$\,=21--23), 
isotopes such as $^{10}$Be, and antiprotons as well, in four papers,
(Shibata et al.\ 2004, 2006, 2007a, 2008),
hereafter referred to as Papers I, II, III and IV, respectively.

 We have applied our model  
 further to the studies of diffuse $\gamma$-rays
 (D$\gamma$'s) (Shibata, Honda, \&
Watanabe 2007b; hereafter Paper~V), and found that all these
components are generally in agreement with each other using the same galactic
parameters, within the uncertainties in the experimental data and 
various kinds of cross-sections used for the numerical calculations.
However, in Paper~V, we use the simulation results for
electron-induced $\gamma$-rays provided by Hunter et al.\ (Bertsch et
al.\ 1993; Hunter et al.\ 1997), where 
the modeling of CR propagation and 
the galactic parameters assumed are somewhat different from ours.
  So we have
yet to see complete internal consistency among all CR components ---
hadrons, electrons and D$\gamma$'s --- using the {\it same} galactic
parameters in our propagation model.

 In the present paper, we extend it to the electron component,  
 based on  the diffusion-halo model proposed by Ginzburg, 
Khazan \& Ptuskin
(1980), taking the reacceleration process into account. 
  However, we focus in the present work on {\it diffuse}
electrons in the steady state without discriminating those produced by
nearby sources from those of distant ones, and present the
intensity of the D$\gamma$'s
produced by them in the energy range,
$E_\gamma$\,=\,30\,MeV--100\,GeV, covered by EGRET and 
FERMI. Comparison  with radio and TeV-$\gamma$ data 
will be reported separately in the near future.

In order to apply our model to the electron component and
electron-induced D$\gamma$'s, we need information on the
interstellar radiation field (ISRF) in addition to the 
interstellar matter (ISM), 
particularly their spatial
gradients for the study of the $(l, b)$-distribution of D$\gamma$'s ($l$:
galactic longitude; $b$: galactic latitude).  
Nowadays the most advanced and standard code for the ISM and ISRF models 
is GALPROP, extensively developed by Strong \& Moskalenko (1998), 
incorporating the latest
survey data in the very wide wavelength range from
ultra-violet to radio. 
In the present work,  we assume empirical
density distributions for the ISM and ISRF, smoothing the 
numerical data given by GALPROP available most recently (Porter et al.\ 2008), 
in order to combine with our analytical solution for electron-induced
 D$\gamma$'s.

In \S\,2, we discuss the interstellar
environment provided by GALPROP, 
focusing on the spatial distribution of both matter
(atomic, molecular, and ionized hydrogen) and photons
(ultraviolet, visible, infrared, mid- and far-infrared, and cosmic microwave
background [CMB] radiations), 
and in \S\,3 we present the relevant 
elementary processes for electrons, focussing on the energy losses
due to ionization, bremsstrahlung, 
synchrotron, 
inverse Compton (IC), and on the energy gain due to the
reacceleration.

In \S\,4, we present the diffusion equation,
 and give its solution explicitly in the steady state, 
$N_e(\vct{r}; E_e)$, where the Klein-Nishina effect is quite
 important in the electron energy spectrum in the high
 energy region, $\gsim$\,10\,GeV. In
\S\,5, we present the emissivity of 
electron-induced $\gamma$'s, 
$q_\gamma(\vct{r}; E_\gamma)$, with
use of realistic spatial distributions of ISM and ISRF as
discussed in \S\,2, and show the numerical results 
 at several observational points, where those of the hadron-induced 
$\gamma$'s are presented as well.
In \S\,6, we first summarize the galactic parameters and their
explicit values expected from the CR data, and then compare our
numerical results of electron flux and D$\gamma$'s 
with recent observational data,
including those most recently obtained by FERMI and H.E.S.S.
Finally in \S\,7, we summarize the
results, and discuss several remaining open questions, while
 we do not touch upon  the so called electron-anomaly.

\begin{figure}[!b]
 \vspace{3mm}
    \includegraphics[width=7.5cm]{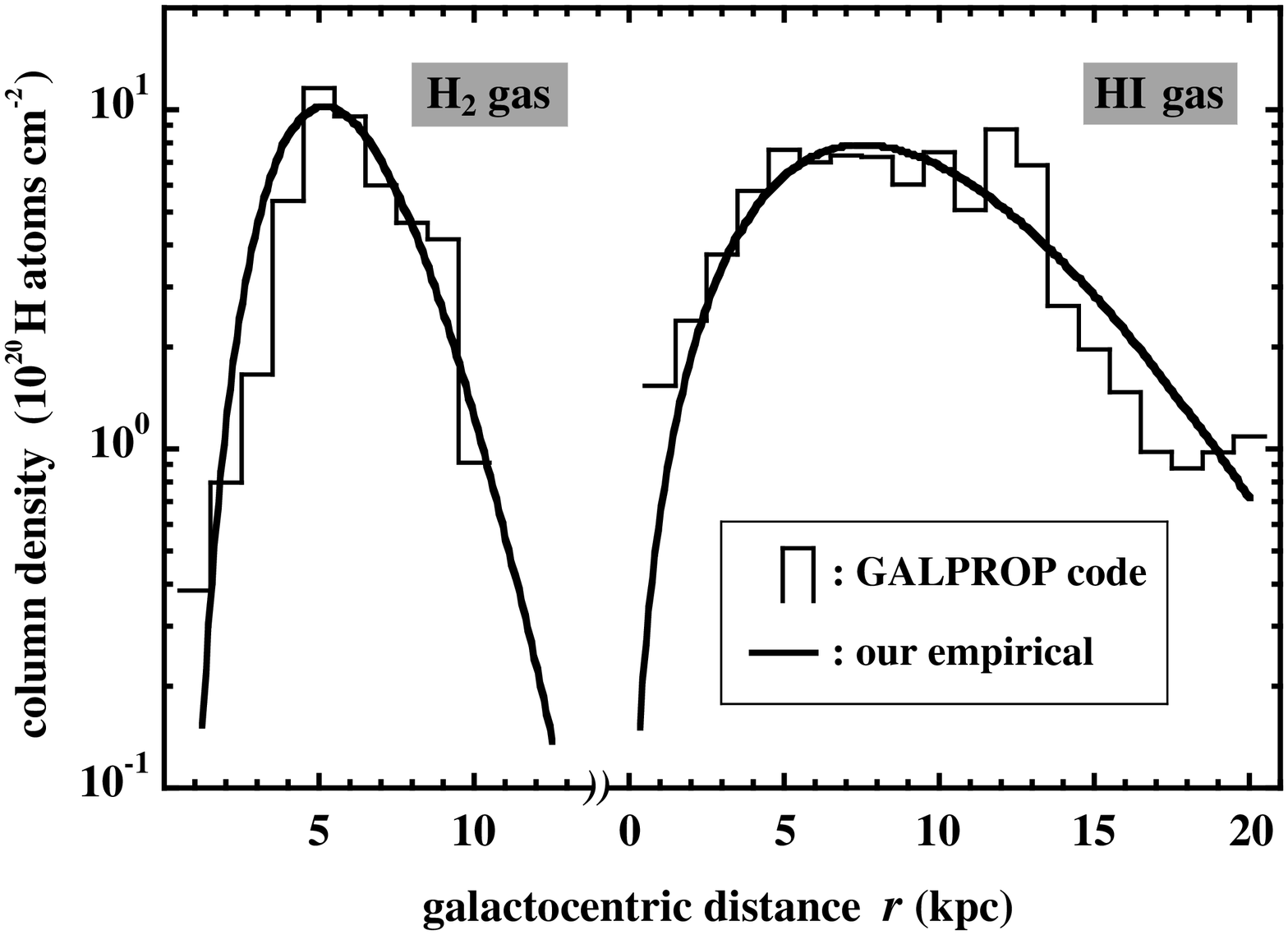}
  \caption{
Column density of interstellar hydrogen. Curves are
 empirical ones given by equation (1) with
 the parameterization summarized in Table 1.
}
\end{figure}

\vspace{-3cm}
{\begin{deluxetable}{rrrrrrrrr}
\tabletypesize{\small}
\tablecaption{
Summary of the numerical values of the coefficients
appearing in equations (1a) and (1b) in units of 
$10^{20}$H\,atoms\,cm$^{-2}$,
 where $\lq \lq (\pm m)$" denotes
 the multiplication of $10^{\pm m}$.
\label{tbl-1}} 
\tablewidth{0pt}
\tablehead{
  \colhead{\  $P_{\tiny \mbox{HI}}^{(0)}$} 
& \colhead{\  $P_{\tiny \mbox{HI}}^{(1)}$} 
& \colhead{\  $P_{\tiny \mbox{HI}}^{(2)}$} 
& \colhead{\  $P_{\tiny \mbox{HI}}^{(3)}$}
& 
& \colhead{$\  P_{\tiny \mbox{H$_2$}}^{(0)}$} 
& \colhead{$\  P_{\tiny \mbox{H$_2$}}^{(1)}$} 
& \colhead{$\  P_{\tiny \mbox{H$_2$}}^{(2)}$} 
& \colhead{$\  P_{\tiny \mbox{H$_2$}}^{(3)}$}  
}
\startdata 
 $3.862(+0)$ & $7.903(-1)$ & $-9.426(-2)$ & $-4.261(+0)$ & & 
 $1.848(+0)$ & $8.339(-1)$ & $-5.560(+0)$ & $2.405(-2)$
 \enddata
 \vspace{-2mm}
\end{deluxetable}

\begin{deluxetable}{lll}
\tabletypesize{\small}
\tablecaption{
Summary of functions for ISM gas density, H\,I and H$_2$, where
 $r$, $r_\odot$\,(=8.5kpc), $z$, and $z_0$ are all
 in units of kpc, and $n_h^{\odot}$ in units of H\,atoms\,cm$^{-3}$.
\label{tbl-2}} 
\tablewidth{0pt}
\tablehead{
  \colhead{$\hspace{-0.1cm}\lq \lq h$"} & \colhead{\ \ $n_h^{\odot}$} 
& \colhead{${\it \Xi}_h(r, z)$} 
} 
\startdata 
\vspace{-1mm}
\\
H\,I &\ \ 0.57 
& 
\hspace{0.5cm}
$\displaystyle 
\frac{1}{0.065 \sqrt{\pi} + 0.160}
\Biggl \{ 0.4\exp{\biggl[-\Bigl( \frac{z}{0.12}\Bigr)^2\biggr]}+
0.2\exp{\biggl[-\Bigl( \frac{z}{0.35}\Bigr)^2\biggr]}+
0.4\exp{\Bigl(-\frac{z}{0.40}\Bigr)}\Biggr \}$
\\ \\
H$_2$ &\ \ 0.53 
&
\hspace{0.5cm}
$\displaystyle 
\frac{1}{0.036 \sqrt{\pi} + 0.2z_0}
\Biggl \{\exp{\biggl[-\Bigl( \frac{z}{0.071}\Bigr)^2\biggr]}+
0.2\frac{z}{z_0}\exp{\Bigl(- \frac{z}{z_0}\Bigr)}\Biggr \}
;\  
z_0(r)= 0.4\cosh \Bigl(\frac{2r}{3r_\odot}\Bigr)$
\\ \\
 \enddata
\end{deluxetable}
}

\vspace{3cm}
\section{Interstellar environment of our Galaxy}

\subsection{Interstellar matter}

First we consider the ISM for two processes,
ionization and bremsstrahlung. In Figure~1 we plot histograms of
column density for H\,I and H$_2$ in the galactic plane (GP) 
given by GALPROP, where we also plot the empirical curves used
in the present work,
$$
\vspace{1mm}
-\ln \rho_{\tiny \mbox{HI}}(r)\hspace{-0.5mm}=\hspace{-0.5mm}P_{\tiny 
\mbox{HI}}^{(0)}\hspace{-0.5mm}+\hspace{-0.5mm}P_{\tiny 
\mbox{HI}}^{(1)}r\hspace{-0.5mm}+\hspace{-0.5mm}P_{\tiny 
\mbox{HI}}^{(2)}\ln {r}\hspace{-0.5mm}+\hspace{-0.5mm}P_{\tiny 
\mbox{HI}}^{(3)}r^{\frac{1}{2}},
\eqno{(\rm 1a)}
$$
$$
\blankline
-\ln \rho_{\tiny \mbox{H$_2$}}(r)\hspace{-0.5mm}=\hspace{-0.5mm}P_{\tiny 
\mbox{H$_2$}}^{(0)}\hspace{-0.5mm}+\hspace{-0.5mm}P_{\tiny 
\mbox{H$_2$}}^{(1)}r\hspace{-0.5mm}+\hspace{-0.5mm}P_{\tiny
 \mbox{H$_2$}}^{(2)}\ln {r}\hspace{-0.5mm}+\hspace{-0.5mm}P_{\tiny
 \mbox{H$_2$}}^{(3)}r^2,
\eqno{(\rm 1b)}
$$
with $r$ in kpc, and $\rho_h$ ($\lq \lq h" \equiv \mbox{H\,I, H}_2$)
in $10^{20}$ H\,atoms\,cm$^{-2}$.
The numerical values of the coefficients are summarized in Table\,1.
However, the choice of above empirical form is not critical,
and other choices may be possible.

The H$_2$ gas is strongly confined to the
GP and its vertical structure is modeled by a gaussian distribution with a width of
approximately 70\,pc, while the H\,I gas lies in a flat layer with a
FWHM of 230\,pc in 
3.5\,kpc $< r < r_{\tiny \mbox{$\odot$}}$\,(=8.5\,kpc), 
and is approximated by the sum of two gaussians and an exponential tail
(Ferriere 2001; Moskalenko et al.\ 2002). 
Taking these situations into account, we assume the following spatial
distribution for the ISM gas density, corresponding to equations (1a) and
(1b),
$$ \hspace{-0.5mm}
\blankline
\frac{{n_h}(\vct{r})}{n_h^\odot} \hspace{-0.5mm}=  \hspace{-0.5mm}
\frac{{\it \Xi_h}(r, z)}{{\it \Xi_h}(r_\odot, 0)}
\frac{\rho_h(r)}{\rho_h(r_\odot)}, 
\ \ (\lq \lq h" \hspace{-0.5mm} \equiv \hspace{-0.5mm} \mbox{H\,I, H}_2),
\eqno{(\rm 2)}
$$
where
$n_{\tiny \mbox{HI}}^\odot$ ($n_{\tiny \mbox{H$_2$}}^\odot$)
is the gas density of H\,I (H$_2$) at the SS with
typically $n_{\tiny \mbox{HI}}^\odot \approx n_{\tiny \mbox{H$_2$}}^\odot
\approx 0.5$ H\,atoms\,cm$^{-3}$.
See Table\,2 for the 
explicit forms of ${\it \Xi}_{\tiny \mbox{\,HI}}$ and  
${\it \Xi}_{\tiny \mbox{\,H$_2$}}$.

For the ionized hydrogen gas, H\,II, we use the two-component model of
Cordes et al.\ (1991), 
$
n_{\tiny \mbox{HII}}(\vct{r}) =
 n_{\tiny \mbox{HII}}^{\scriptsize \mbox{(1)}}(\vct{r})
+ n_{\tiny \mbox{HII}}^{\scriptsize \mbox{(2)}}(\vct{r}),
$
and both components are modeled by a gaussian-type distribution
for the radial structure, and by a simple exponential one
for the vertical structure. The explicit values of the 
two components at the SS, 
[$n_{\tiny \mbox{HII}}^{(1)}(r_\odot),
 n_{\tiny \mbox{HII}}^{(2)}(r_\odot)$], 
are [0.025, 0.013]\,cm$^{-3}$ respectively 
(Cordes et al.\ 1991; Strong et al.\ 1998). 
So the contribution of H\,II is much smaller than those of  
H\,I and H$_2$
and is not important in the present work.\vspace{4mm}

\subsection{Interstellar radiation field}

First we consider the medium --- virtual photons induced by the static
magnetic field --- for the synchrotron process. It is approximately given by an
exponential-type gradient, while the scale height is not yet clear.
Practically, for the study of synchrotron radiation, we need the 
{\it energy density} of virtual photons at 
$\vct{r}$, $\epsilon_{\mbox{\tiny B}}(\vct{r})$, and assume in the present
 work

$$
\epsilon_{\mbox{\tiny B}}(\vct{r}) = 
\epsilon_{\mbox{\tiny B},0}
\exp [-(r/r_{\mbox{\tiny B}} + |z|/z_{\mbox{\tiny B}})],
\eqno{(\rm 3)}
$$
with
$$
\blankline
\epsilon_{\mbox{\tiny B},0} = {B_0^2}/{8\pi},
$$
where $B_0$ is the magnetic field at the galactic center (GC), and 
$\epsilon_{\mbox{\tiny B}, 0}$ is its 
energy density, for instance 
$\epsilon_{\mbox{\tiny B}, 0} \approx 1$\,eVcm$^{-3}$ for 
$B_{0}$\,=\,6\,$\mu$G, and typically
[$2r_{\mbox{\tiny B}}, 2z_{\mbox{\tiny B}}$]\,$\approx$\,[10, 2]\,kpc
(Strong et al.\ 2000).

On the other hand, the photon gas for the IC process is 
somewhat different from those discussed above. Namely, 
we need the number density of the photon gas in the ISRF, 
$n_{\mbox{\scriptsize ph}}(\vct{r}; E_{\scriptsize \mbox{ph}})$,  as a 
function of the target photon energy
 $E_{\scriptsize \mbox{ph}}$ at $\vct{r}$.
Separating it into two parts, 
a $\vct{r}$-dependent energy-density term,
$\epsilon_{\mbox{\scriptsize ph}}(\vct{r})$,
and a $\vct{r}$-independent term, 
$W_{\mbox{\scriptsize ph}}(k)$ with 
$k=E_{\scriptsize \mbox{ph}}/[k_{\mbox{\tiny B}}T_{\mbox{\scriptsize ph}}]$, 
we rewrite $n_{\mbox{\scriptsize ph}}(\vct{r};
 E_{\scriptsize \mbox{ph}})$ as\\
\vspace{-2mm}
$$
\blankline
E_{\scriptsize \mbox{ph}}n_{\mbox{\scriptsize ph}}(\vct{r};
 E_{\scriptsize \mbox{ph}})dE_{\scriptsize \mbox{ph}} =
\epsilon_{\mbox{\scriptsize ph}}(\vct{r})
W_{\mbox{\scriptsize ph}}(k){d\ln k}, 
\eqno{(\rm 4)}
$$
where $k_{\mbox{\tiny B}}$ is the Boltzmann constant, and 
$T_{\mbox{\scriptsize ph}}$ is the characteristic temperature
of the ISRF. 

There are three main radiation sources in the photon gas, 
(i) the 2.7\,K CMB radiation, 
(ii) stellar radiation with wavelengths of 0.1--10\,$\mu$m
(ultraviolet--visible--near-infrared), and 
(iii) re-emitted radiation from dust grains at 10-1000\,$\mu$m 
(mid-to-far--infrared). 

We classify them further into six wavelength bands, 
each labeled with $i$\,=\,0 for (i), 
$i$\,=\,1, 2, 3 for stellar-1, -2, -3 in (ii), and 
$i$\,=\,4, 5 for dust-1, -2 in (iii) (see Fig.\,2).
Needless to say,
there is no spatial gradient in the CMB ($i$\,=\,0), which is distributed 
uniformly in
space, $\epsilon_{\mbox{\scriptsize ph}}(\vct{r})
\equiv \epsilon_{\mbox{\scriptsize ph}}^{(0)} = 0.261\,\mbox{eVcm$^{-3}$}$,
and the normalized spectrum, $W_{\mbox{\scriptsize ph}}^{(0)}(k)$, is
given by the familiar Planck formula with 
$T_{\mbox{\scriptsize ph}}$\,=\,2.73\,K.

On the other hand, for (ii) and (iii) in the 
wavelength range $\lambda$\,=\,0.1--1000\,$\mu$m, the energy density,
$\epsilon_{\mbox{\scriptsize ph}}^{(i)}(\vct{r})$
($i$\,=\,1--5), must depend on $\vct{r}$, and 
$W_{\mbox{\scriptsize ph}}^{(i)}(k)$ is 
unlike the simple CMB spectrum, and is very
complicated. 
In the following discussions, we often omit the suffix $i$ for simplicity
unless otherwise specified.

In the present work,
we assume a gaussian-type distribution in $\ln k$ for $W_{\mbox{\scriptsize ph}}(k)$,
$$
W_{\mbox{\scriptsize ph}}(k)= 
\frac{1}{\sqrt{2\pi}
 \sigma}\,\mbox{e}^{-(\ln k)^2/(2\sigma^2)};\ \ 
k={{\lambda}_0}/{\lambda},
\eqno{(\rm 5)}
$$
so that the mean radiation intensity, $I_{\mbox{\scriptsize ph}}$, 
is given by
$$
\frac{4\pi {\lambda} I_{\mbox{\scriptsize ph}}(\vct{r}; {\lambda})}
{c \epsilon_{\mbox{\scriptsize ph}}(\vct{r})} = 
\frac{1}{\sqrt{2\pi} \sigma}
\exp \Biggl[-\frac{[\ln ({\it \lambda}_0/{\it \lambda})]^2}
{2\sigma^2}\Biggr],
\eqno{(\rm 6)}
$$
where ${\lambda}_0$ is the peak wavelength for each radiation
with 
$k_{\mbox{\tiny B}}T_{\mbox{\scriptsize ph}} =
 2\pi c\hbar/{\lambda}_0$.

\begin{figure}[t]
  \begin{center}
    \includegraphics[width=7.7cm]{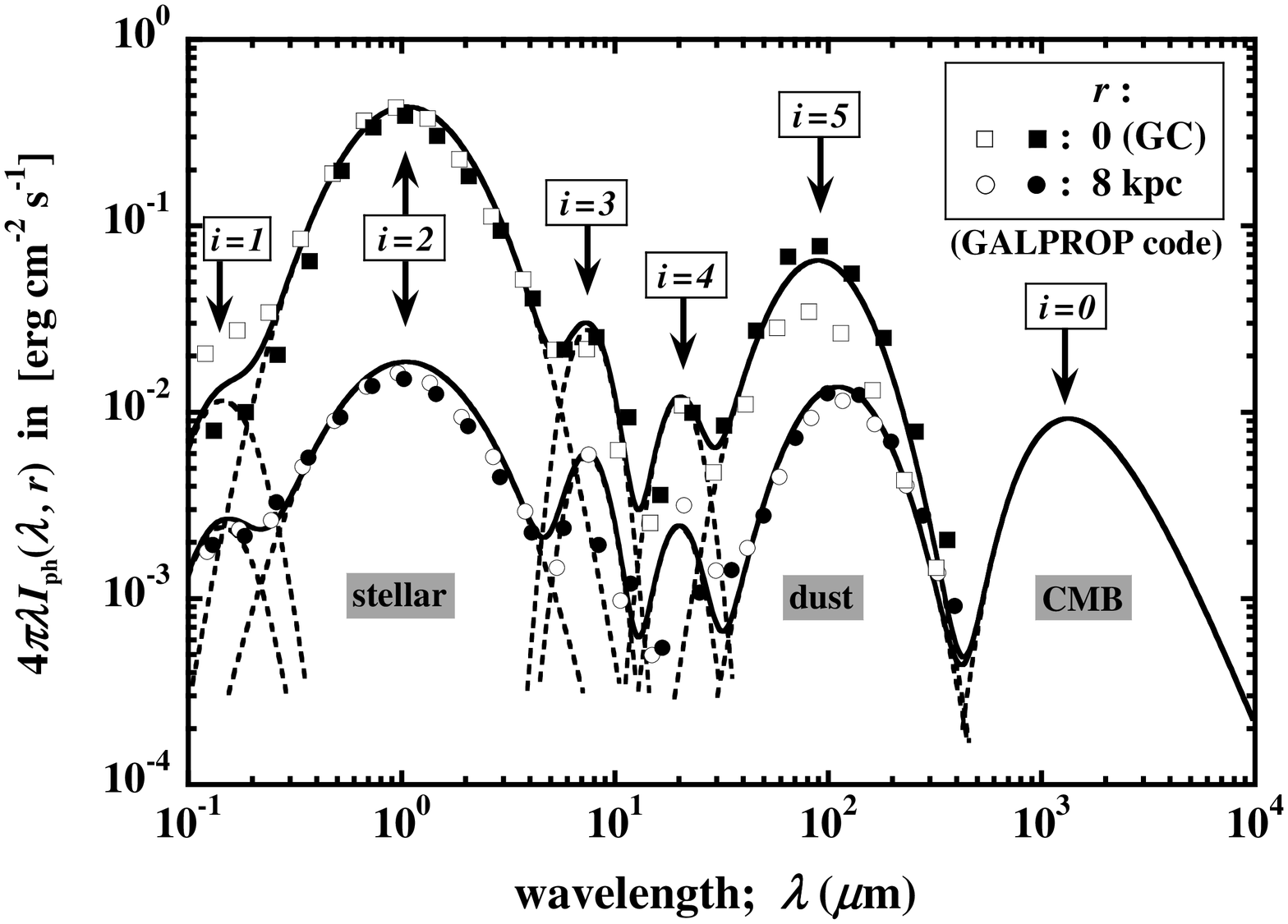}
  \end{center}
  \caption{
Interstellar radiation field (ISRF) at two galactocentric 
  distances  obtained by GALPROP,  $r$\,=\,0 (GC; {\it square symbols}) and 
  8\,kpc (near SS; {\it circle symbols}). 
  Open marks correspond to maximum metalicity gradient, and  filled
  ones to the minimum metalicity gradient.
  Dotted curves are given by equation (6) with parameters summarized in 
  Table~3 for each population $i$, while the solid ones are those
  superposing them.
   CMB radiation ({\it solid curve}) is also shown for reference.
}
\end{figure}

\begin{deluxetable}{crrrrrrl}
\tabletypesize{\small}
\tablecaption{
Summary of the numerical values of [${\lambda}_0^{(i)}$($\mu$m),
$T_i$(K), $\sigma_i$], and those of  
$\epsilon_{\mbox{\scriptsize ph,0}}^{(i)}$(eVcm$^{-3})$ for $r \ge 3$kpc
 in the second line from the bottom, 
 and $\epsilon_{\mbox{\scriptsize ph}}^{(i)}$(eV\,cm$^{-3})$ for $r \le 3$kpc
 in the bottom line, 
 where ${\lambda}_0^{(5)}$, $T_5$(K), and $\sigma_2$ have weak 
 $r$-dependence as shown in remarks with $r$ in kpc, 
 while 
$\epsilon_{\mbox{\scriptsize ph}}^{(i)}$ are independent of $r$ except 
$\epsilon_{\mbox{\scriptsize ph}}^{(2)}$. The numerical value of 
13.9 in
 $\epsilon_{\mbox{\scriptsize ph}}^{(2)}$ corresponds to the energy density 
 of the stellar radiation for the population $i$\,=\,$2$ 
 at GC (see Figure 2). See also equation (7) for 
$\epsilon_{\mbox{\scriptsize ph,0}}^{(i)}$ and 
$\epsilon_{\mbox{\scriptsize ph}}^{(i)}(r)$ for $r \ge 3$kpc.
\label{tbl-3}} 
\tablewidth{0pt}
\tablehead{
   \colhead{}
&
   \colhead{$i=0$}
&  \colhead{$i=1$}
&  \colhead{$i=2$}
&  \colhead{$i=3$}
&  \colhead{$i=4$}
&  \colhead{$i=5$}
&  \colhead{remarks ($r$-dependence)}
} 
\startdata 
$\lambda_0^{(i)}$&
 $1.06(+3)$&  $1.45(-1)$& $1.05(+0)$&
 $7.50(+0)$&  $2.00(+1)$& $9.00(+1)$&\ \ ; 
$\lambda_0^{(5)}(r)\hspace{-0.5mm}= 90\mbox{e}^{r/37.5}$\\
$T_i$&
 $2.73(+0)$&  $9.92(+4)$& $1.37(+4)$&
 $1.92(+3)$&  $7.19(+2)$& $1.60(+2)$&\ \ ; 
$T_5(r)\,\,\,= 160\mbox{e}^{-r/37.5}$\\
$\sigma_i$&------------&
 $3.39(-1)$ & $6.10(-1)$&
 $2.20(-1)$ & $2.19(-1)$& $4.72(-1)$&\ \ ; 
$\sigma_2(r)\,\,\,= 0.61\mbox{e}^{r/87.0}$\\ 
$\epsilon_{\mbox{\scriptsize ph,0}}^{(i)}$&
 $2.61(-1)$&  $5.20(-1)$& $7.92(+0)$&
 $8.10(-1)$&  $3.40(-1)$& $4.10(+0)$&\ \ ; 
 $\epsilon_{\mbox{\scriptsize ph}}^{(i)}(r)\,= 
\epsilon_{\mbox{\scriptsize ph,0}}^{(i)}\mbox{e}^{-r/3.2}$
\\
$\epsilon_{\mbox{\scriptsize ph}}^{(i)}$&
 $2.61(-1)$&  $2.03(-1)$& $1.39(+1)$&
 $3.17(-1)$&  $1.33(-1)$& $1.61(+0)$&\ \ ; 
 $\epsilon_{\mbox{\scriptsize ph}}^{(2)}(r)\,= 13.9\mbox{e}^{-r/2.0}$
 \enddata
\end{deluxetable}

In Figure~2, we present examples of the mean radiation intensity 
(multiplied by $4\pi {\lambda}$) 
for the maximal metalicity gradient ({\it filled symbols}) and 
no metalicity gradient ({\it open symbols}) 
at two radial distances, $r$\,=\,0 ({\it squares}) and 
8\,kpc ({\it circles})
in the GP given by GALPROP, 
where also drawn are curves expected from the right-hand side
of equation (6) for $W_{\mbox{\scriptsize ph}}(k)$,
assuming
$$
\vspace{1mm}
\epsilon_{\mbox{\scriptsize ph}}(\vct{r}) =
\epsilon_{\mbox{\scriptsize ph},0}
\exp [-(r/r_{\mbox{\scriptsize ph}} + |z|/z_{\mbox{\scriptsize ph}})],
\eqno{(\rm 7)}
$$
for $r \ge 3$\,kpc, and see caption of Table~3 otherwise.
In Table~3 we summarize numerical values of
[$\lambda_0^{(i)}, T_i, \sigma_i;\,
\epsilon_{\mbox{\scriptsize ph,0}}^{(i)}$] 
($i$\,=\,0--5) with
$r_{\mbox{\scriptsize ph}}$\,=\,3.2\,kpc irrespective of the population $i$, 
and also presented \blankline are those of 
$\epsilon_{\mbox{\scriptsize ph}}^{(i)}(r)$ for $r \le 3$\,kpc, while they
are independent of $r$ except $\epsilon_{\mbox{\scriptsize ph}}^{(2)}(r)$.

Let us demonstrate the energy density separately for the stellar and the dust
radiation,  $\sum_{i=1}^{3}\epsilon_{\mbox{\scriptsize ph}}^{(i)}$ 
and $\sum_{i=4}^{5}\epsilon_{\mbox{\scriptsize ph}}^{(i)}$ respectively,
against the galactocentric distance $r$ in Figure 3, 
where we plot also
numerical data given by Mathis et al.\ (1983; {\it filled grey symbols}).
Two curves for the stellar emission and the dust re-emission are
drawn by the use of the parameterization summarized in Table~3,
where we do not take the difference in the choice of
metalicity gradient into account, as it is effective only near the
GC for the dust re-emission and is approximately one order of magnitude
smaller than the stellar radiation.

For the latitudinal
scale height, $z_{\mbox{\scriptsize ph}}$, in equation (7), we assume
$z_{\mbox{\scriptsize ph}}$\,$\approx$\,$r_{\mbox{\scriptsize ph}}/8$\,=\,0.4\,kpc,
referring to the speculation by Freudenreich (1998) based on the 
DIRBE (Diffused Infrared Background Experiment) survey, while 
the surveys of the diffuse FIR/sub-mm emission for the latitudinal
direction at various radial distances $r$ are not sufficient
to construct a reliable model.

\begin{figure}[!t]
\vspace{7mm}
  \begin{center}
    \includegraphics[width=7.8cm]{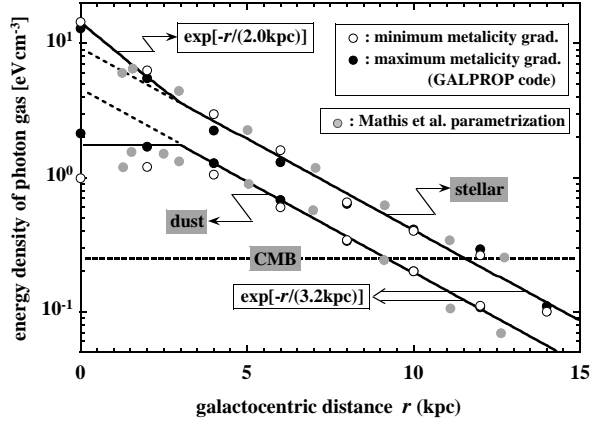}
  \end{center}
  \caption{ISRF energy density as a function of galactocentric distance $r$
  at $z=0$ for the stellar radiation and the re-emission from dust
 grains, each with 
 the maximum metalicity gradient ({\it filled black circles})
  and  minimum metalicity gradient ({\it open circles}), 
  where also shown are those obtained by 
 Mathis et al.\ (1983) ({\it filled grey circles}).
 {\it Solid curves} are 
  the empirical ones obtained by equation (7).}
\end{figure}
\vspace{3mm}

\section{Energy loss and gain}

\subsection{Energy loss in ISM and ISRF}

The energy loss processes for the electron component are dramatically
different from those for 
 the hadronic components, 
with four main processes:
bremsstrahlung ($\equiv \lq \lq$rad"), ionization ($\equiv \lq \lq$ion"),  
synchrotron and IC (together $\equiv \lq \lq$sic").
For the bremsstrahlung (Koch \& Motz 1959; Gould 1969; Ginzburg 1979),
\vspace{1mm}
$$
-\frac{1}{E_e}\biggl\langle \frac{{\it \Delta}E_e}{{\it \Delta} t} 
\biggr\rangle_{\hspace{-0.5mm}\mbox{\scriptsize rad}}
\hspace{-1mm} \simeq 
n (\vct{r}) w_{\mbox{\scriptsize rad}} (E_e)\hspace{-0.5mm}
\biggl[1 +\hspace{0.5mm} {\LARGE O}
\biggl(\hspace{-0.5mm}\frac{n_{\tiny 
\mbox{HII}}}{n}\hspace{-1mm}\biggr)\hspace{-0.8mm}\biggr], 
\eqno{(8)}
$$
with
$$
\vspace{2mm}
n(\vct{r}) =n_{\tiny \mbox{HI}}(\vct{r}) + 
n_{\tiny \mbox{HII}}(\vct{r})+ n_{\tiny \mbox{H$_2$}}(\vct{r}),
\eqno{(9)}
$$
where $w_{\mbox{\scriptsize rad}}(E_e \gg m_ec^2)$ $\equiv$ $w_{\mbox{\scriptsize rad}}^{(\infty)}$ $\approx$ $7.30 \times10^{-16}$ cm$^3$s$^{-1}$,
independent of $E_e$ 
with the complete screening cross-section in the 
high energy region; 
see Appendix~A for the  explicit forms of $w_{\mbox{\scriptsize rad}} (E_e)$, 
and \S\,2.1 for $n_h$, ($\lq \lq h" \equiv \mbox{H\,I, H\,II, H$_2$}$).

Similarly for the ionization, we use the 
Bethe-Bloch formula (Ginzburg 1979)
$$
-\biggl\langle \frac{{\it \Delta}E_e}{{\it \Delta} t} 
\biggr\rangle_{\hspace{-0.5mm}\mbox{\scriptsize ion}}\hspace{-0.5mm}
\simeq 
n(\vct{r}) w_{\mbox{\scriptsize ion}} (E_e)
\biggl[1 + {\LARGE O}
\biggl(\hspace{-0.5mm}\frac{n_{\tiny 
\mbox{HII}}}{n}\hspace{-0.8mm}\biggr)\biggr], 
\eqno{(10)}
$$
with
$$ 
w_{\mbox{\scriptsize ion}}(E_e) = 
w_{\mbox{\scriptsize ion}}^{(0)}\,\Bigl \{\ln [E_e/\mbox{GeV}] + 13.8\Bigr  \},
$$
and
$  w_{\mbox{\scriptsize ion}}^{(0)} = 0.229\times 
\mbox{$10^{-16}$\,cm$^3$s$^{-1}$}$.

On the other hand, the energy losses due to the synchrotron 
 (abbreviated as $\lq \lq$SY" for subscripts appearing in
  the following equations) 
 and IC are rather complicated,
in addition to the energy dependent cross-section of the Klein-Nishina
formula, 
$$
\biggl\langle \frac{{\it \Delta}E_e}{{\it \Delta} t} 
\biggr\rangle_{\hspace{-0.5mm}\mbox{\scriptsize sic}}
= 
\biggl\langle \frac{{\it \Delta}E_e}{{\it \Delta} t} 
\biggr\rangle_{\hspace{-0.5mm}\mbox{\scriptsize SY}}
+
\biggl\langle \frac{{\it \Delta}E_e}{{\it \Delta} t} 
\biggr\rangle_{\hspace{-0.5mm}\mbox{\scriptsize IC}},
\eqno{(\rm 11)}
$$
where
$$
\vspace{1.5mm}
-\frac{1}{w_{\tiny \mbox{T}}E_e^2}
\biggl\langle \frac{{\it \Delta}E_e}{{\it \Delta} t} 
\biggr\rangle_{\hspace{-0.5mm}\mbox{\scriptsize SY}}
= 
\epsilon_{\mbox{\tiny B}}(\vct{r}),
\eqno{(\rm 12a)}
$$
$$
\vspace{2mm}
-\frac{1}{w_{\tiny \mbox{T}}E_e^2}
\biggl\langle \frac{{\it \Delta}E_e}{{\it \Delta} t} 
\biggr\rangle_{\hspace{-0.5mm}\mbox{\scriptsize IC}}
= 
\sum_{i=0}^{5}\epsilon_{\mbox{\scriptsize ph}}^{(i)}(\vct{r})
{\it \Lambda}(E_e, T_i),
\eqno{(\rm 12b)}
$$
with
$w_{\mbox{\tiny T}} = 1.018 \times 10^{-16}\,\mbox{cm}^3\mbox{s}^{-1}$.
See \S\,2.2 and Table~3 for
$[T_i; \epsilon_{\mbox{\scriptsize ph}}^{(i)}(\vct{r})]$ ($i$\,=\,0--5),
and ${\it \Lambda}(E_e, T_i)$ is given by equation (A6)
in Appendix~A, 
which comes from the Klein-Nishina cross-section.
In Figure~4, we present the energy loss divided by $E_e^2$ 
at the SS against
$E_e$ separately for individual (virtual) photon fields
as well as for superposed ones, \vspace{5mm}
$-\langle {\it \Delta}E_e/{\it \Delta} t 
\rangle_{\mbox{\scriptsize sic}}^\odot/E_e^2$,
where we assume $B_{\perp}$\,=\,5\,$\mu$G, corresponding to
$\epsilon_{\mbox{\tiny B}}^\odot$ = 0.93\,eVcm$^{-3}$, for the magnetic 
field, and use 
$\epsilon_{\mbox{\scriptsize ph}}^{(i)}({r}_\odot)$ 
presented in the second line from the bottom of 
Table~3 with $r$\,=\,$r_\odot$ for the photon gas field. 

\begin{figure}[t]
  \begin{center}
    \includegraphics[width=7.7cm]{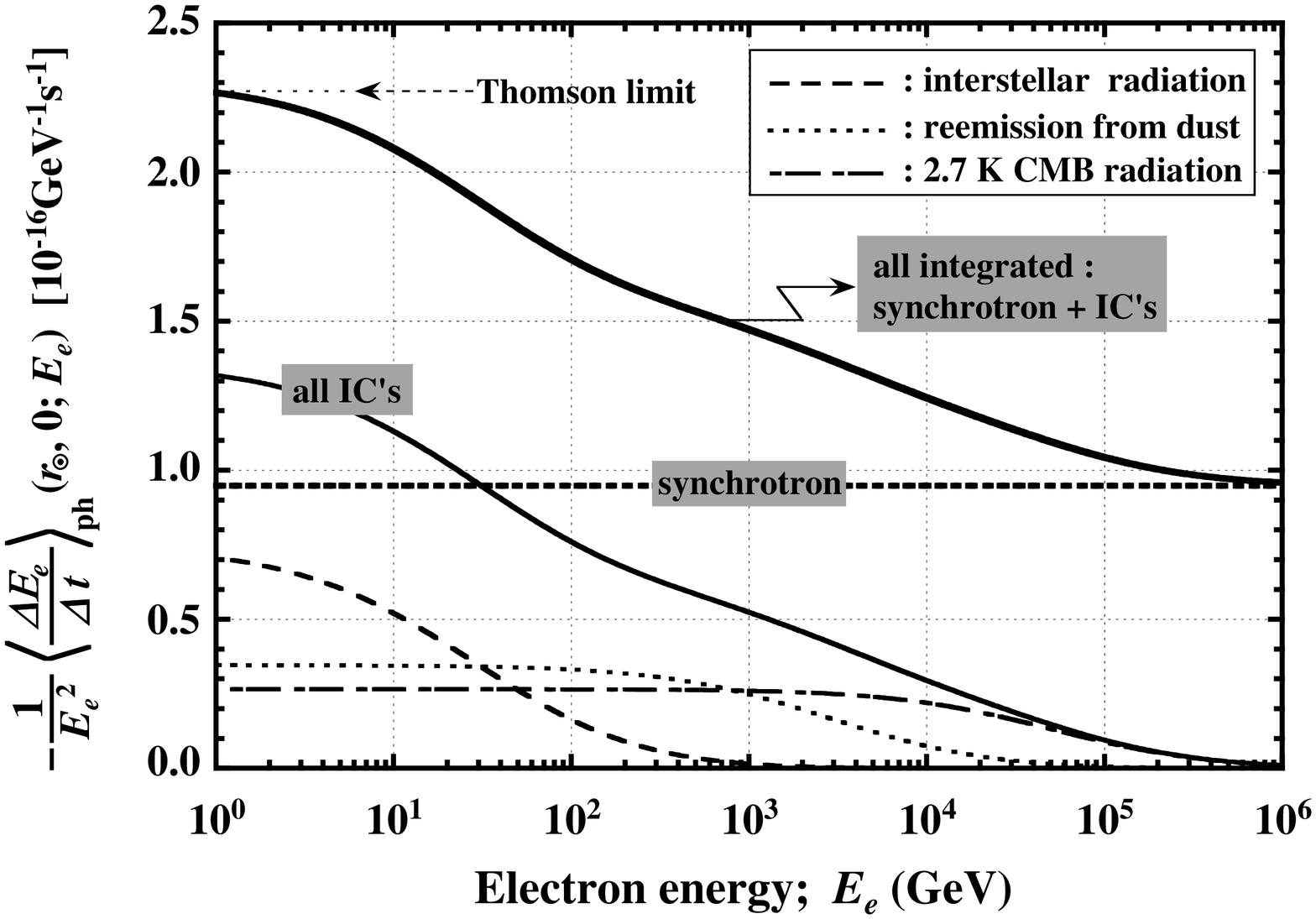}
  \end{center}
  \caption{
Energy losses per unit time of CR electrons in ISRF at 
  SS ($r_\odot$\,=\,8.5\,kpc),  shown separately  
  for four components, synchrotron ({\it heavy dotted line}),
  IC's by stellar radiation ({\it broken curve}),
  by re-emission from dust grains ({\it dotted curve}), and 
  by CMB ({\it broken dotted curve}), together with the sums,
  IC-all $\equiv$ IC-stellar + IC-dust + IC-CMB ({\it thin solid curve}),
  and synchrotron + IC-all ({\it heavy solid curve}).\blankline
}
\end{figure}

For $E_e$\,$\lsim$\,1\,GeV, 
${\it \Lambda}(E_e, T_i)$\,$\approx$\,1, i.e., the Thomson cross-section
is valid, so that 
equation (11) is separable in $\vct{r}$ and $E_e$, leading to a
simple expression, 
${\epsilon}(\vct{r}) w_{\tiny \mbox{T}}E_e^2$, with 
${\epsilon}(\vct{r}) = \epsilon_{\mbox{\tiny B}}(\vct{r})
 + \sum_{i=0}^{5}\epsilon_{\mbox{\scriptsize ph}}^{(i)}(\vct{r})$. 
In practice, we find it is well reproduced by the following form 
over a wide energy range 
$$
\vspace{1mm}
-\biggl\langle \frac{{\it \Delta}E_e}{{\it \Delta} t} 
\biggr\rangle_{\hspace{-0.5mm}\mbox{\scriptsize sic}}
 \simeq 
{\epsilon}(\vct{r}) {w}_{\mbox{\tiny T}}E_e^{2-\delta};\ \ \ 
\delta = 0.075, 
\eqno{(\rm 13)}
$$
while ${\epsilon}(\vct{r})$ depends on $E_e$ very weakly.

In Figure~5 we demonstrate the energy loss of individual processes separately,
those due to $\lq \lq$rad", 
$\lq \lq$ion" and $\lq \lq$sic"
at the SS against the kinetic energy of the electron $E_e$
with $\epsilon_\odot$\,=\,2\,eVcm$^{-3}$ and 
$n_\odot$\,=\,1\,H\,atoms\,cm$^{-3}$, where
we plot the above empirical relationship (13) 
({\it dotted curve}) and the energy gain due to the 
reacceleration ($\equiv \lq\lq$rea"; see next subsection) 
together. One finds that it reproduces satisfactorily the exact one (11)
 with equations (12a) and (12b).

\begin{figure}[t]
  \begin{center}
    \includegraphics[width=7.7cm]{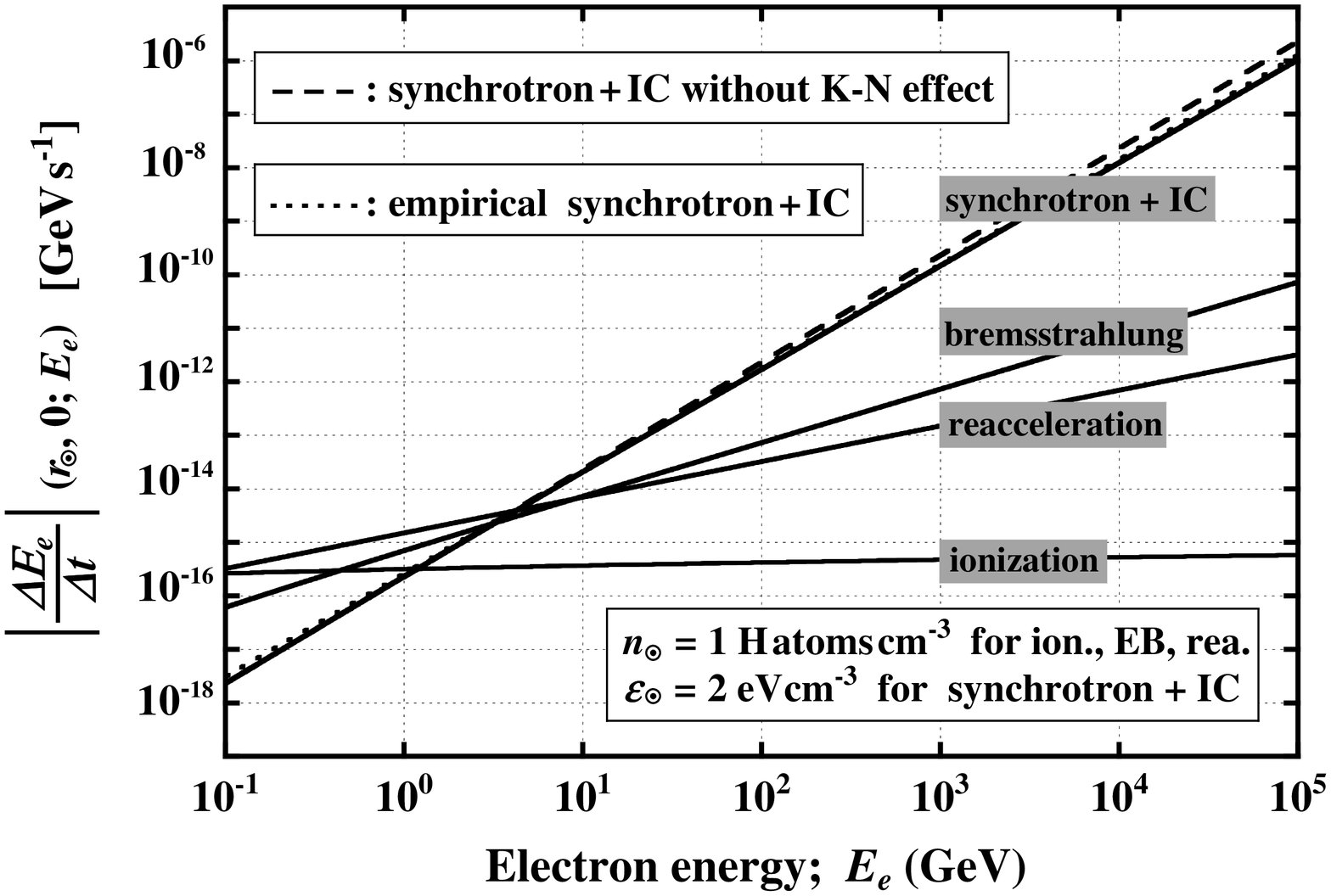}
  \end{center}
  \caption{
Energy losses per unit time of CR electrons in 
   ISRF and ISM at SS ($r = r_\odot$) 
as a function of electron energy, shown
   separately for four processes, synchrotron\,+\,IC, bremsstrahlung, 
   reacceleration, and ionization.
   We present three curves for synchrotron\,+\,IC, with the solid one from 
the Klein-Nishina cross-section, the broken one from 
   the Thomson cross-section,
   and the dotted one from the empirical one.\vspace{-1mm}
}
\end{figure}

\subsection{Energy gain due to the reacceleration}

In Paper~II, we present the energy gain per unit time due to the 
 reacceleration \blankline
$$
\frac{1}{E_e}
\biggl\langle \frac{{\it \Delta}E_e}{{\it \Delta} t} 
\biggr\rangle_{\hspace{-0.5mm}\mbox{\scriptsize rea}}
= {n}(\vct{r}) w_{\mbox {\scriptsize rea}} [E_e/\mbox{GeV}]^{-\alpha},
\eqno{(14)}
$$
with
$$
\blankline
w_{\mbox {\scriptsize rea}}=c\zeta_0;\ \ \zeta_0 \approx
\frac{4}{9} 
\frac{v^{\hspace{0.7mm}2}_{\mbox{\tiny \hspace{-0.4mm}$M$}}}
{{n}_0^* c D_0^*},
\eqno{(15)}
$$
where
$w_{\mbox {\scriptsize rea}}$\,=\,15.0$\times$10$^{-16}$\,cm$^3$\,s$^{-1}$
in the case of, for instance, $\zeta_0$\,=\,50\,millibarn (mbarn), 
corresponding to the choice of a parameter set with
$v_{\mbox{\tiny \hspace{-0.2mm}$M$}}$\,=\,20--30km\,s$^{-1}$
 (Alfv$\acute{\mbox{e}}$n velocity),
${n}_0^*$\,=\,0.06--0.14H\,atoms\,cm$^{-3}$, and
$D_0^*$\,=\,2$\times$10$^{28}$cm$^{2}$s$^{-1}$. The smallness of the gas
density with ${n}_0^* \ll 1$ H\,atoms cm$^{-3}$ indicates that the
 reacceleration process occurs even at some distance from the GP.

The fluctuation in the energy gain due to 
the reacceleration is given (Gaisser 1990; Paper~II) by \blankline
$$
\frac{1}{E_e^2}
\biggl\langle \frac{{\it \Delta}E_e^{\,2}}{{\it \Delta} t} 
\biggr\rangle_{\hspace{-0.5mm}\mbox{\scriptsize rea}}
= \frac{1}{2} 
{n}(\vct{r}) w_{\mbox {\scriptsize rea}}  [E_e/\mbox{GeV}]^{-\alpha}.
\eqno{(16)}
$$

\subsection{Total energy loss and gain}

As discussed in the last two subsections, 
we have the total average energy-loss and the energy-gain per unit
time 
$$ 
- \biggl\langle \frac{{\it \Delta}E_e}{{\it \Delta} t}  
\biggl\rangle_{\hspace{-0.2mm}\mbox{\scriptsize all}}\ = \ 
n(\vct{r}) {\cal W}_n(E_e)   +
\epsilon(\vct{r}) {\cal W}_\epsilon(E_e), 
\eqno{(\rm 17)}
$$
where
$$
\hspace{-0.6mm}
{\cal W}_n(E_e)\hspace{-0.5mm} =\hspace{-0.5mm}
 w_{\mbox{\scriptsize rad}}(E_e)E_e \hspace{-0.1mm}+
{w}_{\mbox{\scriptsize ion}}(E_e) \hspace{-0.1mm}-
{w}_{\mbox{\scriptsize rea}}E_e^{1-\alpha}\hspace{-0.8mm},
\eqno{(\rm 18a)}
$$
and 
$$
\vspace{1mm}
{\cal W}_\epsilon(E_e) \simeq 
w_{\mbox{\tiny T}} E_e^{2-\delta};\ \ 
 \delta = 0.075.
\eqno{(\rm 18b)}
$$

\begin{figure}[t]
  \begin{center}
    \includegraphics[width=7.5cm]{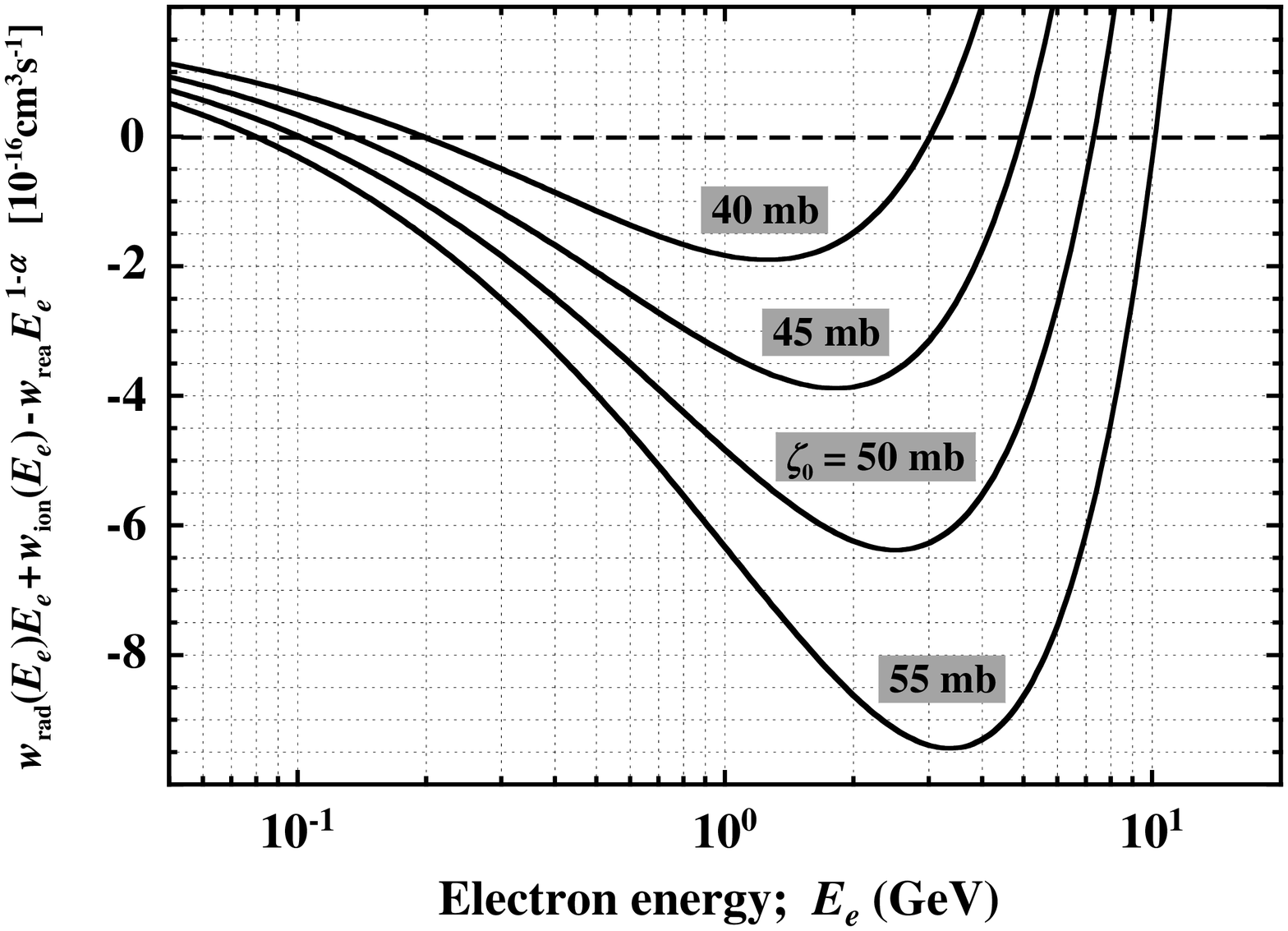}
  \end{center}
  \caption{
 Numerical values of 
${\cal W}_n(E_e) \equiv w_{\mbox{\scriptsize rad}}(E_e)E_e +
{w}_{\mbox{\scriptsize ion}}(E_e) -
{w}_{\mbox{\scriptsize rea}}E_e^{1-\alpha}$
 against $E_e$ for 
 $\zeta_0$\,=\,40--55\,mbarn, where the screening effect for 
 $w_{\scriptsize \mbox{rad}}$ is taken into account.
}
\end{figure}

One might note from equation (18a) that there exist two energies,
${E}_c^{-}$ and ${E}_c^{+}$, 
at which the first term proportional to $n(\vct{r})$ in the right-hand of 
 equation (17) becomes null. In Figure 6, we demonstrate $
{\cal W}_n(E_e)$ against several choices of $\zeta_0$, and find 
[${E}_c^{-}, {E}_c^{+}$] $\approx$ [0.1, 7]\,GeV 
in the case of $\zeta_0$\,=\,50\,mbarn. Namely,
the synchrotron-IC is dominant for $E_e$\,$\gsim$\,${E}_c^{+}$, while 
the reacceleration is effective for 
${E}_c^{-}$\,$\lsim$\,$E_e$\,$\lsim$\,${E}_c^{+}$, and the ionization
 for $E_e$\,$\lsim$\,${E}_c^{-}$.

As discussed in \S\,2, 
the total number density of the ISM gas,
$n(\vct{r})$, and the total energy density of the ISRF, $\epsilon(\vct{r})$,  
have complicated spatial
distributions coming from
local irregularities, which are not yet well established.
On the other hand, 
in our previous papers, we have assumed 
a simple exponential-type form for $n(\vct{r})$, 
smearing out the local irregularities,
$$
\bar{n}(\vct{r}) = \bar{n}_0 \exp [-(r/r_n + |z|/z_n)] ,
\eqno{\rm (19)}
$$
greatly simplifying the complicated distributions given by equations
(1), (2) with Tables 1, 2,
where $\bar{n}_0$ is the ({\it interpolated}) average gas
density with approximately 1.5\,H\,atoms\,cm$^{-3}$
at the GC, and $[r_n, z_n]$ $\approx$ [20, 0.2]\,kpc.

In spite of such a simplification, we have found that our model
reproduces remarkably well the experimental data on 
{\it hadronic components}. This tells us that {\it charged} CR components 
are well mixed during their propagation in the Galaxy over a residence time 
of approximately 10$^7$\,yr, effectively smearing the local inhomogeneous 
structure of the  ISM.
In fact, it is well established that the anisotropy amplitude
of CRs is of the level of at most 10$^{-3}$
at energies of 1--100\,TeV (Sakakibara
1965; Nagashima et al.\ 1989; Cutler \& Groom 1991).
This is the reason why even the simplest leaky-box model and/or the
simplified diffusion model such as, for instance, constant gas density
and constant diffusion coefficient without spatial gradient,
reproduces the CR hadronic components so well (Berezinskii et al.\ 1990).

Now, corresponding to the simplification (19) for $n(\vct{r})$, we assume 
the following simple exponential type form  
for $\epsilon(\vct{r})$  as well
$$
\bar{\epsilon}(\vct{r}) = \bar{\epsilon}_0
 \exp [-(r/r_\epsilon + |z|/z_\epsilon)] ,
\eqno{\rm (20)}
$$
where $\bar{\epsilon}_0$ is the ({\it interpolated}) average energy
density of the ISRF at the GC, and two parameters,  
$r_\epsilon$ and $z_\epsilon$, correspond to the scale heights
for the spatial gradients, almost independent of the energy. 
Typically $[\bar{\epsilon}_0;\ r_\epsilon, z_\epsilon] \approx 
$ [16\,eVcm$^{-3}$; 4\,kpc, 0.75\,kpc] (Ishikawa 2010).

However, while the simplifications given by 
equations (19) and (20) are applied for electrons (and
hadrons), we stress here that
those presented in \S\,2
are actually used for D$\gamma$'s as discussed in \S\,6,
namely $n_h(\vct{r})$ with 
$\lq \lq h" \equiv \mbox{H\,I, H\,II, H$_2$}$
for $n(\vct{r})$,  and 
$\epsilon_{\mbox{\scriptsize ph}}^{(i)}(\vct{r})$ ($i$\,=\,0--5) for 
$\epsilon(\vct{r})$ with weak energy dependences in 
$\epsilon_{\mbox{\scriptsize ph,0}}^{(i)}$ as presented in Table\,3.
This is because
D$\gamma$'s produced by CR hadrons and electrons are directly 
affected by the environment of ISM and ISRF
around the birth site of the produced $\gamma$'s.
\\

\section{Diffusion equation for electron component}

\subsection{Basic equation}

The transport equation for the electron density, 
$N_e({\vct{r}}; E_e, t)$,  
is given by (Berezinskii et al.\ 1990),
$$
\hspace{-1.3mm}
\biggl[\hspace{-0.2mm}\frac{\partial}{\partial t} -
{\it \nabla}\cdot D({\vct{r}};
 E_e){\it \nabla}\hspace{-0.2mm}+\hspace{-0.2mm}
{\it \Delta}_E\hspace{-0.2mm}\biggr] {\cdot} N_e(\vct{r}; E_e, t)
\hspace{-0.3mm}=\hspace{-0.3mm}Q(\vct{r}; E_e, t), 
\eqno{(21)}
$$
with
$$
\blankline
{\it \Delta}_E =   \frac{\partial}{\partial E_e}
\Biggl\langle \frac{{\it \Delta}E_e}{{\it \Delta} t} 
 \Biggr\rangle_{\hspace{-0.2mm}\mbox{\scriptsize all}}
 - \frac{1}{2} \frac{\partial^2}{\partial E_e^2}
\Biggl\langle \frac{{\it \Delta}E_e^{\,2}}{{\it \Delta} t} 
\Biggr\rangle_{\hspace{-0.2mm}\mbox{\scriptsize rea}},
\eqno{(22)}
$$
see equations (19) and (20) for the {\it average} energy-loss (-gain) 
in the all processes, with the replacement of
$[n(\vct{r}), \epsilon(\vct{r})]$  in equation (17) by 
$[\bar{n}(\vct{r}), \bar{\epsilon}(\vct{r})]$, and equation (16) for 
the fluctuation of the energy gain in the 
 reacceleration process respectively.
For the diffusion coefficient and the
source spectrum, we assume  (note $v \approx c$, and $R_e \approx E_e$)
\vspace{1mm}  
$$
\hspace{-1mm}
D({\vct{r}};\hspace{-0.5mm} E_e)\hspace{-0.5mm} = \hspace{-0.5mm}
 E_e^{\alpha} D(\vct{r}), \  \ 
Q({\vct{r}};\hspace{-0.5mm} E_e, t)\hspace{-0.5mm} =\hspace{-0.5mm}
  E_e^{-\gamma} Q(\vct{r};\hspace{-0.2mm} t),
\eqno{(\rm 23)}
$$
with
$$
D(\vct{r}) = D_0 \exp (r/\rD + |z|/\zD),
\eqno{(24a)}
$$
$$
\blankline
Q(\vct{r}; t) = Q_0(t) \exp [-(r/\rQ + |z|/\zQ)].
\eqno{(24b)}
$$
In Table~4, we summarize parameters related to the scale heights,
$\rD, \zD, \ldots\ $, which often appear in the present paper.

\begin{deluxetable}{llll}
\tabletypesize{\small}
\tablecaption{
Summary of parameters often appearing in our propagation model, 
 classifying them into two groups, one related to the
 gas density of ISM, $\bar{n}(\vct{r})$, and another to the
 energy density of ISRF, $\bar{\epsilon}(\vct{r})$.
\label{tbl-4}} 
\tablewidth{0pt}
\tablehead{
  \colhead{Parameters for ISM } &
  \colhead{Typical values for ISM} &
  \colhead{Parameters for ISRF} &
  \colhead{Typical values for ISRF} 
} 
\startdata 
\vspace{-1mm}
\\ 
 $\displaystyle \nu = \frac{1}{1\,+\,\zD/z_n}$ &
 $\nu = 0.04-0.06$ &
$\displaystyle \kappa = \frac{1}{1\,+\,\zD/z_\epsilon}$ &
 $\kappa = 0.15-0.21$
\\ \\
 $\displaystyle U_\nu = 2\sqrt{\nu+\nu^2}$ &
 $U_\nu = 0.4-0.5$ &
 $\displaystyle U_\kappa = 2\sqrt{\kappa+\kappa^2}$ &
 $U_\kappa = 0.8-1.2$
\\ \\
$ \displaystyle \frac{1}{\bar{r}_n} = 
\frac{1}{2}\biggl(\frac{1}{\rD}\,+\,\frac{1}{r_n}\biggr)$ &
$\bar{r}_n = [20-40]$kpc  & 
$ \displaystyle \frac{1}{\bar{r}_\epsilon} = 
\frac{1}{2}\biggl(\frac{1}{\rD}\,+\,\frac{1}{r_\epsilon}\biggr)$ &
$\bar{r}_\epsilon = [5-10]$kpc
\\ \\ 
$ \displaystyle \frac{1}{\bar{z}_n} = 
\frac{1}{2}\biggl(\frac{1}{\zD}\,+\,\frac{1}{z_n}\biggr)$ &
$\bar{z}_n = [0.3-0.5]$kpc            & 
$ \displaystyle \frac{1}{\bar{z}_\epsilon} = 
\frac{1}{2}\biggl(\frac{1}{\zD}\,+\,\frac{1}{z_\epsilon}\biggr)$ &
$\bar{z}_\epsilon = [1.0-1.4]$kpc
\\ \\
 $\displaystyle \omega_\nu = \biggl(
 \frac{1}{\zQ} - \frac{1}{2z_n}\biggr)\bar{z}_n$ &
 $\omega_\nu =0.8-1.2$ &
$\displaystyle \omega_\kappa = \biggl(
 \frac{1}{\zQ} - \frac{1}{2z_\epsilon}\biggr)\bar{z}_\epsilon$ &
 $\omega_\kappa = 4.5-5.5$
\\ \\
 \enddata
\end{deluxetable}

Now, remembering ${\cal W}_\epsilon(E_e)$\,$\gg$\,${\cal W}_n(E_e)$ in the
high energy (HE) region, 
say, $E_e$\,$\gsim$\,${E}_c^{+}$ ($\approx$\,7\,GeV), 
and vice versa in the low energy (LE) region, $E_e$\,$\lsim$\,${E}_c^{+}$, 
the energy loss given by equation (17) is written as
\begin{displaymath}
\hspace{3cm}
- \Biggl\langle \frac{{\it \Delta}E_e}{{\it \Delta} t}  
\Biggl\rangle_{\hspace{-0.2mm}\mbox{\scriptsize all}}  
\simeq  \left\{\begin{array}{ll}
\hspace{-2mm}  \bar{\epsilon}(\vct{r}){\cal W}_\epsilon(E_e) + 
{\LARGE O}[\bar{n}(\vct{r}) {\cal W}_n(E_e)] ;
  \ \mbox{$E_e\,\gsim\, {E}_c^{+}$},\hspace{2.2cm} ({\rm 25a})
\\ 
\\
 \hspace{-2mm}\bar{n}(\vct{r}) {\cal W}_n(E_e) +
{\LARGE O}[\bar{\epsilon}(\vct{r}){\cal W}_\epsilon(E_e)] ;
 \ \mbox{$E_e\, \lsim\, {E}_c^{+}$},\hspace{2.2cm} ({\rm 25b})
                       \end{array} 
                       \right.
\vspace{1mm}
\end{displaymath}
so that in the following discussion, we give first 
the solution of the diffusion
equation (21) in the HE region, regarding 
$\bar{n} {\cal W}_n$ as a perturbative term, 
where we can neglect the fluctuation term
due to the reacceleration.
  Next we give the solution 
in the LE region, regarding 
$\bar{\epsilon}{\cal W}_\epsilon$ as a
perturbative term by contrast, which 
is completely the same as the former one after replacing 
[$\bar{\epsilon}, {\cal W}_\epsilon$] with
[$\bar{n}, {\cal W}_n$] (and vice versa), while we have to take the
fluctuation term, 
$\langle {\it \Delta}E_e^{\,2} \rangle_{\mbox{\scriptsize rea}}$, 
into account in this case.

Thus for the steady state ($\partial/\partial t = 0$), 
the solution of equation (21) in the HE region is devided into three
$$
\blankline
N_{e,\epsilon} \simeq
N_{e,\epsilon}^{(0)} + 
\tilde{N}_{e,n}^{(0)} + 
N_{e,\epsilon}^{(1)},
\eqno{\rm (26a)}
$$
where the first term is a principal one coming from
$\bar{\epsilon}{\cal W}_\epsilon$, 
 the second term corresponds to the perturbative term from 
$\bar{n}{\cal W}_n$, and the third term 
 to the fluctuation due to the reacceleration given by the second term
 of the right-hand in equation (22), while it is negligible
 in practice, $N_{e,\epsilon}^{(1)} \approx 0$. 

The solution in the LE region
 is similarly given by replacing the suffix $\epsilon$ with $n$ 
(and vice versa), but we can not neglect the fluctuation term $N_{e,n}^{(1)}$
 in contrast,
$$
N_{e,n} \simeq
N_{e,n}^{(0)} + 
\tilde{N}_{e,\epsilon}^{(0)} + 
N_{e,n}^{(1)}.
\eqno{\rm (26b)}
$$

The first term in equation (26a) is written immediately as
$$
N_{e,\epsilon}^{(0)}({\vct{r}}; E_e) = 
\int_0^{\infty} \! \!
{\it \Pi}_\epsilon^{(0)} ({\vct{r}}; y)f_\epsilon^{(0)}(y; E_e)dy,
\eqno{(27)}
$$
where
${\it \Pi}_\epsilon^{(0)}$ and $f_\epsilon^{(0)}$ satisfy,
$$
\hspace{-0.5mm}
\Bigl[\bar{\epsilon}(\vct{r}) c\frac{\partial}{\partial y}
\hspace{-0.5mm} -\hspace{-0.5mm}
{\it \nabla}\!\cdot\!D({\vct{r}}){\it \nabla}\hspace{-0.5mm} \Bigr]{\cdot}
{\it \Pi}_\epsilon^{(0)} ({\vct{r}}; y)\hspace{-0.5mm}
=\hspace{-0.5mm} Q(\vct{r})\hspace{-0.5mm} \delta(y),
\eqno{\rm (28a)}
$$
$$
\vspace{2mm}
\Bigl[ c E_e^{\alpha} \frac{\partial}{\partial y} - 
\frac{\partial}{\partial E_e}{\cal W}_\epsilon (E_e)
 \Bigr] {\cdot} f_\epsilon^{(0)}(y; E_e) = 0, 
\eqno{\rm (28b)}
$$
with $f_\epsilon^{(0)}(0; E_e) = E_e^{-\gamma -\alpha}$. 
\vspace{2mm}
\subsection{Solution in the steady state}

It is possible to solve exactly equation (28a) with use of the
procedure presented in Paper~I, after replacing 
$\bar{n}(\vct{r})$ by $\bar{\epsilon}(\vct{r})$, and we
present here only the critical term related to $(r, z;\ y)$,
omitting constant terms such as $Q_0$ and $\bar{\epsilon}_0$
(see Appendix~B for the full form),
$$
{\it \Pi}_\epsilon^{(0)} (\vct{r}; y) \propto
 \exp [-\bar{s}_r y - |z|/\zD], 
\eqno{\rm (29)}
$$
$$
\vspace{2mm}
\bar{s}_r  \simeq 
      \frac{D_r}{\bar{\epsilon}_r c \zD^2}
 \biggl(1 + \frac{1}{\kappa}\biggr);\ \
 \kappa = \frac{1}{1+\zD/z_\epsilon},
\eqno{\rm (30)}
$$
with $D_r\equiv D(r,0)$, $\bar{\epsilon}_r\equiv \bar{\epsilon}(r,0)$.
As ${\it \Pi}_\epsilon^{(0)}$ 
is of the form of  $\mbox{e}^{-\bar{s}_r y}$, 
the Laplace transform of $f_\epsilon^{(0)}$ with respect to $y$, 
$F_{r,\epsilon}^{(0)}(E_e)$, is sufficient for our purpose 
to obtain the electron density,
$$
\vspace{-1mm}
F_{r,\epsilon}^{(0)}(E_e) = 
\int_0^\infty \mbox{e}^{-\bar{s}_r y} f_\epsilon^{(0)}(y; E_e)dy,
$$
thus we have immediately from equation (28b)
$$
F_{r,\epsilon}^{(0)}(E_e) = \frac{c}{{\cal W}_\epsilon(E_e)}
\int_{E_e}^{\infty}\hspace{-2mm}
dE_0 E_0^{-\gamma} \mbox{e}^{-{Y}_{r,\epsilon}(E_e, E_0)}, 
\eqno{\rm (31)}
$$
with
\vspace{-3mm}
$$
\vspace{1mm}
{Y}_{r,\epsilon}(E_e, E_0)
 = c \bar{s}_r \int_{E_e}^{E_0}\! 
\frac{E^\alpha}{{\cal W}_\epsilon(E)} dE.
$$

In the HE limit, $E_e \gg 1$\,GeV, using equation (18b), we find 
$$
\vspace{2mm}
\displaystyle 
F_{r,\epsilon}^{(0)}(E_e) \simeq 
\frac{cE_e^{-(\gamma+1-\delta)}}
{(1-\alpha-\delta)w_{\tiny \mbox{T}}}
\Biggl[1 - \frac{\gamma +\alpha-\delta -2}
{c \bar{s}_r E_e^{\alpha}/w_{\tiny \mbox{T}}}
 + \ldots \Biggr],
$$
giving a spectral index with $\gamma + 1- \delta$, 
where $\delta\,(=0.075)$ comes from the effect of 
the Klein-Nishina cross-section.
Practically, however, 
it must be softer than the above index because of the 
exponential cutoff with e$^{-E_e/E_{\scriptsize \mbox{cut}}}$ in the 
electron injection spectrum somewhere around 20\,TeV 
(Reynolds \& Keohane 1999; Hendrick \& Reynolds 2001; Yamazaki et al.\ 2006).

Now the principal term, $N_{e,\epsilon}^{(0)}$,
 in equation (26a) for the electron density
 in the HE region,
$E_e$\,$\gsim$\,$E_c^{+}$, is given by\vspace{-1mm}
$$
N_{e,\epsilon}^{(0)}(\vct{r}; E_e) \propto
F_{r,\epsilon}^{(0)}(E_e)\, \mbox{e}^{-|z|/\zD}, 
\eqno{(32)}
$$
 while the perturbative term, $\tilde{N}_{e,n}^{(0)}$,
 is obtained by the use of the iteration method as presented in Appendix~B1,
 giving 
$\tilde{N}_{e,n}^{(0)}/N_{e,\epsilon}^{(0)}$\,$\sim$\,10\% with the first
 iteration 
 for $E_e$\,$\gsim$\,1\,GeV at SS as shown in Figure~24a. 
  In practice, we perform only the first iteration, neglecting the second
 and higher iterations. Full form of 
$N_{e,\epsilon}(\vct{r}; E_e)$ is given by equation (B3).

The numerical procedure in the LE region is similar to 
  that in the HE region mentioned above  by 
  replacing the suffix $\lq \lq \epsilon$" 
 with $\lq \lq n$" (and vice versa),
 while we have to take into account the third term in equation (26b), 
$N_{e,n}^{(1)}$, corresponding to the fluctuation.
 We find again that the perturbative term, $\tilde{N}_{e,\epsilon}^{(0)}$,
 is obtained by the use of the iteration method as presented in Appendix~B2,
 giving 
$\tilde{N}_{e,\epsilon}^{(0)}/N_{e,n}^{(0)}$\,$\sim$\,10\% with the first
 iteration 
for $E_e$\,$\lsim$\,10\,GeV, as shown in Figure~24b.

\begin{figure}[t]
\vspace{-1mm}
  \begin{center}
    \includegraphics[width=7.75cm]{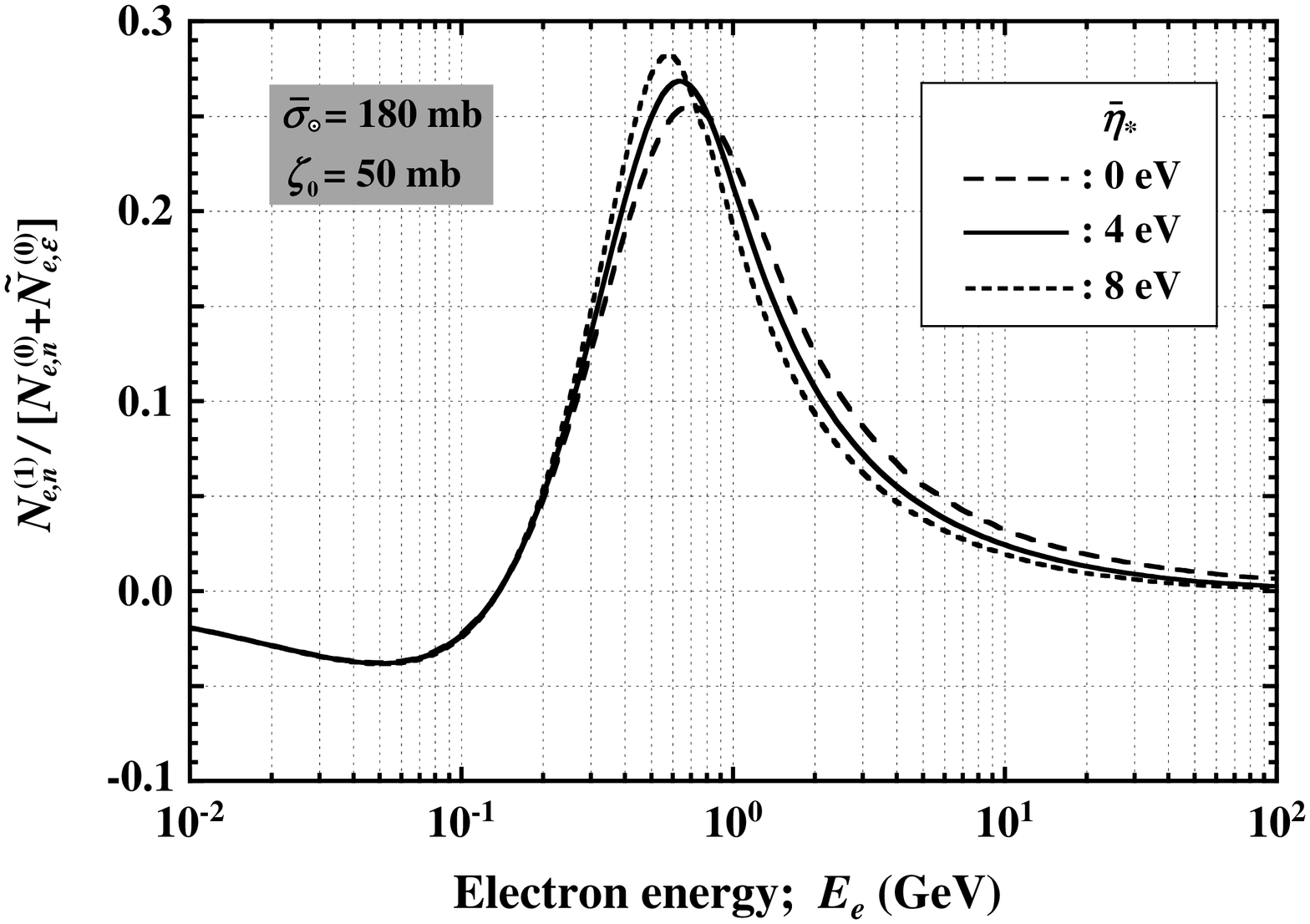}
  \end{center}
  \caption{
 Contributions of the fluctuation in the reacceleration
 with $[\bar{\sigma}_\odot, \zeta_0]$\,=\,[180, 50]\,mbarn at SS for
 several sets of the free parameter $\bar{\eta}_*$
 given by equation (B6).
}
\end{figure}

On the other hand, the numerical procedure in the fluctuation effect due to the
 reacceleration is a little bit cumbersome, which is presented 
in Appendix~B3. We give an example of the ratio, 
${N}_{e,n}^{(1)}/[N_{e,n}^{(0)}+\tilde{N}_{e,\epsilon}^{(0)}]$, at SS 
 for the first iteration in Figure 7,  where 
$\bar{\eta}_*$ is the {\it effective} ratio of the energy density
 to the gas density defined by equation (B6),
 approximately with 2eV. One finds that 
it is significant around 0.3--1.5\,GeV 
in the case of $\zeta_0$\,=\,50\,mbarn,  
boosting the solution without the fluctuation, 
$N_{e,n}^{(0)}+\tilde{N}_{e,\epsilon}^{(0)}$, by approximately 25\%. 
 So we perform only the first iteration also for
$N_{e,n}^{(1)}$ in the LE region, as the contribution coming from
 the second and higher iterations
 is at most of the magnitude of a few \% or less (Ishikawa 2010).
 Full form of $N_{e,n}(\vct{r}; E_e)$ is given by equation (B15).

Finally, 
we give the electron density covering all energies
so that it continues smoothly at the energy $E_c$ between
the HE and LE regions at the SS $(\vct{r}=\vct{r}_\odot)$, 
with $E_c \approx E_c^{+}$ in practice, 
but not always $E_c = E_c^{+}$, \vspace{2mm}
\begin{displaymath}
\hspace{3.7cm}
\frac{N_e(\vct{r}; E_e)}{N_{e, 0}} =    
 \left\{ \begin{array}{ll} \vspace{2mm}
 {\displaystyle \frac{F_{r, \epsilon}(E_e)}{F_{\odot, \epsilon}(E_c)}
\mbox{e}^{-|z|/\zD}}: 
 \ \  \mbox{for $E_e\,\ge\,E_c$}, \hspace{3.15cm} (\rm 33a)
\vspace{-2mm} 
\\
\vspace{-2mm}
\\
{\displaystyle  \frac{F_{r, n}(E_e)}{F_{\odot, n}(E_c)}
\mbox{e}^{-|z|/\zD}}:
 \ \  \mbox{for $E_e\,\le\,E_c$}, \hspace{3.1cm} (\rm 33b)
                       \end{array} 
                       \right.
\end{displaymath}

with
$$
\blankline
\frac{\partial}{\partial E_e}\frac{F_{\odot, \epsilon}(E_e)}
{F_{\odot, \epsilon}(E_c)}\Biggl|_{E_e=E_c}
=  \frac{\partial}{\partial E_e}\frac{F_{\odot, n}(E_e)}
{F_{\odot, n}(E_c)}\Biggl|_{E_e=E_c},
$$
see equations (B4) and (B13) for $F_{r, \epsilon}(E_e)$
and $F_{r, n}(E_e)$ respectively, and $N_{e, 0}$ is
determined by the normalization with the experimental data
as discussed in \S\,5.

\begin{figure}[!b]
  \begin{center}
    \includegraphics[width=7.6cm]{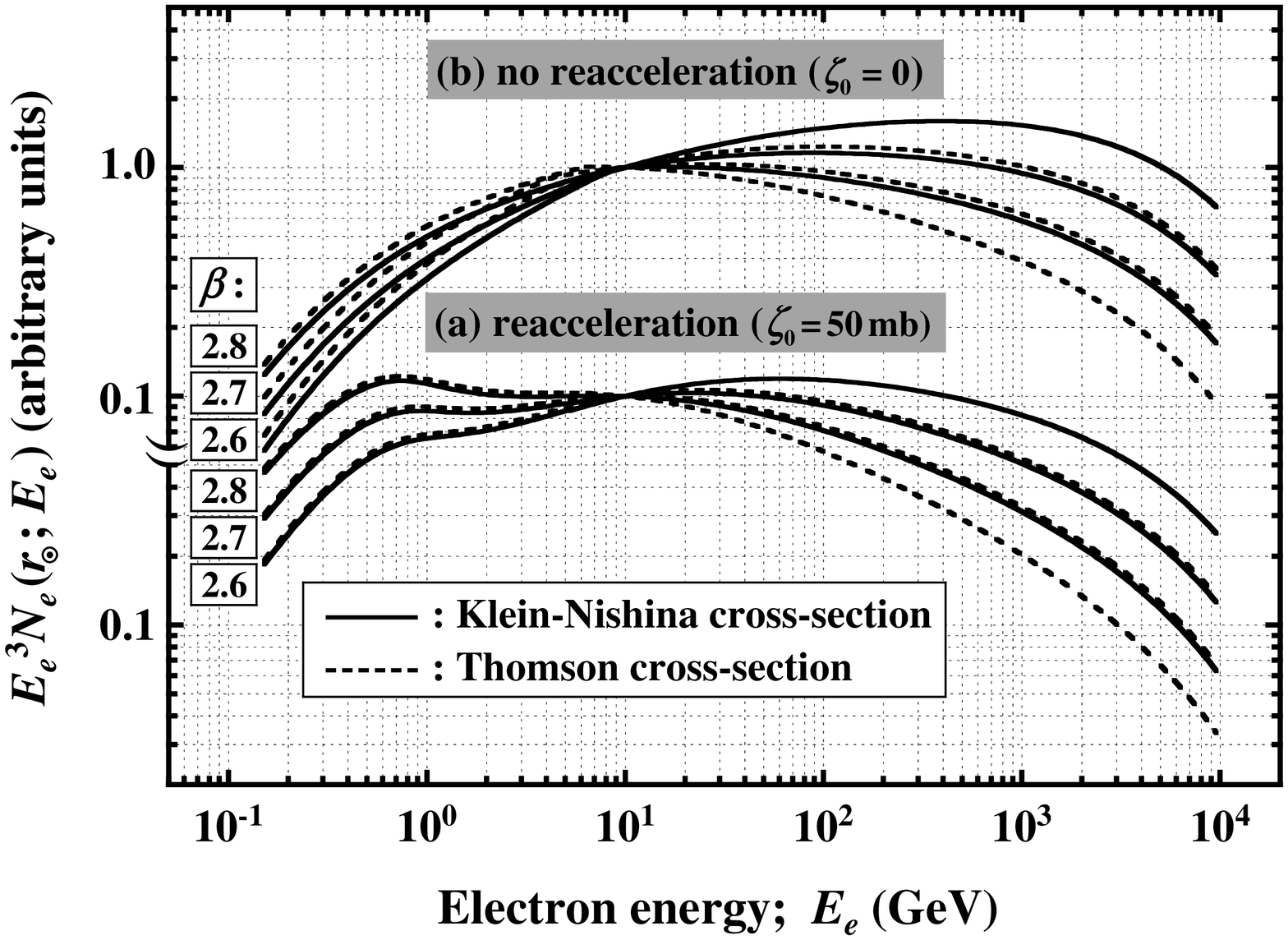}
  \end{center}
  \caption{
Numerical results of the electron density, $N_e(r_\odot; E_e)$,
 at the SS ($r=r_\odot$) in the case of 
(a) reacceleration with
[$\alpha; \zeta_0,\bar{\sigma}_\odot$]\,=\,
 [$\frac{1}{3}$; 50\,mbarn, 180\,mbarn], and 
(b) no reacceleration with [$\frac{1}{2}$; 0, 90\,mbarn], 
for $\beta$\,=\,2.6, 2.7, and 2.8,
 where the vertical axis is multiplied by
 $E_e^3$, and normalized to $E_e$\,=\,10\,GeV. We show results for 
 two cross-sections in IC process from the Thomson ({\it dotted curves}) 
 and Klein-Nishina ({\it solid curves}) formulae.
}
\end{figure}

\subsection{Numerical results}

In Figure~8 we show the numerical results of 
$N_e(\vct{r}_\odot; E_e)/N_{e,0}$ in two cases, 
(a) [$\zeta_0,\bar{\sigma}_{\tiny \mbox{$\odot$}}$, 
$\bar{s}_{\tiny \mbox{$\odot$}}$]\,=\,
[50, 180, 30\,eV$^{-1}$]mbarn with $\alpha=\frac{1}{3}$ 
(reacceleration with Kolmogorov-type spectrum in 
 hydromagnetic turbulence), and 
(b) [0, 90, 15\,eV$^{-1}$]mbarn with $\alpha=\frac{1}{2}$ 
(no reacceleration with Kraichnan-type spectrum)  for 
$\bar{\epsilon}_\odot/\bar{n}_\odot$\,=\,2\,eV with 
$\beta$ $(\equiv \gamma+\alpha)$ = 2.6, 2.7 and 2.8, see equation (30) with 
$r$\,=\,$r_\odot$ for $\bar{s}_{\tiny \mbox{$\odot$}}$, where
we assume $E_{\scriptsize \mbox{cut}}$\,=\,20\,TeV 
(Reynolds \& Keohane 1999; Hendrick \& Reynolds 2001; Yamazaki et al.\ 2006) 
in the electron injection spectrum with
$E_e^{-\gamma}\mbox{e}^{-E_e/E_{\scriptsize \mbox{cut}}}$, 
and the results show the
use of both Klein-Nishina  ({\it solid curves}) and Thomson
({\it dotted curves}) cross-sections.

We find two critical points in Figure~8.
First, those by the former cross-section give approximately 40--50\%
(20--30\%) larger than those by the latter at 1\,TeV (100\,GeV), where the
density is normalized at $E_e$\,=\,10\,GeV, leading to 
significantly harder spectra than those with the Thomson cross-section, 
as expected. Similar results are also recently reported  by
 Delahaye et al.\ (2010), 
while their main purpose is to study the nearby sources of electron 
and the positron excess problem as well, which are outside the range of the
present paper.

Second, the reacceleration effect
 is significant in the energy region less than 
10\,GeV as compared to the curves without the reacceleration process.  
Unfortunately,
however, it is difficult to observe such a signal in the {\it direct}
experimental data on the electron component because of the modulation
effect in the low energy region $\lsim$\,5\,GeV, which masks the 
electron flux boosted by the reacceleration, see Figure~13, even if 
 it occurs actually. 

\begin{figure}[!t]
\vspace{-2mm}
    \includegraphics[width=7.8cm]{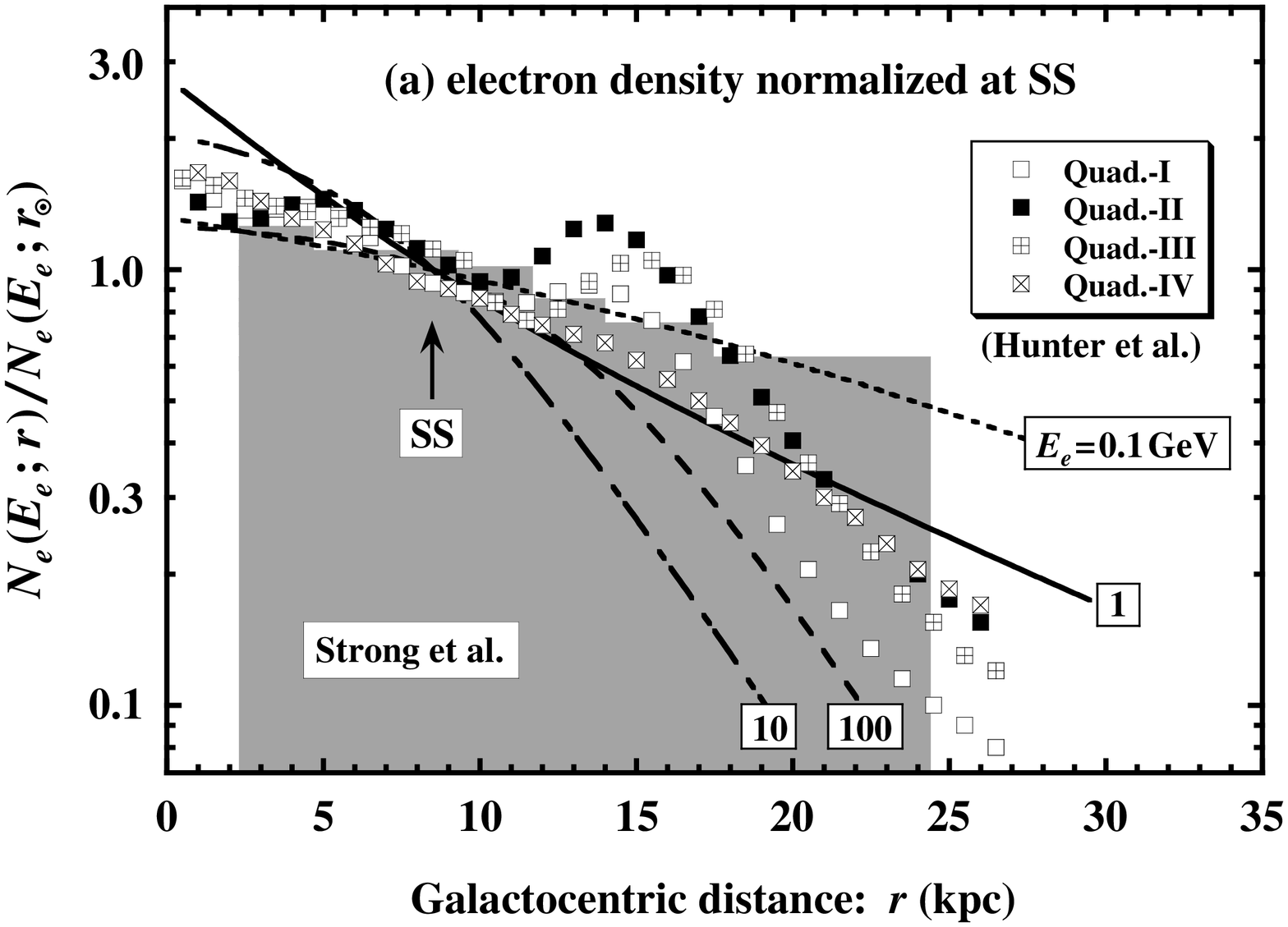}
    \includegraphics[width=7.8cm]{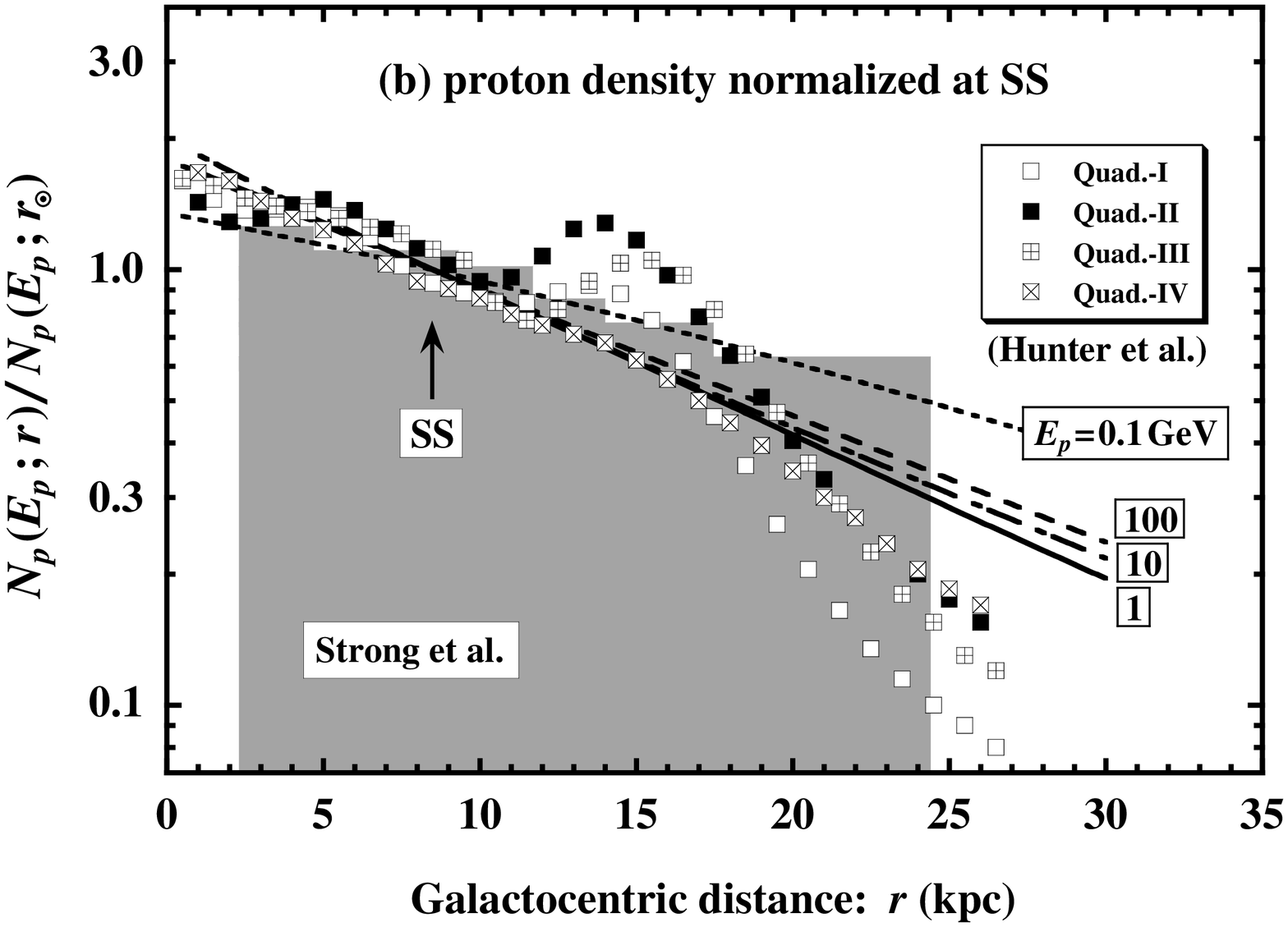}
  \caption{
CR densities of (a) electrons, $N_e(r; E_e)$,
 and (b) protons, $N_p(r; E_p)$, as a function of the
 galactocentric distance $r$  for several energies,
 where the vertical axis is normalized to the density at SS
 ($r$\,=\,$r_\odot$).
 We present also those given by Hunter et al.\ (1997) 
 (four kinds of {\it square symbols}), and by 
 Strong et al.\ (1988) ({\it thin filled histograms}), where 
 the CR densities are averaged azimuthally in 
 each galactocentric quadrant by Hunter et al.\,(1997), 
 assuming that they are 
 coupled to the gas density of ISM independent of the energy,
 while azimuthal symmetric $\gamma$-ray emissivity is
 used by Strong et al.\ (1988).}
\end{figure}

Next we examine the spatial dependence of the electron density. First 
it attenuates exponentially with the latitudinal distance $z$ from the
GP with the latitudinal scale height of the diffusion coefficient,
 $\zD$\,=\,2--4\,kpc,
 independent of the
energy $E_e$. This is the same result as in the case of the proton
density, $N_p(\vct{r}; E_p)$ (Paper~V), 
namely the ratio of electron
density to the proton density is independent of $z$.

Contrary to the latitudinal behavior, 
the longitudinal behavior of the CR densities,
$N_e(\vct{r}; E_e)$ and $N_p(\vct{r}; E_p)$, are somewhat 
complicated, both of which depend on the
energy, and appear implicitly in the form of $\bar{s}_r$
and $\bar{\sigma}_r$ 
(see Paper~I for $\bar{\sigma}_r$ and its physical meaning). 
We present these in
Figure~9 against the radial distance $r$ for the (a) electron and (b)
proton components, both normalized
at the SS for four energies, 0.1, 1, 10, and 100\,GeV, 
with $\beta$\,=\,2.7, where the scale
heights are set as
$[\bar{r}_n, \bar{r}_\epsilon]$\,=\,[30, 8]\,kpc  and  
$[\zD; z_n, z_\epsilon]$\,=\,[3; 0.2, 0.75]\,kpc
(see Table\,4 for $\bar{r}_n$ and $\bar{r}_\epsilon$). We plot
the results of Hunter et al.\ (1997; 
{\it square symbols}) and
Strong et al.\ (1988; {\it thin filled histogram}) together, 
where the former are based on the assumption
that the CR density is coupled to the density of ISM, and plotted
separately for four galactocentric quadrants, I, II, III, and IV.
We find that the radial dependence of the electron density,
$N_e(\vct{r}; E_e)$, is much stronger than that of
the proton density, $N_p(\vct{r}; E_p)$, in the energy region of
1--100\,GeV as expected, while the other two authors 
assume no spatial dependence in the energy spectrum, namely
the {\it shape} of the energy spectrum at the SS is 
the same everywhere in the Galaxy.

\begin{deluxetable}{ll}
\tabletypesize{\small}
\tablecaption{
Summary of the production cross-sections of $\gamma$-rays
in the bremsstrahlung  and the IC processes with 
$x = E_\gamma/E_e$, where $E_e$ is 
the incident energy of electron, and $E_\gamma$ is the energy of
the produced $\gamma$'s, and $E_{\mbox{\scriptsize ph}}$
 the energy of the target photon 
before electron scattering. For the bremsstrahlung process, we present 
the cross-section in the case of only one-electron atoms ($Z=1$),
see Gould (1969) for two-electron atoms ($Z=2$).
\label{tbl-5}}
\tablewidth{0pt}
\tablehead{
 \colhead{  bremsstrahlung (EB)} \ \ \ \ \ \ \ 
& \colhead{ inverse Compton (IC)} 
} 
\startdata 
\vspace{-2mm}
\\
$\sigma_{\mbox{\tiny EB}}(E_e, E_\gamma)dE_\gamma =
\sigma_{\mbox{\tiny EB}}^{(0)}\,
\phi_{\mbox{\tiny EB}}(x, E_\gamma){\displaystyle \frac{dx}{x}}$
& \ 
$\sigma_{\mbox{\tiny IC}}(E_e, E_\gamma; E_{\mbox{\scriptsize ph}})dE_\gamma =
\sigma_{\mbox{\tiny IC}}^{(0)}\,
\phi_{\mbox{\tiny IC}}(x, q){\displaystyle \frac{dx}{X}}$
\\ \\ 
$\sigma_{\mbox{\tiny EB}}^{(0)} = 
{\displaystyle 4\alpha_f Z(Z+1)
 \Bigl(\frac{e^2}{m_ec^2}\Bigr)^2;\ \ \alpha_f=\frac{1}{137}}$
 & \ 
$\sigma_{\mbox{\tiny IC}}^{(0)} = 
{\displaystyle 3\sigma_{\mbox{\tiny T}}}
= 8\pi {\displaystyle \biggl(\frac{e^2}{m_ec^2}\biggr)^2}$
\\ \\
$\phi_{\mbox{\tiny EB}}(x, E_\gamma) =
 \Bigl\{1 + (1-x)^2\Bigr\}\phi_1(\chi)
-\displaystyle \frac{2}{3}(1-x)\phi_2(\chi)$\ \ \ 
& \ 
$\displaystyle 
\phi_{\mbox{\tiny IC}}(x, q) = 2q\ln q 
+(1-q)\biggl(1+2q+\frac{1}{2}\frac{x^2}{1+x}\biggl)$
\\ \\
$\displaystyle 
\phi_1(\chi) = 1 + \int_\chi^1\!
\phi_0(y)\biggl(1-\frac{\chi}{y}\biggr)^2\frac{dy}{y}$
& \ 
$\displaystyle 
q \equiv q(x, X) = \frac{x}{1-x} \frac{1}{X}$
\\ \\
$\displaystyle 
\phi_2(\chi) = \frac{5}{6} + \int_\chi^1\! \phi_0(y)
\biggl\{1+3\frac{\chi^2}{y^2}\biggl(1+ \ln
 \frac{\chi^2}{y^2}\biggr)-4\frac{\chi^3}{y^3}\biggr\} \frac{dy}{y}$
& \ 
$\displaystyle X \equiv X(E_e, E_{\mbox{\scriptsize ph}})
 = \frac{k}{{\it \Theta}_e^2}
 = \frac{E_{\mbox{\scriptsize ph}}E_e}{[m_e c^2/2]^2}
$
\\ \\
$\displaystyle 
\phi_0(y) = 1 - \frac{1}{[1+y^2/(2\alpha_fZ)^2]^2}$
&\ 
$\displaystyle 
 k \equiv k(E_{\mbox{\scriptsize ph}}, T_{\mbox{\scriptsize ph}}) = 
\frac{E_{\mbox{\scriptsize ph}}}
{k_{\mbox{\scriptsize B}}T_{\mbox{\scriptsize ph}}}$  
\\ \\
$\displaystyle 
\chi \equiv 
 \chi (x, E_\gamma) = 
 \frac{x^2}{1-x} \frac{m_ec^2}{2E_\gamma}$
& \ 
$\displaystyle 
 {\it \Theta}_e \equiv {\it \Theta}_e(E_e, T_{\mbox{\scriptsize ph}}) = 
\frac{m_e c^2/2}{\sqrt{k_{\mbox{\scriptsize B}}
T_{\mbox{\scriptsize ph}}E_e}}
$   
\\ \\
 \enddata
\end{deluxetable}

\section{Electron-induced $\gamma$-ray spectrum}

For convenience in the following discussion, we summarize two 
cross-sections 
in Table~5, $\sigma_{\mbox{\tiny EB}}(E_e, E_\gamma)$ and
$\sigma_{\mbox{\tiny IC}}(E_e, E_\gamma; E_{\mbox{\scriptsize ph}})$,
 each for the bremsstrahlung (abbreviated as $\lq \lq$EB"
 for subscript attached here and in the following)  and the
IC processes respectively, where $E_{\mbox{\scriptsize ph}}$
 is the energy of target photon
before scattering. In these cross-sections, we take into account the
screening effect for the bremsstrahlung (Koch \& Motz 1959; Gould 1969),
 and the Klein-Nishina cross-section 
(Jones 1965, 1968; Blumenthal \& Gould 1970) for IC. 
In the following discussion, we put $F_r(E_e) 
 \equiv F_{r, \epsilon}(E_e)$ for $E_e \ge E_c$, and 
$F_r(E_e) \equiv F_{r, n}(E_e)$ for $E_e \le E_c$ 
in equation (33) for simplicity.

First we consider the emissivity of $\gamma$'s from the bremsstrahlung
 at the position $\vct{r}$, which is immediately written down as

$$
\blankline
q_{\mbox{\tiny EB}} (\vct{r}; E_\gamma) =\hspace{-1mm}
\int_{E_\gamma}^{\infty} \hspace{-1mm}N_e(\vct{r}; E_e) [n(\vct{r}) c 
\sigma_{\mbox{\tiny EB}}(E_e, E_\gamma)]dE_e, 
\eqno{\rm (34)}
$$
where the electron density, $N_e(\vct{r}; E_e)$, is given by 
equation (33). 
For the numerical calculation of equation (34), we
need the {\it absolute} electron density at $\vct{r}$. 
To do so, we use the observational data on the electron intensity at the
SS, $dI_e^\odot/dE_e$, which is related to the electron density by\vspace{2mm}
$$
\blankline
\frac{dI_e^{\tiny \mbox{$\odot$}}}{dE_e}(E_e) = \frac{c}{4\pi} N_e(\vct{r}_\odot; E_e).
$$

In practice, we normalize the electron density at $E_s$ = 10\,GeV \ 
with use of the most recent data (see Fig.\,14),  where the
solar modulation effect is negligible, 
$$
cN_s^{\tiny \mbox{$\odot$}} \equiv 
cN_e(\vct{r}_{\tiny \mbox{$\odot$}}; E_s) =
 2.26\,\mbox{m$^{-2}$s$^{-1}$GeV$^{-1}$}, 
$$
corresponding to $dI_e^{\odot}/dE_e = 
0.180$\,m$^{-2}$sr$^{-1}$s$^{-1}$ GeV$^{-1}$
 at $E_e = 10$\,GeV in Figure 14, 
while  $E_s$\,=\,100 GeV (per nucleon) for
the hadron-induced $\gamma$'s ($\pi^0 \rightarrow 2\gamma$) 
 with 
$cN_p(\vct{r}_{\tiny \mbox{$\odot$}}; E_s)$ =
 6.16\,{m$^{-2}$s$^{-1}$GeV$^{-1}$} (Paper V). One should keep in mind that
 the uncertainty in the normalization is of the magnitude as large as 10\%.

Thus taking care of the terms related to $\vct{r}$, 
we have 
$$
\blankline
\frac{q_{\mbox{\tiny EB}} (\vct{r}; E_\gamma)}
{n(\vct{r}) w_{\mbox{\tiny EB}}^{\mbox{\tiny (0)}}
 N_s^{\tiny \mbox{$\odot$}}} = \mbox{e}^{-|z|/\zD} \! \!
\int_0^1 \! \phi_{\mbox{\tiny EB}}(x, E_\gamma)
\frac{F_r(E_x)}{F_{\tiny \mbox{$\odot$}}(E_s)}\frac{dx}{x^2},
\eqno{\rm (35)}
$$
for $E_\gamma \ge E_c$ with $E_x = E_\gamma/x$,
and $w_{\mbox{\tiny EB}}^{\mbox{\tiny (0)}}$\,=\,$c \sigma_{\mbox{\tiny EB}}^{\mbox{\tiny (0)}}$ = 1.39$\times$10$^{-16}$cm$^3$s$^{-1}$ for the hydrogen gas ($Z$\,=\,1),
where one should take care of the energy range $E_e \le E_c$
in the case of $E_\gamma \le E_c$. 

Next we consider the emissivity of $\gamma$'s coming from the IC
process, which is somewhat complicated, as there are several kinds of
target photons with different energy density as well as with different
scale heights in the spatial gradient. Here we present a result only,
taking into account the six wavelength bands in
$\epsilon_{\scriptsize \mbox{ph}}^{(i)}(\vct{r})$ ($i$\,=\,0--5)
(see eq. [7] and Table~3),\vspace{2mm}
$$
\blankline
\frac{q_{\mbox{\tiny IC}}^{(i)} (\vct{r}; E_\gamma)}
{\epsilon_{\scriptsize \mbox{ph}}^{(i)}(\vct{r}) w_{\mbox{\tiny T}}
 N_s^{\tiny \mbox{$\odot$}}} = \mbox{e}^{-|z|/\zD} \! \! 
\int_0^1 \! 
{\it \Phi}_{\mbox{\tiny IC}}^{(i)}(x, E_\gamma)
\frac{F_r(E_x)}{F_{\tiny \mbox{$\odot$}}(E_s)}\frac{dx}{x^2},
\eqno{\rm (36)}
$$
where
${\it \Phi}_{\mbox{\tiny IC}}^{(i)}(x, E_\gamma)$ is given by equation
(C2), see Appendix~C for the details.

Let us show the numerical results for two cases of emissivity in
Figure~10,\ \ 
(a) $r/r_{\tiny \mbox{$\odot$}} = 0.5, 1, 2$ with $z=0$
in the GP, and 
(b) $z=0.2, 0.4, 0.6$\,kpc with $r=r_{\tiny \mbox{$\odot$}}$ 
normal to the GP at SS, assuming 
$\beta \equiv \gamma + \alpha$ =\,2.7, 
where we present separately those coming from 
$\pi^0$ ({\it solid curves}), EB ({\it broken curves}), 
and IC ({\it dotted curves}). One
finds that EB-$\gamma$'s and $\pi^0$-$\gamma$'s are comparable around
50\,MeV, and IC-$\gamma$'s and $\pi^0$-$\gamma$'s around two
energies, $\sim$\,20\,MeV and $\sim$\,1\,TeV.

See Paper~V for the emissivity originating in $\pi^0$, 
$q_{\pi^0} (\vct{r}; E_\gamma)$, while
we use more realistic gas density, $n(\vct{r})$, in the present paper. 
 Note also in $q_{\pi^0}$ 
that the semi-empirical production cross-section of
 $\gamma$'s, $\sigma_{pp\rightarrow \gamma}(E_p, E_\gamma)$, in
 proton-proton collision we use is valid over very wide energy ranges,
 1\,GeV--1\,PeV, reproducing nicely various kinds of physical quantities
 such as psuedo-rapidity, energy spectrum, multiplicity, etc, 
 obtained by both the accelerator and CR experiments 
 with local target layer (Suzuki, Watanabe \& Shibata 2005). 

\begin{figure}[!t]
    \includegraphics[width=7.7cm]{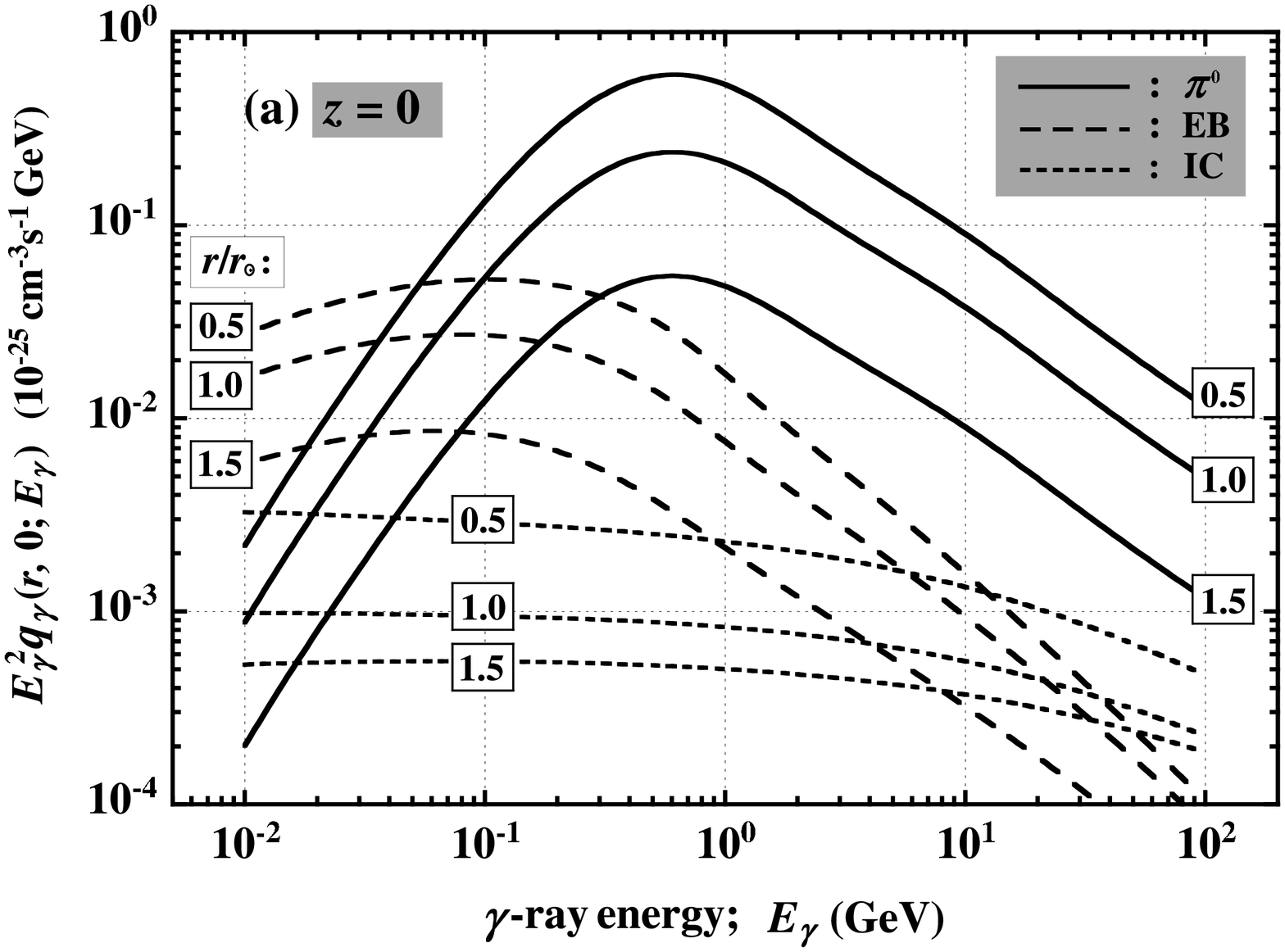}
    \includegraphics[width=7.7cm]{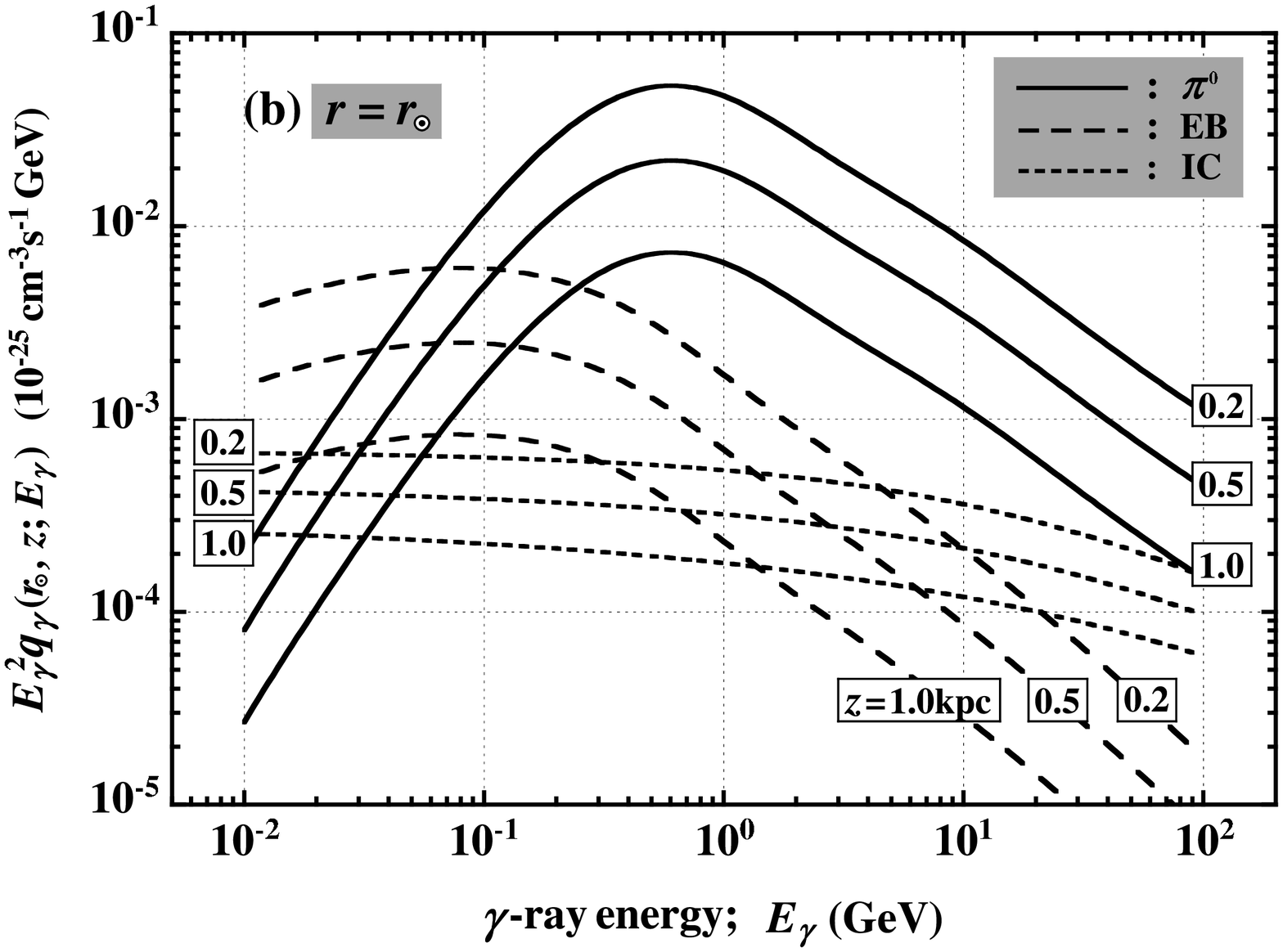}
  \caption{
Emissivity of $\gamma$'s (a) at three radial distances,
  $r/r_\odot$\,=\,0.5, 1.0, 1.5 in the galactic plane ($z=0$),
  and (b) at three latitudinal distances, $z$\,=\,0.2, 0.4, 0.6\,kpc
  with $r$\,=\,$r_\odot$, where the three components 
  for $\gamma$'s emission,
  $\pi^0$ ({\it solid curves}), EB ({\it broken curves}), and
  IC ({\it dotted curves}) are shown separately.
}
\end{figure}

Once we have the emissivity of $\gamma$'s induced by the interaction
between the electrons and the media of ISM and ISRF, 
we can obtain immediately the intensity
of $\gamma$'s observed at the SS ($\vct{r}=\vct{r}_{\! \mbox{\tiny $\odot$}}$),
  coming from the direction
$\vct{\theta}(l, b)$
$$
\frac{d^3I^{\mbox{\tiny $\odot$}}_{\gamma}(\vct{\theta}; E_{\gamma})}
 { dE_{\gamma} dl d(\sin b)}  = 
\frac{1}{4\pi}\int_0^{\infty}\! 
q_{\gamma}(\vct{r}; E_{\gamma})ds,
$$
with\vspace{-3mm}
$$
q_{\gamma}(\vct{r}; E_{\gamma}) = 
q_{\mbox{\tiny EB}}(\vct{r}; E_{\gamma}) + 
\sum_{i=0}^{5}q_{\mbox{\tiny IC}}^{(i)}(\vct{r}; E_{\gamma}),
$$
where the integration with respect to $s$ 
is performed along the arrival direction of $\gamma$'s,
$\vct{\theta}(l, b)$, at the SS, and  
$\vct{r} (r, z)$ is bound to $(s;\, l, b)$ as follows,
$$
r(s;\, l, b) = 
\sqrt{r^2_{\hspace{-0.5mm}\mbox{\tiny $\odot$}}  + s^2\cos ^2 b - 
2r_{\hspace{-0.5mm}\mbox{\tiny $\odot$}} s \cos b\hspace{0.5mm} \cos l},
$$
$$ 
z(s;\, b) = s \sin b.
$$

\section{Comparison\,with\,the\,observational data}

\subsection{Critical parameters}

We assume that the
source distribution of electron component, $Q(\vct{r}; E_e)$,
  is the same as that of the
hadronic component except for the cutoff electron energy, for instance 
$E_{\scriptsize \mbox{cut}} \approx 20$\,TeV,
 with the supernova remnants as the main energy supply, while 
the pulsars and pulsar wind nebulae might contribute to them as well,
 particularly to positrons and electrons (for instance, Delahaye et al.\ 2010).  So the galactic parameters used in the present work are essentially the
 same as those appearing in Papers~I--V, and we summarize them
 briefly in the following.

The recent observational data on the energy spactra of CR hadronic
 components
 give indices with $2.74\pm0.08$  for proton (Derbina et al.\ 2005), 
and with a common value of $\sim$2.7 for nuclei between the oxygen and iron 
 (M\"{u}ller 2009), whereas there still remains uncertainty
  for helium, for instance with $2.68\pm0.05$ by JACEE (Asakimori
 et al.\ 1998) in contrast to $2.78\pm0.20$ by RUNJOB (Derbina et al.\ 2005).
 Note that PAMELA (Picozza et al.\ 2007) 
reports recently a common index of 2.73 in both the proton and helium spectra, 
albeit the energy region is limited below 500\,GeV. 
Any way, the spectrum index $\beta$ of proton,  
 must lie well within 2.7--2.8 in the high energy region at the SS, 
 which is the most effective element 
  for the hadron-induced D$\gamma$'s. 
 See Paper~V for the contribution of helium and nuclei to D$\gamma$'s, 
  which is taken into account by introducing
 the enhancement factor with 1.53.
So in the present paper 
we use the critical parameter $\beta$ in place of $\gamma$
(source index of the energy spectrum)
with $\beta = \gamma + \alpha$, and 
consider three values of $\beta$; 
2.6, 2.7, and 2.8, each for
 $\alpha = \frac{1}{3}$ (Kolmogoroph-type spectrum) and
 $\frac{1}{2}$ (Kraichnan-type spectrum).

 There are three galactic parameters, [$D(\vct{r})$,
$\bar{n}(\vct{r})$, $Q(\vct{r})]$, in our approach to
 the CR propagation, and 
six scale heights for longitudinal and latitudibal
 directions correspoding to each one,  $[\rD, r_n, \rQ]$ and 
 $[\zD, z_n, \zQ]$, respectively. In practice, however,
 explicit parameters needed to compare with the
 experimental data appear in two critical ones alone, 
$\bar{\sigma}_r$ and $\zeta_0$,
 besides [$\alpha$, $\beta$] mentioned above,
 while the parameter, $\bar{\mu}_r \equiv 2\zD/\sqrt{\bar{\tau}_0D_r}$, 
  is also  important for the study of the CR isotopes
 ($\bar{\tau}_0$: normalized life time of an isotope with 10$^6$yr).

 For electron components, the additional parameter newly appears,  
$\bar{s}_r$, given by equation (30),  physical meaning of which
 is essentially the same as $\bar{\sigma}_r$; i.e., 
  while the inverse of $\bar{\sigma}_r$ gives 
 the average path length, $\bar{x}_r$, 
in units of cm$^{-2}$ in ISM as discussed in
 Paper~I, that of $\bar{s}_r$ corresponds to the
 average path length, $\bar{y}_r$,
in units of eVcm$^{-2}$ in ISRF,
 namely the total amount of  photon-gas energy that CR has passed
 through the ISRF.

Now from equation (33), one should remark that there appear only 
 {\it three} critical parameters,
[$\zeta_0$, $\bar{\sigma}_\odot$, $\bar{s}_\odot$],
 in $F_{r, \epsilon}(E_e)$ and $F_{r, n}(E_e)$  needed to  
 compare with the observational data, aside from 
   two critical indices, [$\alpha$, $\beta$],
 note that 
  various galactic parameters such as the diffusion constant, 
   gas density, energy density, their scale heights, etc are 
 all involved implicitly in these three ones.

\begin{figure}[!b]
    \includegraphics[width=7.7cm]{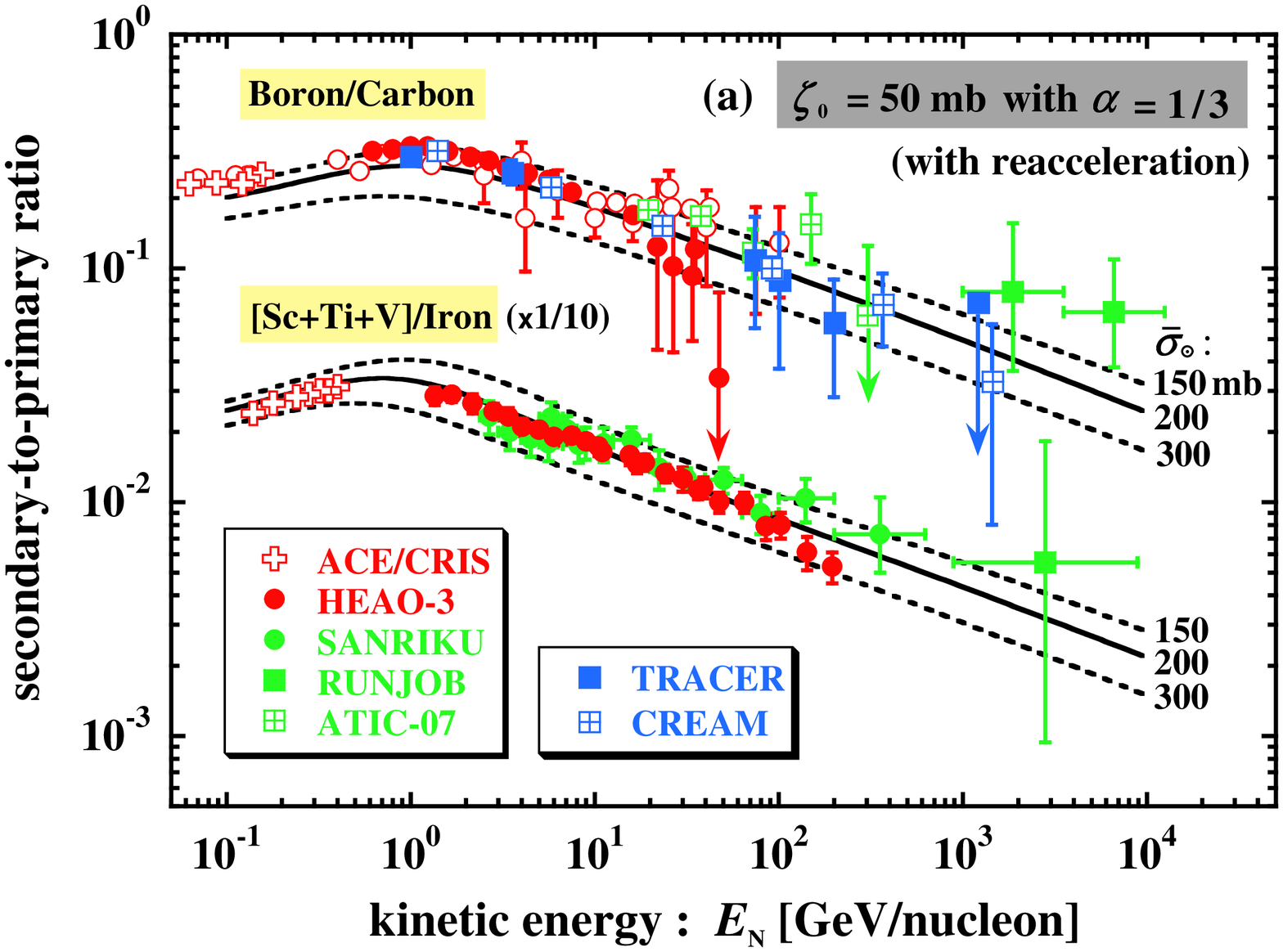}
    \includegraphics[width=7.7cm]{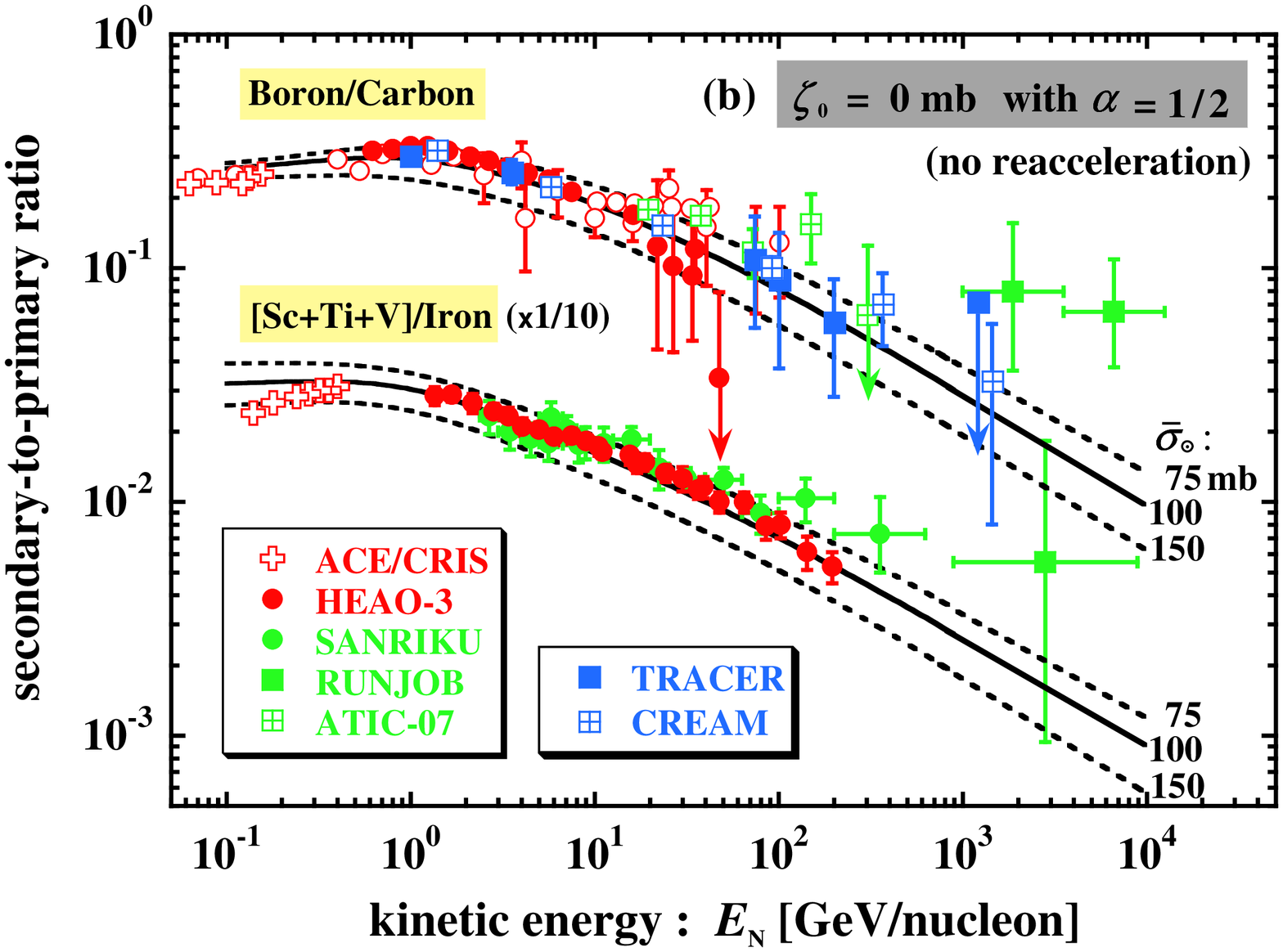}
  \caption{
Energy dependence of the secondary-to-primary 
ratio for boron/carbon and
 sub-iron/iron. See Paper~II and references therein for the 
 experimental data,  while CREAM (Ahn et al.\ 2008) 
and TRACER (M\"{u}ller 2009) data are newly plotted.
 Numerical curves are demonstrated for two cases;
 (a) reacceleration with
 ($\alpha,\zeta_0$)\,=\,($\frac{1}{3}$, 50\,mbarn)
 and (b) no reacceleration with ($\frac{1}{2}$, 0).
}
\end{figure}
\vspace{1mm}

\subsection{Charged components}

\subsubsection{Hadron components}

  \begin{figure}[!b]
\blankline
    \includegraphics[width=7.7cm]{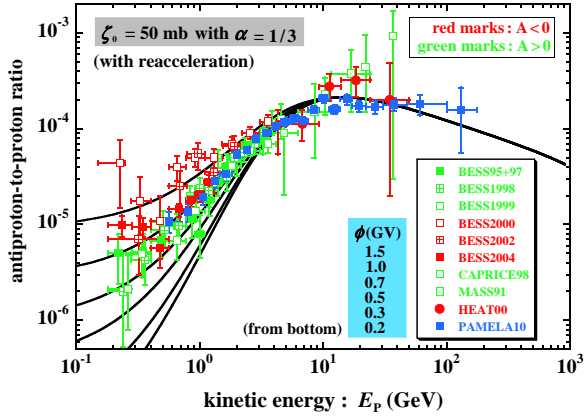}
  \caption{
Energy dependence of the antiproton-to-proton ratio,
 where we assume the reacceleration model with 
$\alpha = \frac{1}{3}$ for 
  six modulation parameters, 0.2, 0.3, 0.5, 0.7, 1.0, and 1.5\,GV,
 and $A < 0$ ($A > 0$) corresponds to positive (negative) 
polarity state in heliospheric magnetic field, although the
 present calculations do not take the effect into account.
 See Paper IV and the references therein for the data other than
 PAMELA (Adriani et al.\ 2010).
}
\end{figure}

\begin{figure}[t]
    \includegraphics[width=7.6cm]{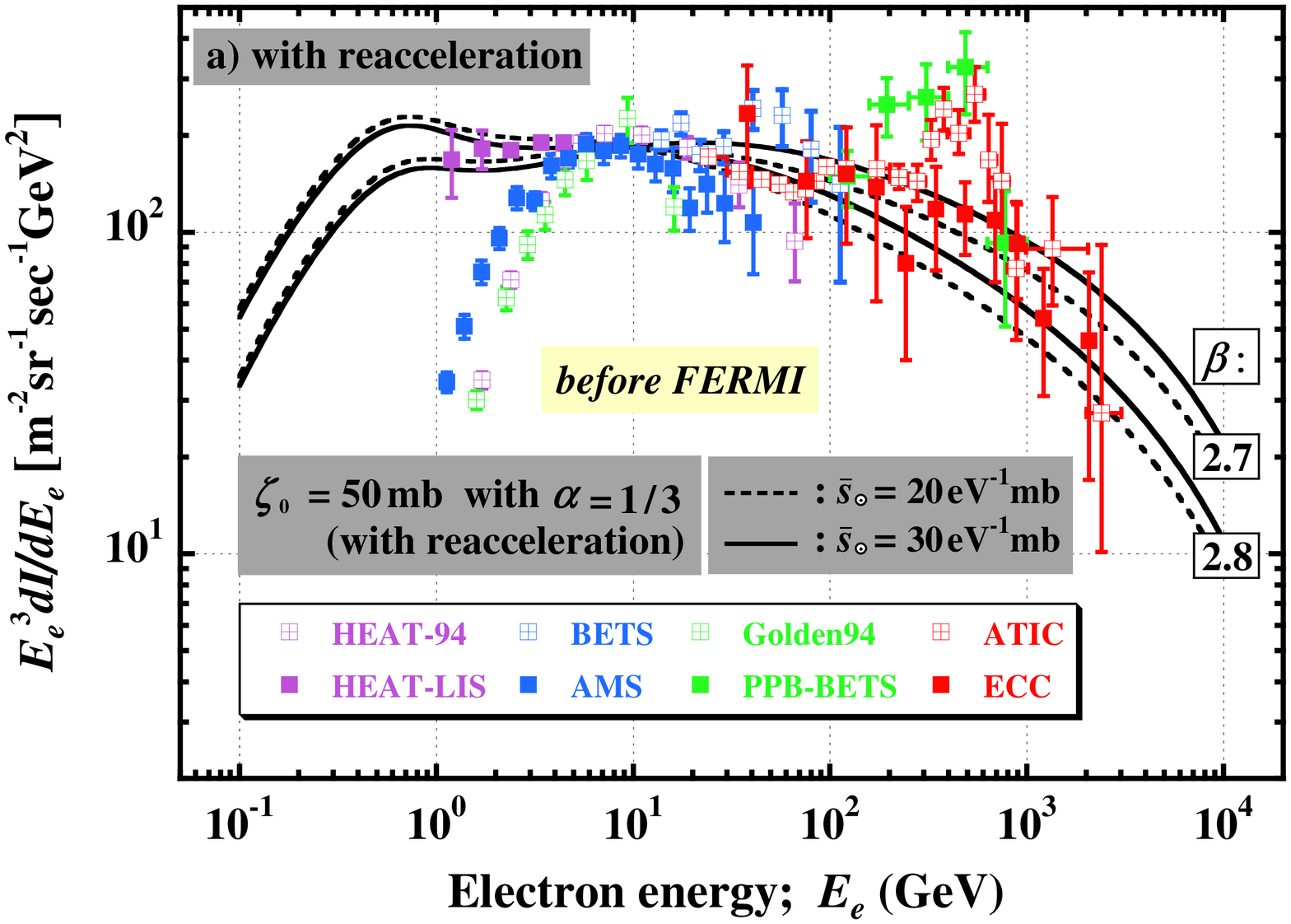}
    \vspace{5mm}
    \includegraphics[width=7.6cm]{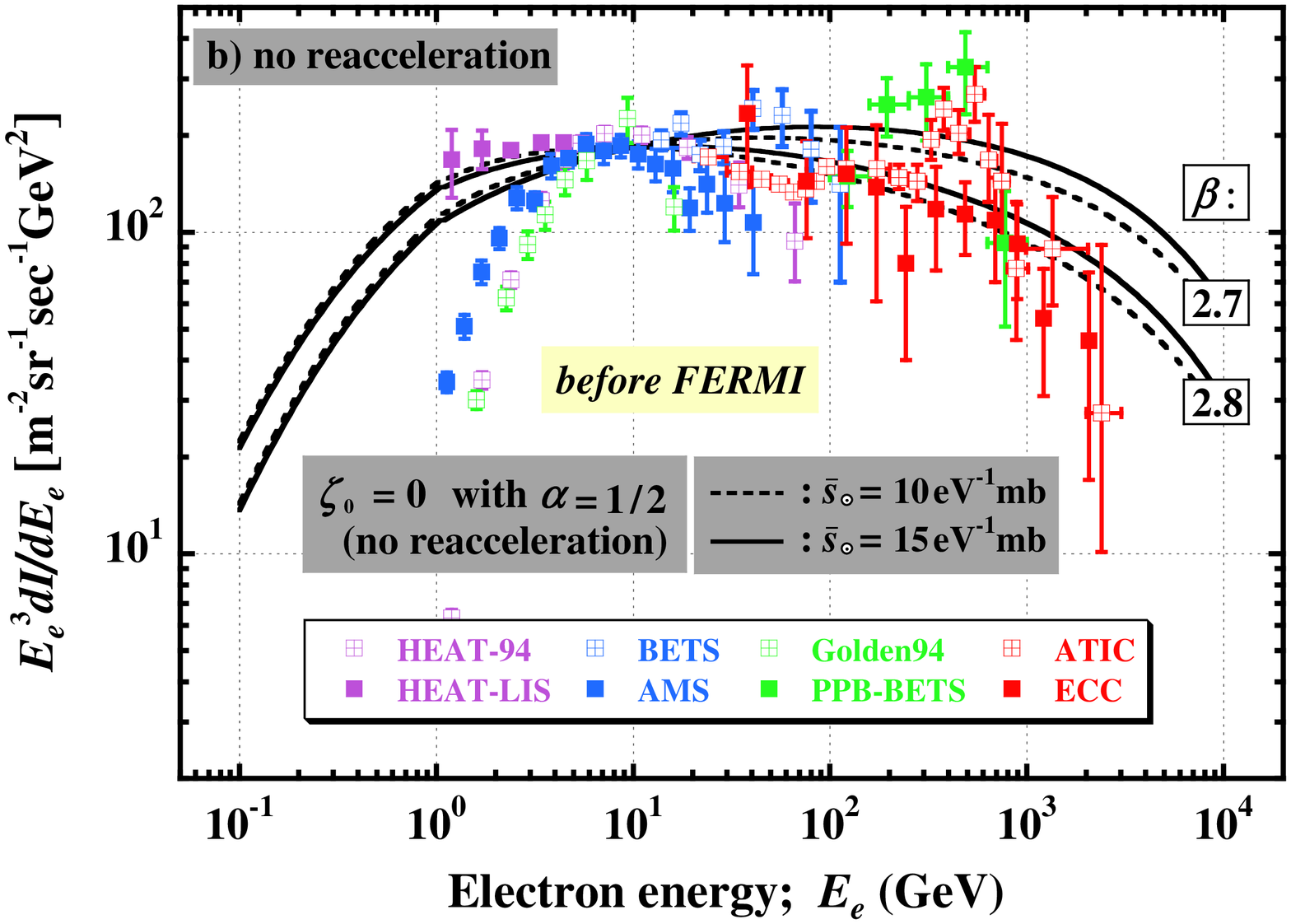}
  \caption{
Electron energy spectra in two models, (a) with reacceleration
  and (b) no reacceleration,
  compared with the measurements, where the vertical axis is 
  multiplied by $E_e^3$. All numerical values are 
  normalized to $E_e$\,=\,10\,GeV, and 
  several sets of ($\beta$, $\bar{s}_r$) are assumed.
  See text for references for individual experimental data.
}
\end{figure}

As we have presented the experimental results on CR hadron components
in the past papers (Papers I--IV), we give here only three kinds of 
secondary-to-primary
ratio with new data, B/C, sub-Fe/Fe, and $\bar{p}/p$,
 that have since become available. See Paper III for the
 secondary unstable nuclei, while new data are still not
 available.

 In Figure~11, we present B/C and
sub-Fe/Fe, plotted together with new ones from CREAM 
(Ahn et al.\ 2008) and TRACER (M\"{u}ller 2009), where
we plot also RUNJOB (Derbina et al.\ 2005) data for reference, 
while the data quality is
rather poor with large atmospheric correction. 
We compare our numerical results with the data
for two models, (a) reacceleration with the set of 
[$\zeta_0,\bar{\sigma}_{\tiny \mbox{$\odot$}}$]\,=\,[50, 150--300]\,mbarn,
for $\alpha$\,=\,$\frac{1}{3}$, 
and (b) no reacceleration with 
[0, 75--150]\,mbarn, for $\alpha$\,=\,$\frac{1}{2}$.
It is still not clear which model 
reproduces the experimental data more satisfactorily. 
As is well known, the advantage of the former
explains naturally the drop of the ratio in the lower energy region
around ACE/CRIS (Davis et al.\ 2000) without assuming an ad
hoc drop in the path length distribution.

Next we present $\bar{p}/p$ in Figure~12, plotted together 
with new data from PAMELA (Adriani et al.\ 2010), where we present 
 numerical curves with several sets of modulation parameters, 
0.2-1.5\,GV, for the reacceleration model shown in Figure 11a.
 One finds that our result is in good agreement with the PAMELA
 in the high energy region around 100\,GeV, where the modulation 
 effect is absolutely negligible.

  \begin{figure}[!b]
    \includegraphics[width=7.7cm]{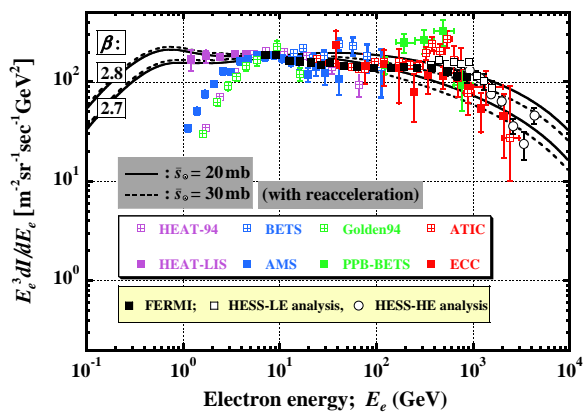}
  \caption{Same as Figure~13a, but with FERMI (Abdo et al.\ 2009, 2010b) 
 and H.E.S.S.\ (Aharonian et al.\ 2009), where
 drawn are numerical curves with the reacceleration 
 shown in Figure 13a.}
\end{figure}
\subsubsection{Electron component}

Let us present the electron data separately before and after 
FERMI, where $\lq \lq$electron" denotes both electron and positron. 
First in Figure~13 we present the electron energy spectrum 
before FERMI, where 
the experimental data are presented for those reported 
in the period from 1994 to 2008 alone 
(Golden et al.\ 1994; Kobayashi et al.\ 1999; 
DuVernois et al.\ 2001; Torii et al.\ 2001, 2006;
Aguilar et al.\ 2002;  Chang et al.\ 2008), and also plotted are
the  data ({\it filled purple squares}) for reference 
after applying a demodulated correction to HEAT data, HEAT-LIS, 
(DuVernois et al.\ 2001)
using the force-field approximation with the modulation 
parameter of 755\,MV (670\,MV) for the 1994 (1995) data.

The numerical curves are normalized at 10\,GeV with 
two indices, $\beta$\,=\,2.7, 2.8, 
assuming two models, (a) reacceleration and (b) no reacceleration
each with the same parameter sets as those used in Figure~11, 
while we assume additionally two cases of $\bar{s}_\odot$,
 [20, 30]\,mbarn for the reacceleration (a), and [10, 15]\,mbarn for 
no reacceleration (b).
Aside from the prominent spectral features around 500\,GeV 
appearing in ATIC (Chang et al.\ 2008) and PPB-BETS (Torii et al.\ 2006) data, 
our model with the reacceleration reproduces the data well in the 
higher energy region, $\gsim 10$\,GeV, in Figure~13a, 
where the solar modulation effect is small. 
On the other hand, the model without reacceleration in Figure\,13b 
is somewhat difficult to fit to the demodulated HEAT-LIS data.

Now, in Figure~14 we present the most recent data
obtained by FERMI (Abdo et al.\ 2009, 2010b) and 
H.E.S.S.\ (Aharonian et al.\ 2009)
together with those presented in Figure~13, where numerical
 curves are the same as shown 
in Figure~13a. 
We find that both FERMI and H.E.S.S.\ data 
do not exhibit the prominent bump around 500\,GeV
reported by ATIC and PPB-BETS, with both giving a spectrum 
falling with energy as $E^{-3}$ up to 1\,TeV, which is not 
inconsistent with emulsion chamber data (Kobayashi et al.\ 1999) 
within the statistical errors.
Looking Figure~14, however, we find that 
FERMI and H.E.S.S.\ data seem to deviate systematically from 
numerical  curves with an enhancement by 
20--30\% around 500\,GeV, indicating still some additional local 
sources of high energy CR electrons,
 which will be discussed again in \S\,7.

\vspace{2mm}
  
\subsection{Diffuse $\gamma$-ray component}
\vspace{1mm}
\subsubsection{Isotropic background $\gamma$-rays}

D$\gamma$'s near the GP are mainly 
hadron-induced  ($\pi^0$\,$\rightarrow$\,$2\gamma$) and 
electron-induced  (EB + IC). In addition to these two components, 
we have isotropic background $\gamma$'s (BGs)
with various origins such as extragalactic sources (EGs), unidentified
sources, instrumental sources,
dark matter (DM), etc, so that the BGs depend
on individual detectors with different sensitivity in
energy and the angular resolution, while depending on the propagation
model as well. Therefore it is not easy task to estimate the extragalactic
  D$\gamma$, while its origin is one of the fundamental problems 
 in astrophysics, studied in so many papers with  various candidates;
   unresolved blazers (e.g.\ Stecker \& Salamon 1996;
 Chiang \& Mukherjee 1998; M\"{u}cke \& Pohl 2000), 
intergalactic shocks produced by
 the assembly of large-scale structures (e.g.\ Loeb \& Waxman 2000; 
 Totani \& Kitayama 2000; Miniati et al.\ 2000; Gabici \& Blasi 2003), 
 dark matter annihilation (e.g.\ Bergstr\"{o}m 2000; Ullio et al.\ 2002; 
Ahn et al.\ 2007), etc. In the present paper, however, we use the 
 acronym 
 $\lq \lq$BGs" all together for D$\gamma$'s other than those induced by
  $\pi^0$, EB (bremsstrahlung) and IC, 
while acknowledging EGRET and FERMI teams have 
 estimated very carefully the EG-$\gamma$ intensity.

  \begin{figure}[!t]
    \includegraphics[width=7.7cm]{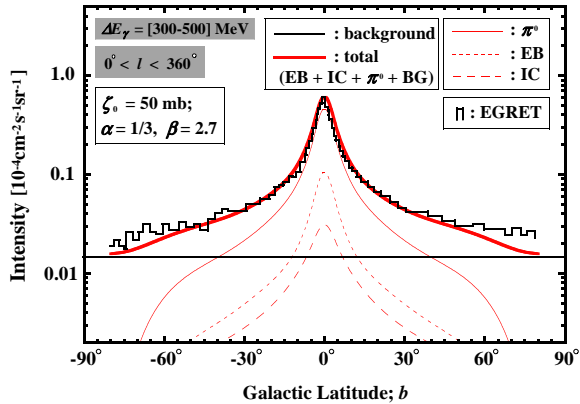}
  \caption{
 An example of the estimation of the BG 
 ({\it black horizontal line}) using the latitudinal D$\gamma$'s 
 data from EGRET (Hunter et al.\ 1997) 
 with the energy interval of 300--500\,MeV averaged over 
 the whole radial direction, $l$\,=\,0$^\circ$--\,360$^\circ$.
}
\end{figure}

In Figure~15, we present an example of EGRET data ({\it histogram}; source
 subtracted) (Hunter et al.\ 1997) together with numerical curves 
on the latitudinal distribution averaged over
full longitude ranges, 0$^\circ$--360$^\circ$ with the energy interval
300--500\,MeV, where we give the contributions of D$\gamma$'s
separately from $\pi^0$ ({\it solid red}), 
EB ({\it dotted red}), IC ({\it broken red}), 
BG ({\it solid black}), and total flux, 
$\pi^0$+EB+IC+BG ({\it heavy solid red}),   
assuming $[\zeta_0, \bar{\sigma}_{\tiny \mbox{$\odot$}}, 
\bar{s}_{\tiny \mbox{$\odot$}}]$\,=\,[50,\,180,\,30\,eV$^{-1}$]mbarn
 with $[\alpha, \beta]$\,=\,[$\frac{1}{3}$, 2.7]. Here we draw a
horizontal line for BG by the use of 
 the least square method so that 
the histogram is well reproduced, where
 the fitting is applied for $|b| \le 60^{\circ}$ as there remain 
 considerable uncertainties in the latitudinal distribution 
for both the ISM and ISRF {\it far distant} from the GP, see 
${\it \Xi}_h(r, z)$ ($\lq \lq h" \equiv \mbox{H\,I, H}_2$) 
in equation (2) and
 the scale height $z_{\scriptsize \mbox{ph}}$ in equation (7).
It is remarkable that the numerical curve is in good agreement with
the data not only in shape, but also in absolute value,
 except the high latitude around the galactic pole.

In Figure~16, we summarize the intensity of BGs obtained by
  past works, Kappadath et al.\ (1996) for COMPTEL, 
Sreekumar et al.\ (1998) for EGRET, Strong et al.\ (2004) for
 EGRET (revised), and Abdo et al.\ (2010a) for FERMI, where
  also plotted are those estimated in this work (see Figs.\ 15 and 17)
 for the reference,
 six points ({\it open circles}) for EGRET 
and one point ({\it filled circle}) for FERMI.
 We draw a broken line 
 given by Abdo et al.\ with $dI_{\tiny {\mbox{BG}}}/dE_\gamma
 = 9.6 \cdot 10^{-3}
\times E_\gamma^{-2.41}$ in units of [cm$^{-2}$s$^{-1}$sr$^{-1}$MeV$^{-1}$]
 with $E_\gamma$ in MeV,
 and a solid curve with 
$dI_{\tiny {\mbox{BG}}}^*/dE_\gamma = dI_{\tiny {\mbox{BG}}}/dE_\gamma
 \times [1- \exp{(-0.33E_\gamma^{0.4})}]$, used in the
 present work, 
 slightly modifying the FERMI result in the low energy region,
 while the modification does not affect any change for the
 results.
 Significant difference
 between EGRET and FERMI, with the former giving much harder
 spectrum than the latter, might be 
due to a diffrent model in CR propagation as well as 
those in ISM and ISRF.

  \begin{figure}[!t]
   \vspace{-1.5mm}
    \includegraphics[width=7.7cm]{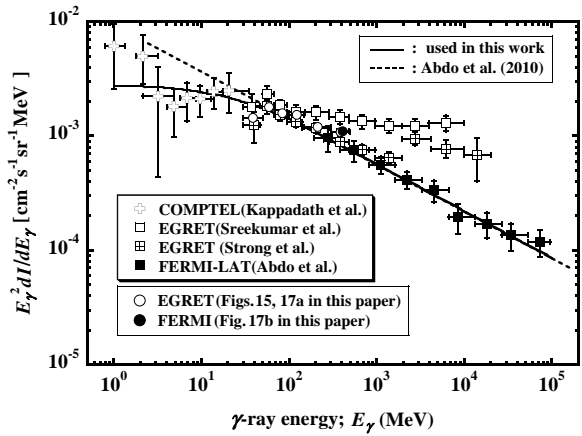}
  \caption{
 The BG spectrum  
 obtained by COMPTEL (Kappadath et al.\ 1996),
 EGRET (Sreekumar et al.\ 1998), EGRET (revised; Strong et al.\ 2004), 
 and FERMI (Abdo et al.\ 2010a), where also plotted are those
 estimated by the present work using EGRET and FERMI data 
 (see Figs.\ 15 and 17). 
 Dotted line is given by Abdo et al., and the solid curve is
 used in the present work (see text), modifying it slightly in the
  low energy region.
\blankline
}
\end{figure}

\subsubsection{Spatial distribution}

We present two examples of 
the latitudinal distributions for EGRET and FERMI (Porter 2009) 
with the energy interval around [300--500]\,MeV in 
Figures 17a and 17b respectively, together with our
numerical results taking the BG contribution ({\it broken-dotted lines}) 
into account mentioned
above, $dI_{\tiny {\mbox{BG}}}^*/dE_\gamma$, 
where plotted are three curves for each figure with $\beta$\,=\,2.6
({\it green}), 2.7 ({\it red}), and 2.8 ({\it blue}). One finds the
agreement between the data and the curves is excellent except
 the high latitude $|b|$\,$\gsim$\,$60^\circ$.
In these calculations, we take the angular resolution (PSF) effect
with the energy dependence into account, for instance, with 
7$^\circ$ (HWHM) at 30--50\,MeV  (Hunter et al.\ 1997).

  \begin{figure}[!t]
    \includegraphics[width=7.7cm]{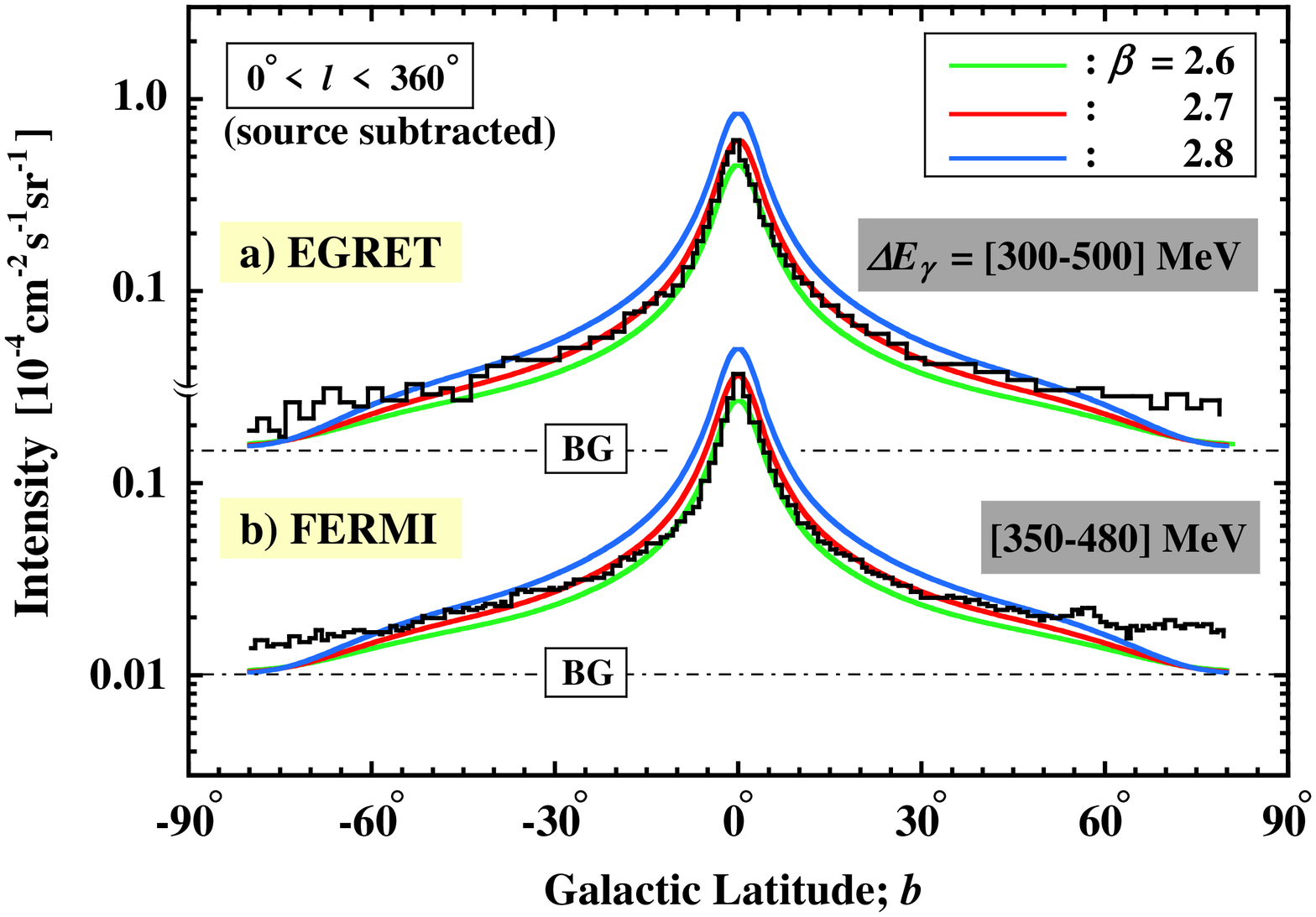}
  \caption{
Latitudinal distributions of D$\gamma$'s 
  obtained by a) EGRET (Hunter et al.\ 1997) with the energy 
   range 300--500\,MeV, and b) FERMI (Porter 2009) with the energy
  range 350--480\,MeV, both averaged over the
 the whole radial direction, $l$\,=\,$0^\circ$--$360^\circ$, 
  where drawn are three curves with $\beta$\,=\,2.6 ({\it green}), 
  2.7 ({\it red}), and 2.8 ({\it blue}), 
  taking the BG contributions into account.
\blankline
}
\end{figure}
  \begin{figure}[!h]
    \includegraphics[width=7.8cm]{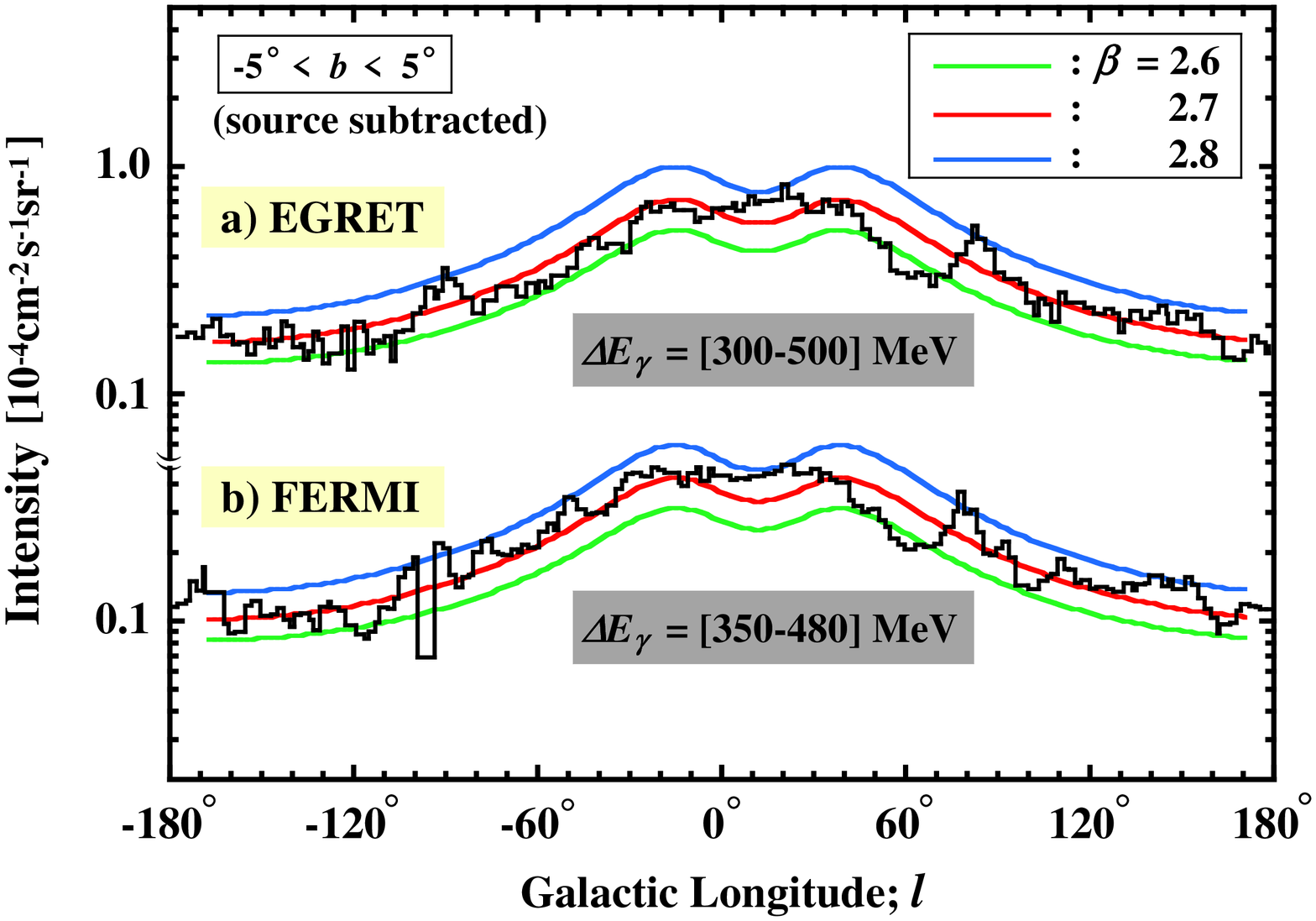}
  \caption{
Same as Figure 17, but for 
longitudinal distribution with the same condition, 
where BG contributions are not 
 presented.
}
\end{figure}

Corresponding to the latitudinal distributions as shown in 
Figures~17a and 17b, we demonstrate the longitudinal distributions 
 near the GP in Figures~18a and 18b, where numerical curves are 
  shifted  by $\Delta l = +10^\circ$ in both EGRET and FERMI 
so that experimental data are reproduced more satisfactorily.
Again we find the 
numerical results are in nice agreement with the data in 
both shape and absolute value, and consistent with  $\beta \sim 2.7$.

\begin{figure}[!t]
  \begin{center}
    \includegraphics[width=7.7cm]{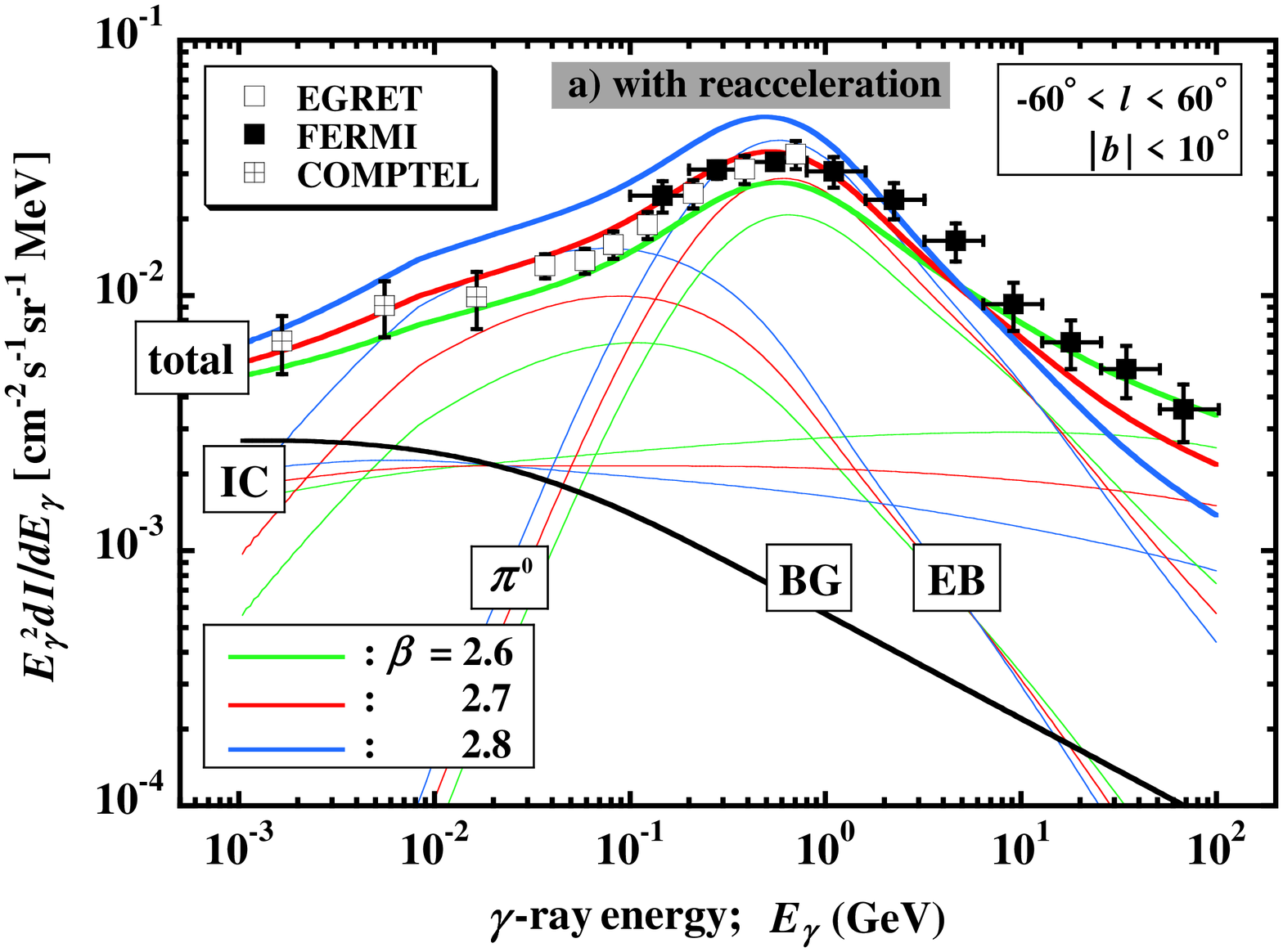}
    \includegraphics[width=7.7cm]{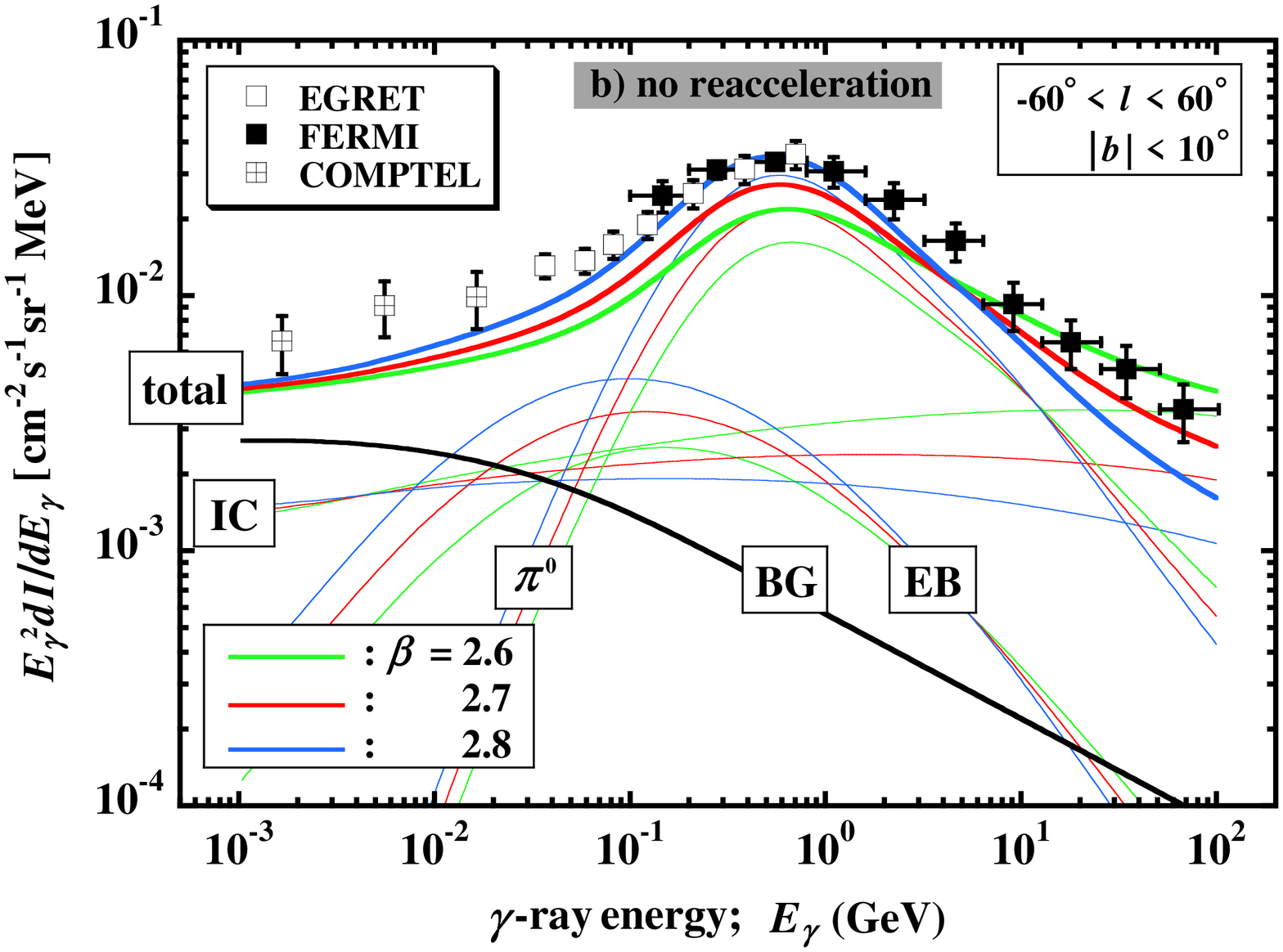}
  \end{center}
  \caption{
Differential energy spectra of D$\gamma$'s 
 averaged over the whole radial direction with $|b| < 10^\circ$ 
obtained by
  COMPTEL (Kappadath et al.\ 1996), EGRET (Hunter et al.\ 1997) and 
 FERMI (Abdo et al.\ 2010b). Numerical curves are demonstrated for
 two cases, a) with the reacceleration and b)
 no reacceleration, each presented  separately for 
 individual components.
\blankline}
\end{figure}

\begin{figure}[!t]
    \includegraphics[width=7.5cm]{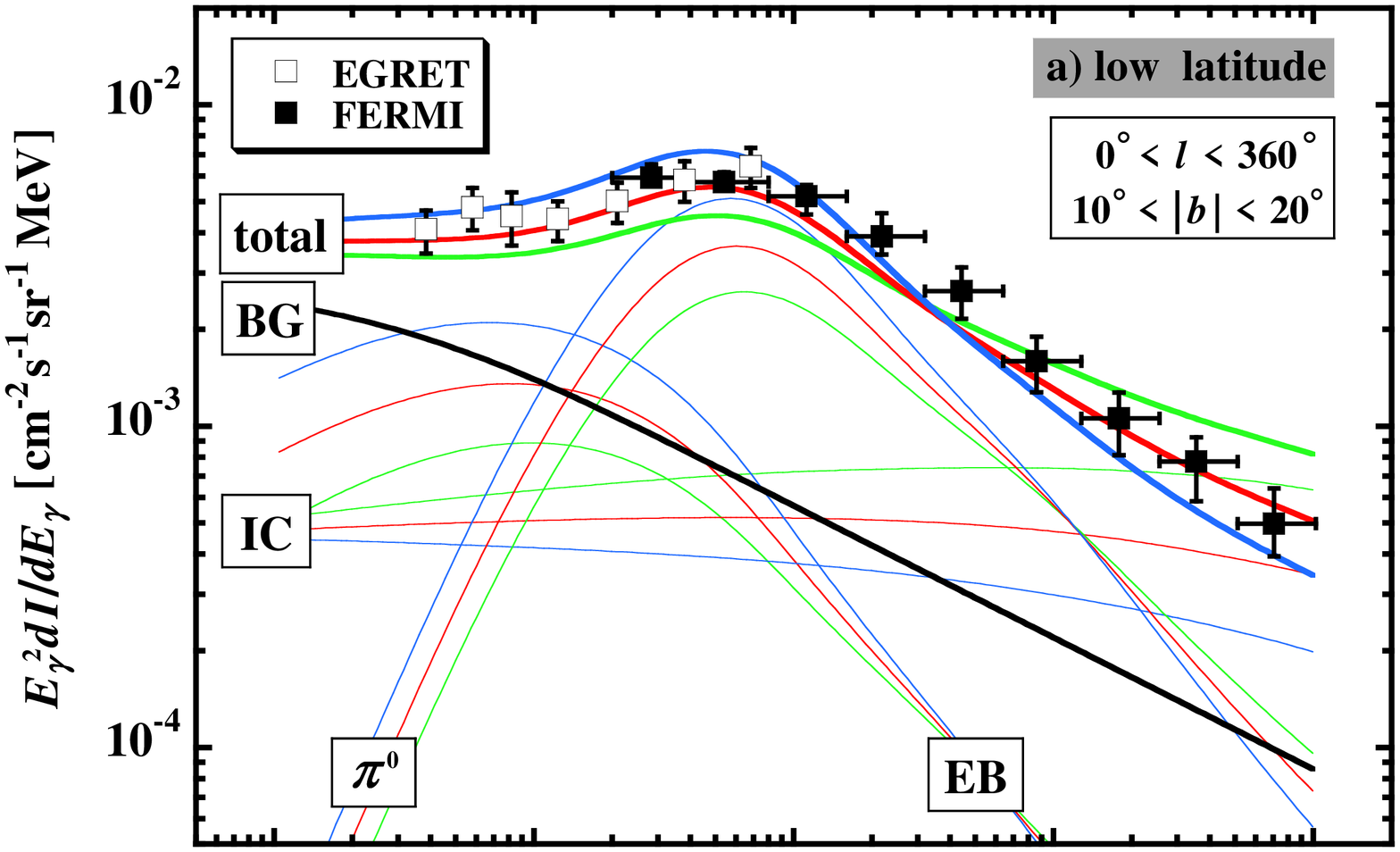}
    \includegraphics[width=7.5cm]{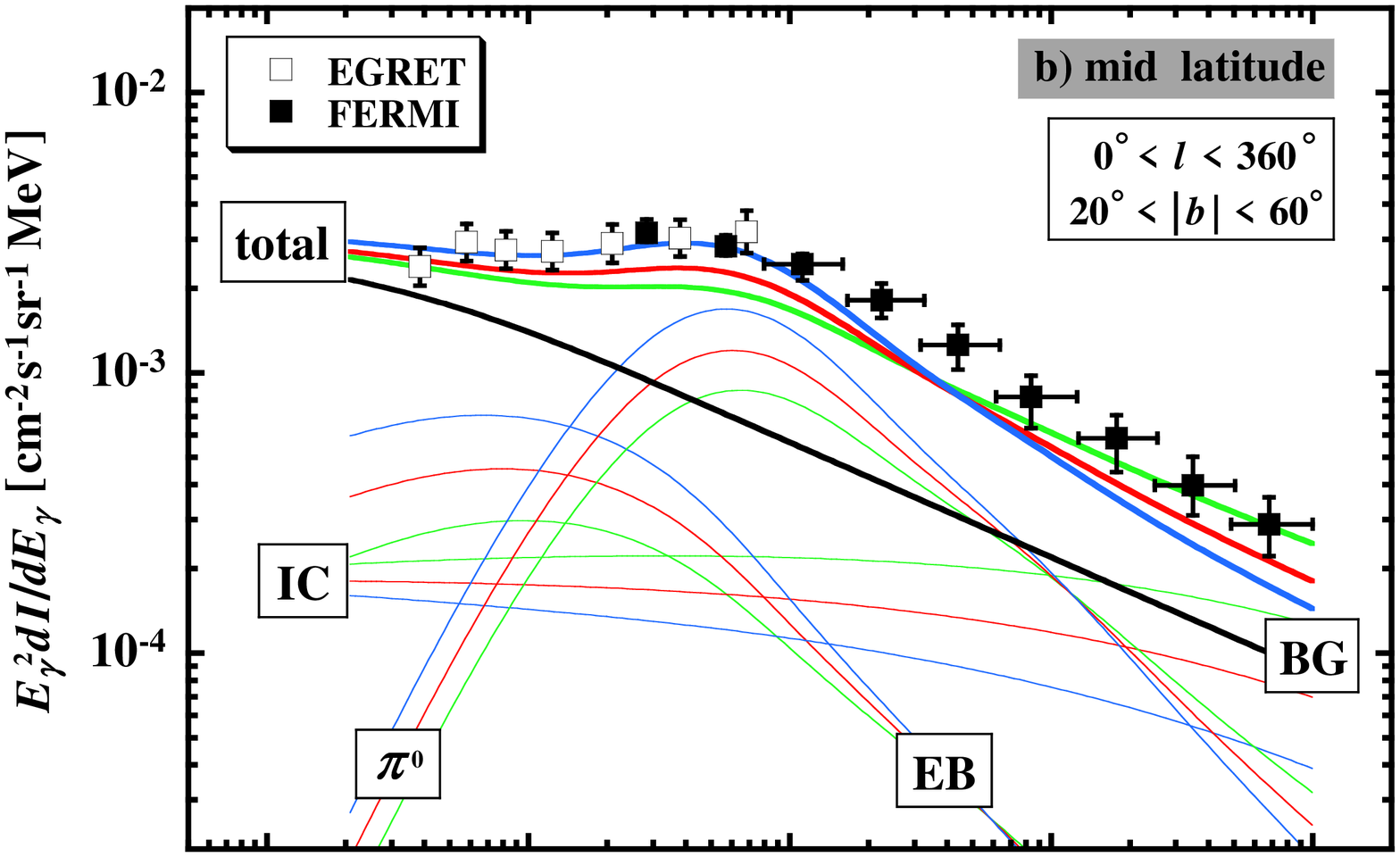}
    \includegraphics[width=7.5cm]{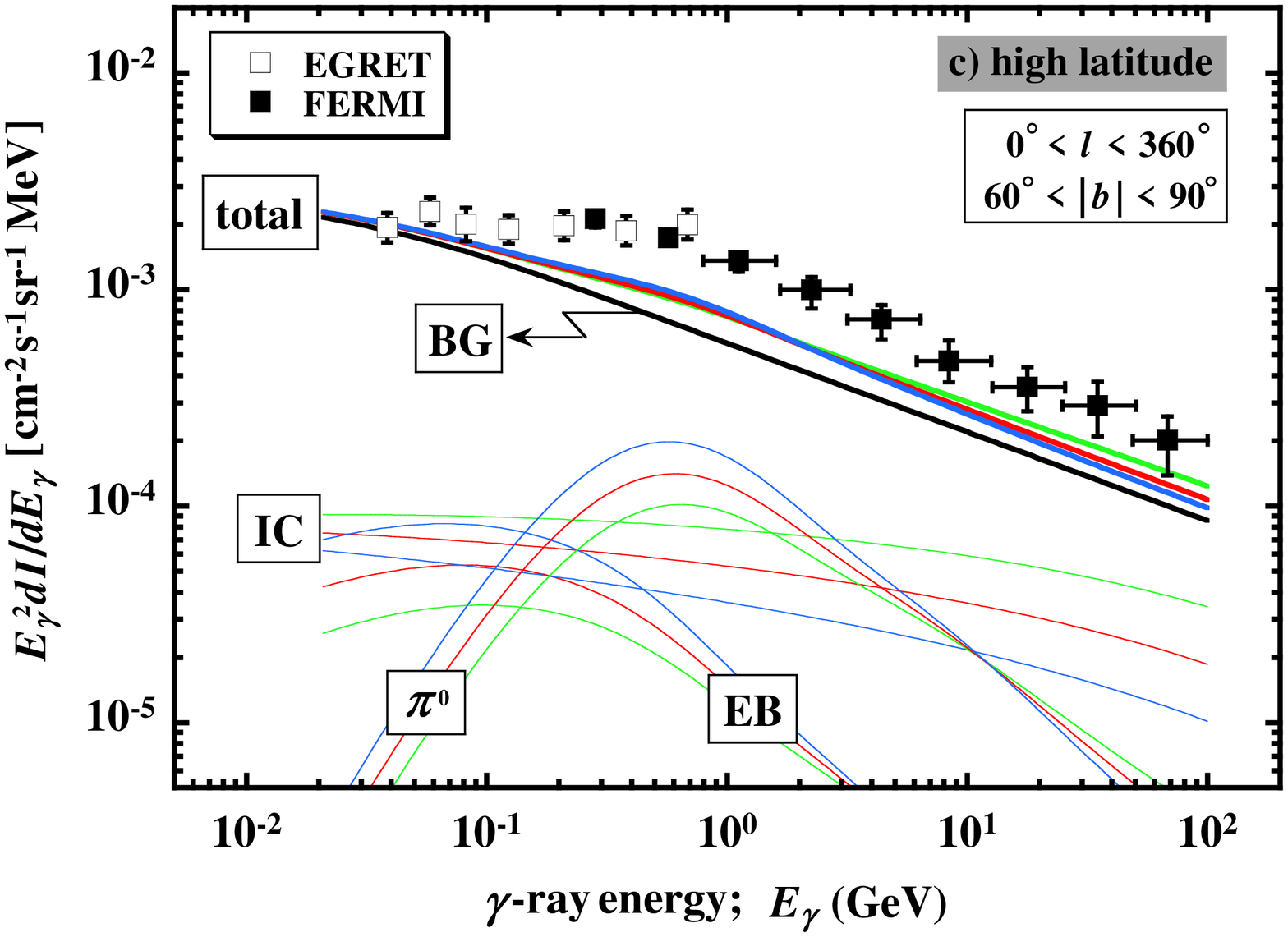}
  \caption{Same as Figure 19a, but those averaged over
 three different ranges in the galactic latitude, a) low latitude with 
 $10^\circ < |b| < 20^\circ$, b) mid latitude with
 $20^\circ < |b| < 60^\circ$, and c) high latitude with 
 $60^\circ < |b| < 90^\circ$. Parameter sets for the numerical
 calculations are the same as those used in Figure 19a with the
 reacceleration.}
\end{figure}

\begin{figure}[!t]
    \includegraphics[width=7.5cm]{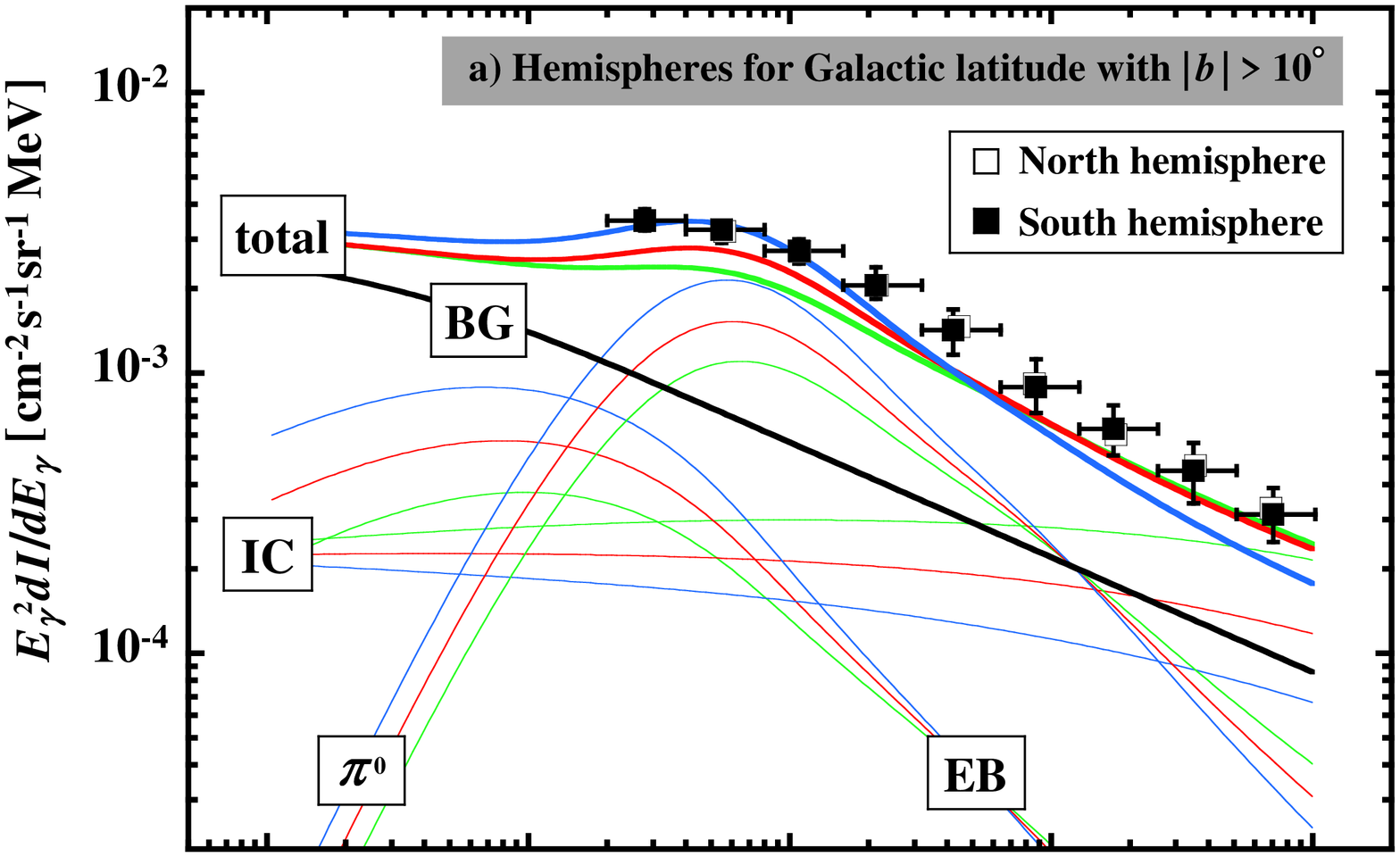}
    \includegraphics[width=7.5cm]{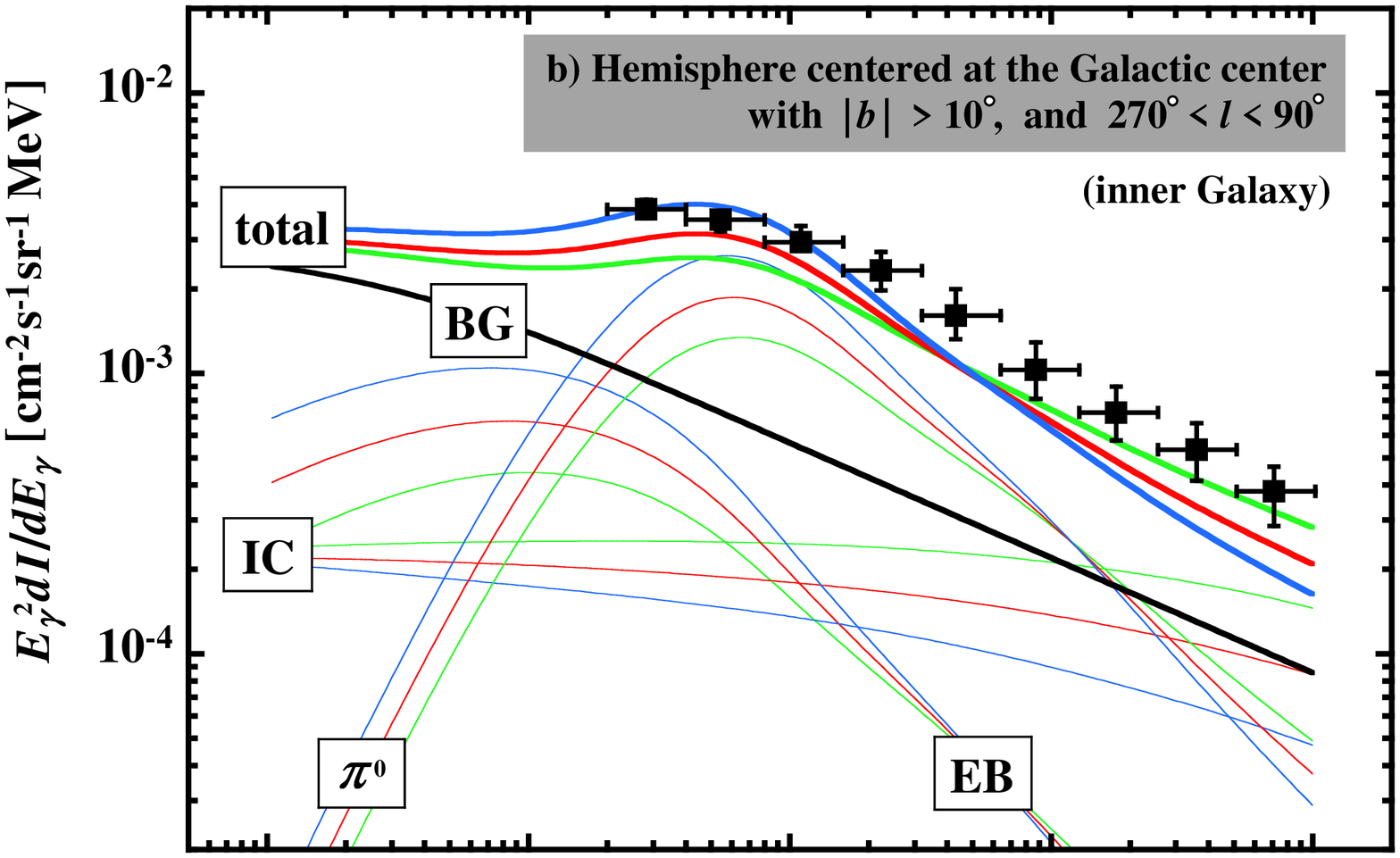}
    \includegraphics[width=7.5cm]{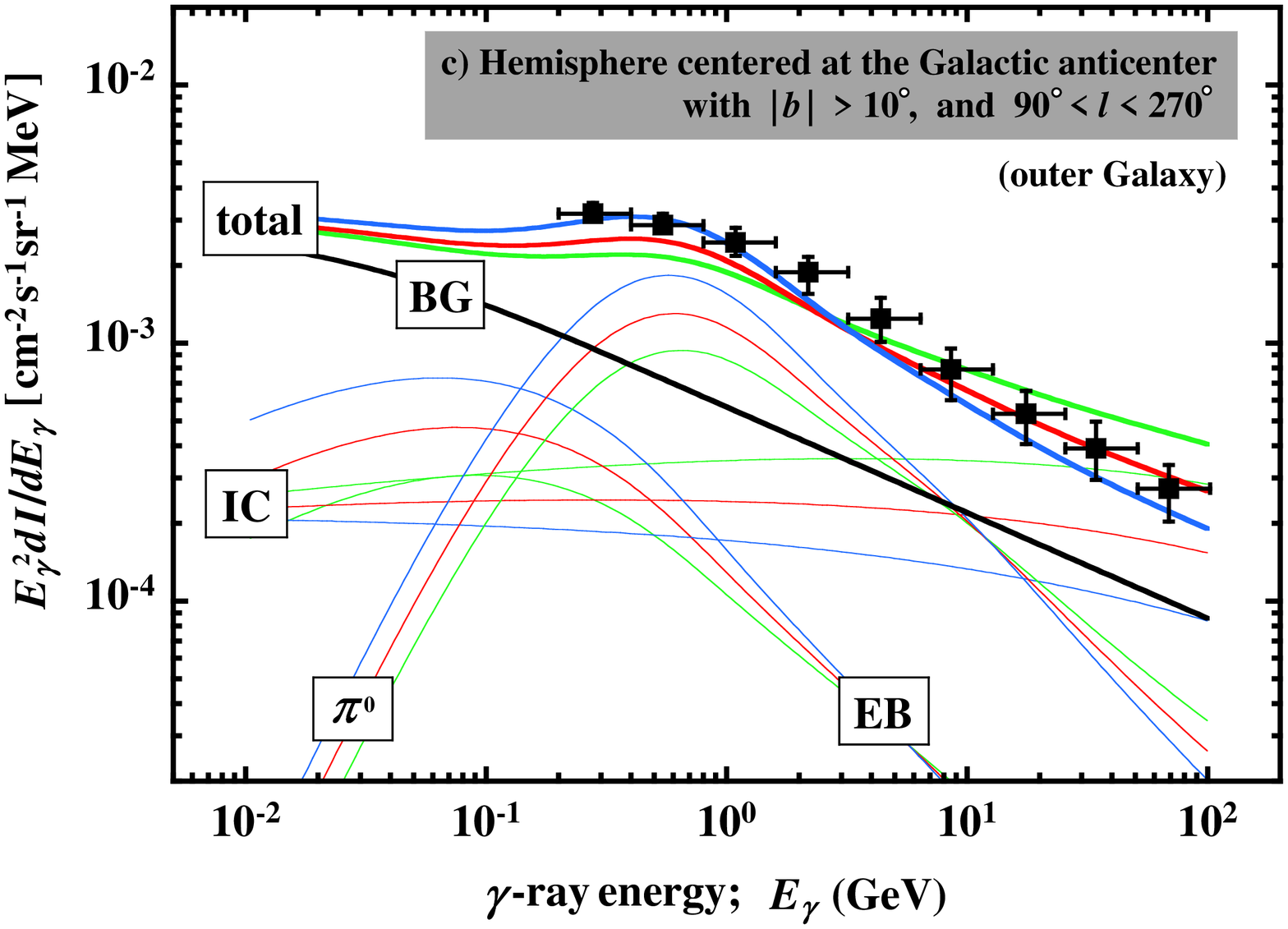}
  \caption{
 Same as Figure 20, but those averaged over
   different hemispheres on the sky for the galactic latitudes with
 $|b| > 10^\circ$, which are  centered at
 a) the north galactic pole ({\it open square}) and the 
 south galactic pole ({\it filled square}), b) the galactic center
 with $270^\circ < l < 90^\circ$ (inner Galaxy), and c) the anticenter 
 $90^\circ < l < 270^\circ$ (outer Galaxy).}
\end{figure}

\subsubsection{Energy spectrum}

First, in Figure 19a we present the energy
spectrum of D$\gamma$'s averaged over the field of view with
 $-60^\circ$\,$<$\,$l$\,$<$\,$60^\circ$ and $|b|$\,$<$\,$10^\circ$, 
where numerical curves with the reacceleration are 
also presented separately for those coming from $\pi^0$,  EB, and IC 
(all with {\it colored thin solid curves}), 
BG ({\it heavy black solid curve}), and total
 ({\it colored heavy solid curves}).
 Here and in the following we omit  EGRET data in GeV region, 
 because of the instrumental problem in detection of 
$\gamma$'s (Stecker et al.\ 2008).
One finds that the curve with $\beta$\,=\,2.7 ({\it red}) is in good
agreement with the data in the energy region below 1\,GeV, while 
they deviate slightly from the curve  
above 1\,GeV, with approximately 20\% enhancement.

Figure~19b reproduces Figure~19a, but for curves 
without reacceleration, corresponding to Figures~11b and 13b (see also
Fig.\,8b). The fit is not as 
good as for the reacceleration model, particularly in the 
low energy region, $E_\gamma$\,$\lsim$\,200\,MeV, 
 with $\sim$\,40\% enhancement, while with $\sim$\,20\% 
 in the high energy region, $E_\gamma$\,$\gsim$\,1GeV,
 giving nearly the same enhancement as in the case of (a) with 
 the reacceleration.

Second, in Figures 20 and 21, we present the energy
spectra of D$\gamma$'s for different sky views
(Abdo et al.\ 2010a; see also supplementary material at
 http://link.aps.org/supplement-al/10.1103/PhysRevLett.101101).
Figure 20 shows those averaged over independent 
galactic latitude ranges covering 
 low, mid and high galactic latitudes, 
a) $10^\circ$\,$<$\,$|b|$\,$<$\,$20^\circ$, 
b) $20^\circ$\,$<$\,$|b|$\,$<$\,$60^\circ$ and 
c) $60^\circ$\,$<$\,$|b|$\,$<$\,$90^\circ$ respectively. 
 Figure 21 shows those averaged over different hemishperes,
 which are, a) centered at the north ($b \ge 0^\circ$; 
{\it open squares}) and
 south ($b \le 0^\circ$; {\it filled squares}) galactic poles, 
b) the galactic center
 ($270^\circ \le l \le 90^\circ$), and c) anticenter
 ($90^\circ \le l \le 270^\circ$), all with the galactic
 latitudes excluding $|b| < 10^\circ$. In these figures,
 we subtract $\gamma$'s coming from point sources based on the
 FERMI catalog.

 It is remarkable in Figure 20 that EGRET and FERMI data agree pretty well 
  with each other, overlapping nicely
 around 0.2--1GeV, in all latitude ranges.
One finds that the numerical curves 
with $\beta=$\,2.7 reproduce generally well 
both the EGRET and FERMI data in Figures 20 and 21 but 20c,
 taking account of the uncertainties in various galactic parameters,
 particularly in those related to the ISM and ISRF.

\begin{figure}[!t]
  \begin{center}
    \includegraphics[width=7.7cm]{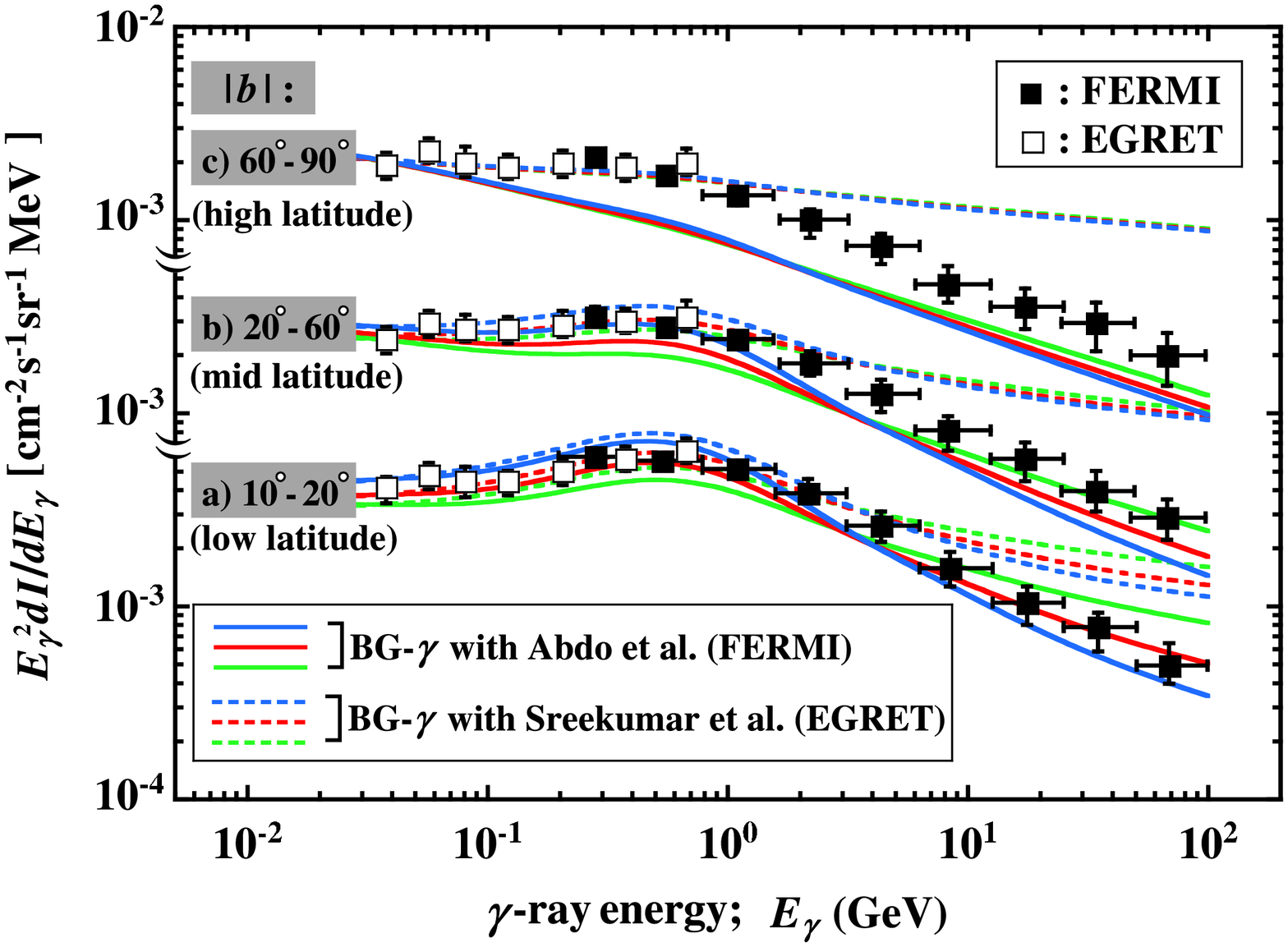}
  \end{center}
  \caption{Same as Figure 20, but two cases of numerical
 curves, a) with the BG-$\gamma$ given by FERMI (Abdo et al.\ 2010a)
 ({\it solid curves}),  
 and b) with the BG-$\gamma$ by EGRET (Sreekumar et al.\ 1998)
 ({\it dotted curves}).
}
\end{figure}

 On the other hand, in Figure 20c for the high latitude, $|b| > 60^\circ$,
we have a noticeable enhancement in FERMI with approximately 70\%
 as compared to the numeical curves.
To see the deviation more clearly, we present them all together
 in Figure 22, where we show additionally numerical curves 
 ({\it dotted colors}) using EGRET-BG 
 obtained by Sreekumar (1998) for reference (see Fig.\ 16). 
  Figure 23 reproduces Figure 21 with numerical curves using 
EGRET-BG ({\it dotted colors})
 in addition to those using FERMI-BG ({\it solid colors}).

 One finds the spectrum shapes with EGRET-BG are quite different from the
 data in the high energy region, although the enhancement 
 is rather improved in the energy region $\lsim$\,1\,GeV, 
which is discussed again in the next 
 section.

\begin{figure}[!t]
  \begin{center}
    \includegraphics[width=7.7cm]{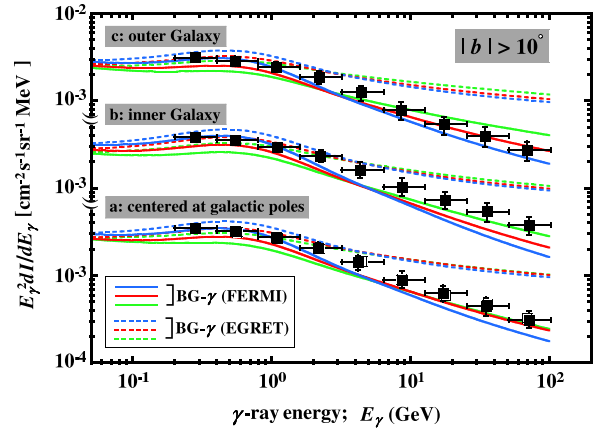}
  \end{center}
  \caption{Same as Figure 21, but two cases of numerical
 curves, a) with the BG-$\gamma$ given by FERMI (Abdo et al.\ 2010a)
 ({\it solid curves}),  
 and b) with the BG-$\gamma$ by EGRET (Sreekumar et al.\ 1998)
 ({\it dotted curves}).
}
\end{figure}

\section{Discussion and summary}

We have studied the diffusion-halo model with stochastic
reacceleration, comparing it with the most recent data on hadronic,
electronic and D$\gamma$ components. We have two particular
interests: to find an unified model for
the CR acceleration and propagation from the viewpoint of
astrophysics, and to search for a signal of novel sources
such as PBH and/or DM from the viewpoints of particle physics and
cosmology.  Both are of course closely connected with each other in
the sense that the knowledge of the former is decisive in
confirming the latter. While several groups
(Torii et al.\ 2006; Chang et al.\ 2008) have
reported the possibility of annihilation and/or decay of DM particles, 
giving a significant bump in electron flux around 500\,GeV, 
FERMI (Abdo et al.\ 2009) and H.E.S.S.\ (Aharonian et al.\ 2009)
give  a rather flat spectrum up to 1\,TeV without the prominent excess.
In the present paper, however, we have focussed our interest rather
conservatively on the internal consistency among various
CR components from the view point of astrophysics, leaving the
puzzle of the possible electron/positron-excess to further observations and 
mutual cross-checks in data analysis among individual
groups.

In our past works on the hadronic component, we concluded that the
diffusion-halo model with the reacceleration with the parameter set,
$[\zeta_0,\,\bar{\sigma}_{\tiny \mbox{$\odot$}}]$\,=\,[50, 180]\,mbarn
with $[\alpha,\,\beta]$\,= [$\frac{1}{3}$, 2.7--2.8], is
in harmony with the CR hadron data presently available.
The most recent data on the B/C ratio by CREAM and TRACER (Fig.\ 11)
as well as on the $\bar{p}/p$ ratio by PAMELA (Fig.\ 12) 
also support the present model.  However,
it is worth mentioning here that our interpretation for the
energy dependence of the B/C ratio is somewhat different from that by
CREAM (Ahn et al.\ 2008) and TRACER (M\"{u}ller 2009). 

They claim that the index $\alpha$
favors 0.5--0.6 instead of $\frac{1}{3}$, resulting in a rapid
decrease with energy for the interstellar propagation path length.
In contrast to their interpretation, 
we would like to point out that the value of 0.5--0.6 is not
{\it fundamental}, but is rather {\it accidental} 
due to the reacceleration
effect, namely it is boosted upward around the GeV region by the energy
gain, resulting coincidentally in
the soft slope with 0.5--0.6 in the energy
region 1--100\,GeV.  
The intrinsic one must be $\frac{1}{3}$ (Kolmogorov-type
for wave number spectrum in hydromagnetic turbulence), leading to
1) a natural drop in path length distribution in the low energy region
$\lsim$\,1\,GeV without introducing an artificial break there,
as originally proposed by Simon et al.\ (1986), and 2) a reasonable 
 amplitude in the anisotropy of CR's with the level of $10^{-3}$
 in TeV region nowadays established experimentally.
 
We apply the diffusion-halo model with and without the stochastic
reacceleration for the electron and D$\gamma$ components.
Apart from the electron-anomaly around 500\,GeV, we find that
the parameter set with the reacceleration,
$[\alpha, \beta; \zeta_0$, $\bar{\sigma}_{\tiny \mbox{$\odot$}}]$,
expected from hadron component reproduces rather well both the 
 spectrum shape and the absolute value in both  the 
electron (Fig.\ 13a) and D$\gamma$ (Figs.\ 17--21) components, 
assuming the additional parameter $\bar{s}_{\tiny \mbox{$\odot$}}$ 
with 20--30\,eV$^{-1}$mbarn. Physical meanings of the numerical
 set with 
$[\zeta_0,\,\bar{\sigma}_{\tiny \mbox{$\odot$}}]$\,=\,[50, 180]\,mbarn
 are discussed in Papers I--III in connection with the 
 diffusion constant $D_{\tiny \mbox{$\odot$}}$, gas density 
$\bar{n}_{\tiny \mbox{$\odot$}}$,
 their scale heights, $\zD, z_n$, etc, giving reasonable values
  matched with the observtional data.

Let us consider the physical meaning of
20--30\,eV$^{-1}$mbarn in $\bar{s}_{\tiny \mbox{$\odot$}}$.
The relation between $\bar{\sigma}_{\tiny \mbox{$\odot$}}$ and
$\bar{s}_{\tiny \mbox{$\odot$}}$ is given by
$$
\vspace{2mm}
\frac{\bar{s}_{\tiny \mbox{$\odot$}}}{\bar{\sigma}_{\tiny \mbox{$\odot$}}}
 = \frac{\bar{n}_{\tiny \mbox{$\odot$}}}
{\bar{\epsilon}_{\tiny \mbox{$\odot$}}}
\frac{1+1/\kappa}{1+1/\nu} = \frac{\bar{n}_{\tiny \mbox{$\odot$}}}
{\bar{\epsilon}_{\tiny \mbox{$\odot$}}}
\frac{2+\zD/z_\epsilon}{2+\zD/z_n},
$$
see \S\,4.2 for $\bar{s}_{\tiny \mbox{$\odot$}}$ and 
$\bar{\sigma}_{\tiny \mbox{$\odot$}}$ with $r$\,=\,$r_{\tiny \mbox{$\odot$}}$, 
and Table~4 for $\nu$ and $\kappa$. Namely, it is  
closely related to the ratio of the energy density 
$\bar{\epsilon}_{\tiny \mbox{$\odot$}}$ to the
gas density $\bar{n}_{\tiny \mbox{$\odot$}}$ at the SS, 
for the {\it smeared} energy density, {\it smeared} gas density
respectively,
and three latitudinal scale heights, $z_\epsilon$, $z_n$ and
$\zD$.
As discussed in \S\S\,2 and 3.1, 
we have ${n}_{\tiny \mbox{$\odot$}}=
n_{\tiny \mbox{HI}}^\odot + n_{\tiny \mbox{H$_2$}}^\odot + 
n_{\tiny \mbox{HII}}^\odot =
 1.14$H\,atoms\,cm$^{-3}$, and 
${\epsilon}_{\tiny \mbox{$\odot$}}=
\epsilon_{\tiny \mbox{B}}^\odot + \epsilon_{\scriptsize \mbox{ph}}^\odot 
 \approx 2.3$\,eV\,cm$^{-3}$, leading to 
${\epsilon}_{\tiny \mbox{$\odot$}}/{n}_{\tiny \mbox{$\odot$}} \approx$
 2\,eV. Remembering that the scale heights used in the present paper are
[$z_n, z_\epsilon; \zD$] = [0.2, 0.75; 3.0]\,kpc,  we find  
$\bar{s}_{\tiny \mbox{$\odot$}} = 32$\,eV$^{-1}$mbarn for
$\bar{\sigma}_{\tiny \mbox{$\odot$}} = 180$\,mbarn, giving  a
consistent result, while the latter with 180\,mbarn 
is expected from the relation, 
$\bar{\sigma}_{\tiny \mbox{$\odot$}}$\,$\simeq$\,$D_{\tiny \mbox{$\odot$}}/[\bar{n}_{\tiny \mbox{$\odot$}}c\zD z_n]$  with a reasonable set 
$[D_{\tiny \mbox{$\odot$}}, \bar{n}_{\tiny \mbox{$\odot$}}]$\,=\,[3$\times$10$^{28}$\,cm$^2$\,s$^{-1}$, 1\,cm$^{-3}$] and 
$[\zD, z_n]$\,=\,[3, 0.2]\,kpc as discussed in Papers I, II.

As mentioned above, 
the electron spectra currently available are {\it generally} 
 in agreement with those expected from the hadron spectra, 
 considering the uncertainties inherent in both the experimental data
 and the numerical parameters, but not quite satisfactory,  
with the FERMI data giving the excess by 20--30\% around several hundred
 GeV compared to the
 numerical results as seen in Figure 14. It might be related to the
 positron excess around 10--100\,GeV observed by PAMELA (Adriani et al.\ 2009),
 indicating some nearby sources 
 and/or exotic ones from DM annihilation or decay,
 while beyond the subject of the present paper.
  In fact most recently Delahaye et al.\ (2010)  show that 
 the electron spectra with FERMI, HESS and PAMELA 
  are reproduced rather well by the standard astrophysical processes,
 assuming two sources separately, the distant and local nearby ones, 
 whereas they stress that there remain too large
 theoretical uncertainties to build 
a standard model for CR electrons. So it is critical to study the 
  D$\gamma$'s and diffuse radio emissions simultaneously in order to reduce 
  the uncertainties inherent in the galactic parameters assumed
 for the numerical calculations. 

We compared our
numerical results on the energy spectrum of D$\gamma$'s with EGRET and
FERMI data for several sets of the field of view (Figs.\ 17--21), 
 and found that overall, the CR data, hadron
(Fig.\,6 in Paper V) and electron (Fig.\,14) components, reproduce 
rather satisfactorily D$\gamma$'s for both EGRET and FERMI, 
considering the fact that we have uncertainties with at least 10--20\% 
 in the galactic parameters assumed here as well as 
in the flux normalization of the hadron and electron components.
  Small enhancements of D$\gamma$'s in GeV region (Figs.\ 19, 20), 
albeit they are still within the uncertainties, may indicate those 
from  nearby sources such as the supernova remnants, pulsars,
 and pusar wind nebulae. 

We found, however, that FERMI data give the significant excess with
 approximately 70\% or more in the {\it high latitude} (Fig.\ 22), 
 well beyond the uncertaities, against the 
 numerical results in GeV region. This result may indicate a signature of  
 very large electron-halo far distant from the GP, with, for instance,
 as large as 25kpc (Keshet et al.\ 2004), and/or something else coming from
 the cosmological origin.
 We are also concerned if the
 exess here discussed relates to those 
 appearing in the electron spectrum between 100 and 1000GeV 
 observed by FERMI and HESS (Fig.\ 14) and in the positron spectrum around
 several tens GeV by PAMELA.
 To make clear the correlation between these excesses, D$\gamma$'s in
 the high latitude and the electrons/positrons around several 
tens to hundred GeV, crucially important
 is the anisotropy study for the high energy electron, 
 which will be discussed elsewhere in the near future.

Finally we briefly argue the electron spectrum 
obtained by FERMI 
from the observational point of view, aside from the prominent bumps
 indicated by ATIC and PPB-BETS.
 FERMI is indeed excellent in the 
observation for $\gamma$-rays, we have some concerns about the 
separation of electrons from hadrons as well as their energy 
determination in the high energy region,  
while acknowledging the team have studied very carefully the 
reliability from various kinds of checks, with both beam tests 
and the simulational analyses.

Nevertheless, one should keep in mind that FERMI is 
not {\it purely-direct} observations for electrons, 
but {\it quasi-direct} ones in
the sense that electron events are selected by {\it statistical} 
analysis based on simulations for the spread of electron showers, 
where a small number of electrons are statistically 
selected from a large
proton background. In contrast to these quasi-direct experiments, 
the PAMELA apparatus consists of a permanent magnetic spectrometer 
with a silicon tracking system, providing good identification 
between electrons  and positrons, 
though limited to a maximum detectable rigidity (MDR) of 100\,GV.

Anyway, we await further studies and mutual cross-checks
among the groups from various points of view to get a firm conclusion
for the electron-excess around 500GeV, 
while not so prominent as given by ATIC and PPB-BETS.
 So results from
the AMS program (Bindi 2009), AMS-02,  will be of particular interest.
This program aims at high precision measurements
of CR (both electron and hadron) and $\gamma$-ray fluxes from
a few hundred MeV
to a few TeV using a super-conducting magnet\footnote{After submitting 
 the present paper, we find that they decided to use the
 permanent magnet in place of the super-conducting magnet
(Kounine 2010).},  
with the space shuttle launch scheduled for September 2010.
 We also look forward to 
D$\gamma$'s data from  ground-based telescopes currently 
operating such as H.E.S.S., MAGIC, and VERITAS, as well as the
CTA-program now under consideration
(Caballero et al.\ 2008), the threshold energies of which are now overlapping
with the FERMI satellite data.

\acknowledgments

We are very grateful to P.G.\ Edwards 
(CSIRO Astronomy and Space Science) 
for his careful reading of the manuscript and valuable comments.

\appendix 

\begin{center}
APPENDIX A \\ 
  ENERGY LOSS OF ELECTRONS IN ISM and ISRF
\end{center}

The energy-loss rate due to 
the bremsstrahlung in the gas density $n(\vct{r})$ is given by
$$
\vspace{-3mm}
-\biggl\langle \frac{{\it \Delta}E_e}{{\it \Delta} t} 
\biggr\rangle_{\hspace{-0.5mm}\mbox{\scriptsize rad}}
=
 \int_0^{E_e} E_\gamma [n(\vct{r}) c
\sigma_{\mbox{\tiny EB}}(E_e, E_\gamma)]dE_\gamma
= n(\vct{r}) w_{\mbox{\scriptsize rad}}(E_e)E_e, 
\eqno{(\rm A1)}
$$
where
$$
\vspace{-3mm}
w_{\mbox{\scriptsize rad}}(E_e) = w_{\mbox{\tiny EB}}^{(0)}
 \int_0^1 \phi_{\mbox{\tiny EB}}(x, E_ex)dx,
$$
with
$$
\vspace{1mm}
w_{\mbox{\tiny EB}}^{(0)} = c\sigma_{\mbox{\tiny EB}}^{(0)}
 = 4c\alpha_f  Z(Z+1)
 \Bigl(\frac{e^2}{m_ec^2}\Bigr)^2
=
1.39 \times 10^{-16} \mbox{cm}^3\mbox{s}^{-1},
$$
for hydrogen atoms, and see the left-hand side of Table 5 for 
$\phi_{\mbox{\tiny EB}}(x, E_\gamma)$.

For $E_e \gg m_e c^2$, we can use the complete screening cross-section, 
leading to the well-known result
$$
-\biggl\langle \frac{{\it \Delta}E_e}{{\it \Delta} t} 
\biggr\rangle_{\hspace{-0.5mm}\mbox{\scriptsize rad}}
=
n(\vct{r}) w_{\mbox{\scriptsize rad}}^{(\infty)} E_e;\ \ \ 
w_{\mbox{\scriptsize rad}}^{(\infty)}  =
 7.30 \times 10^{-16} \mbox{cm}^3\mbox{s}^{-1}.
$$

On the other hand the energy-loss rate due to the IC is given, 
taking into account 
the energy spectrum of
the target photon at $\vct{r}$, 
$n_{\mbox{\scriptsize ph}}(\vct{r}; E_{\mbox{\scriptsize ph}})$,
by
$$
-\biggl\langle \frac{{\it \Delta}E_e}{{\it \Delta} t} 
\biggr\rangle_{\hspace{-0.5mm}\mbox{\scriptsize IC}}
= \int_0^{\infty}\! \! dE_{\mbox{\scriptsize ph}} 
\int_{E_{\mbox{\scriptsize ph}}}^{E_{\mbox{\tiny M}}}\! \!
E_\gamma \,[n_{\mbox{\scriptsize ph}}(\vct{r}; E_{\mbox{\scriptsize ph}})
c \sigma_{\mbox{\tiny IC}}
(E_e, E_\gamma; E_{\mbox{\scriptsize ph}})]d E_\gamma,
\eqno{(\rm A2)}
$$
with
$$
\vspace{0.5mm}
E_{\mbox{\tiny M}} \equiv E_{\mbox{\tiny M}}(E_e, 
E_{\mbox{\scriptsize ph}}) 
 = E_e \frac{X}{1+X};\ \ \ X \equiv 
X(E_e, E_{\mbox{\scriptsize ph}}) =
 \frac{4E_{\mbox{\scriptsize ph}}E_e}{(m_e c^2)^2},
\eqno{(\rm A3)}
$$
see the right-hand side of Table 5 for 
 $\sigma_{\mbox{\tiny IC}}(E_e, E_\gamma; E_{\mbox{\scriptsize ph}})$.
Here we omit the suffix $i$ introduced in \S\,2.2 for simplicity.
For $E_e \gg m_ec^2$, 
equivalently $X \gg 1$, one finds a reasonable result, 
$E_{\mbox{\tiny M}} \approx E_e$, leading to 
$E_{\mbox{\scriptsize ph}} \le E_\gamma \le E_e$.

From equation (4) in the text 
$$
\vspace{0.4mm}
n_{\mbox{\scriptsize ph}}(\vct{r}; E_{\mbox{\scriptsize ph}})
dE_{\mbox{\scriptsize ph}} =  
\Biggl[\frac{\epsilon_{\mbox{\scriptsize ph}}(\vct{r})}
{k_{\mbox{\tiny B}} T_{\mbox{\scriptsize ph}}}\Biggr]
 W_{\mbox{\scriptsize ph}}(k)\frac{dk}{k^2};\ \ \ 
k = \frac{E_{\mbox{\scriptsize ph}}}{k_{\mbox{\tiny B}}
 T_{\mbox{\scriptsize ph}}},
$$
where $W_{\mbox{\scriptsize ph}}(k)$ is the Planck function 
for the 2.7\,K CMB,
and the gaussian function given
by equation (5) for the stellar radiation and
the re-emission from the dust grains.

The integration with respect to $E_\gamma$ is given (Jones 1965, 1968) by,
 (see eq.\ [A3] for $X$) 
$$
\vspace{-0.5mm}
 \int_{E_{\mbox{\scriptsize ph}}}^{E_{\mbox{\tiny M}}}\! \!
E_\gamma 
\sigma_{\mbox{\tiny IC}}
(E_e, E_\gamma; E_{\mbox{\scriptsize ph}}) d E_\gamma \approx 
\sigma_{\mbox{\tiny IC}}^{(0)} E_e S_{\mbox{\tiny IC}}(X)/{X^2}, 
$$
with
$$
\sigma_{\mbox{\tiny IC}}^{(0)} = 3\sigma_{\mbox{\tiny T}} = 
2.00 \times 10^{-24} \mbox{cm}^2,\ \ 
(\sigma_{\mbox{\tiny T}}:\ \mbox{Thomson cross-section}), 
$$
and
$$
\blankline
S_{\mbox{\tiny IC}}(X) = \biggl(\frac{X}{2}+6+\frac{6}{X}\biggr) \ln(1+X) - 
\frac{11X^3/12 + 6X^2 +9X +4}{(1+X)^2} - 2 -
2\int_0^{X} \frac{\ln (1 +t)}{t} {dt}.
$$
One should note that the approximation used above is 
only $E_e \gg m_ec^2$, readily satisfying the condition
in the energy region of interest,
$E_e$\,$\gsim$\,10\,MeV.

Now, we have the energy-loss rate due to the IC scattering in a
compact form after integrating over the energy 
$E_{\mbox{\scriptsize ph}}$ of the target photon
in equation (A2),
$$
\blankline
-\biggl\langle \frac{{\it \Delta}E_e}{{\it \Delta} t} 
\biggr\rangle_{\hspace{-0.5mm}\mbox{\scriptsize IC}}
=  \epsilon_{\mbox{\scriptsize ph}}(\vct{r})
{w_{\mbox{\tiny T}}}
{\it \Lambda}(E_e, T_{\mbox{\scriptsize ph}})E_e^2,
\eqno{(\rm A4)}
$$
where $E_e$ is in units of GeV and $\epsilon_{\mbox{\scriptsize ph}}$
in eVcm$^{-3}$, and 
$$
w_{\mbox{\tiny T}} = 
\frac{4}{3}
\frac{c\sigma_{\mbox{\tiny T}}\times 10^{-9}}
{[m_e c^2/\mbox{GeV}]^2}
 = 1.018 \times 10^{-16}\mbox{cm}^3\mbox{s}^{-1},
\eqno{(\rm A5)}
$$
$$
{{\it \Lambda}(E_e, T_{\mbox{\scriptsize ph}})} \equiv  
{{\it \Lambda}({\it \Theta}_e)}
=
 9 \int_0^\infty S_{\mbox{\tiny IC}}(X) 
W_{\mbox{\scriptsize ph}}({\it \Theta}_e^2 X){dX}/{X^4},
\eqno{(\rm A6)}
$$
with
$$
{\it \Theta}_e  \equiv {\it \Theta}_e(E_e,  T_{\mbox{\scriptsize ph}}) = 
\frac{m_e c^2/2}
{\sqrt{(k_{\mbox{\tiny B}}T_{\mbox{\scriptsize ph}}) E_e}}.
\eqno{(\rm A7)}
$$

The above discussions are applicable also for the synchrotron radiation, 
since it is 
caused by the collision between an electron and the
virtual photon induced by the magnetic field.
Practically, however, 
we have the condition $\hbar \omega_c {\it \Gamma}_e \ll m_ec^2$
 (${\it \Gamma}_e$: Lorentz factor of electron) 
with $\omega_c = eH_\perp/m_ec^2$,
and we can use the Thomson scattering cross-section, namely
${\it \Lambda}(E_e, T_{\mbox{\scriptsize ph}})
 \rightarrow 1$. Hence we obtain equation (12a).

\appendix 
\blankline
\begin{center}
APPENDIX B\\ CONTRIBUTION OF PERTURBATIVE TERMS IN THE TRANSPORT EQUATION
\end{center}

\begin{center}
B1. HIGH ENERGY REGION  $E_e\ \gsim \ E_c^{+}$
\end{center}

Since  we can neglect the fluctuation due to the
reacceleration in the HE region,
we take here the second term in equation (25a) alone, 
omitting the second term in equation (22). 
The transport equation for the electron density in the HE region,
 $N_{e, \epsilon}^{(0)}({\vct{r}}; E_e, t)$,
without the perturbative term, is given by 
$$
\vspace{-1mm}
\biggl[
\frac{\partial}{\partial t} - 
{\it \nabla}\cdot D({\vct{r}}; E_e){\it \nabla}  
 - \bar{\epsilon}(\vct{r}) \frac{\partial}{\partial E_e}
{\cal W}_\epsilon(E_e)
 \biggr] {\cdot} N_{e, \epsilon}^{(0)}({\vct{r}}; E_e, t)
 = Q({\vct{r}}; E_e, t).
$$

As discussed in \S\,4.1, we regard $N_{e, \epsilon}^{(0)}({\vct{r}}; E_e, t)$
as the solution of the first order approximation
for equation (21), so that 
we have the following equation with the perturbative term, 
$\bar{n}(\vct{r}) [{\cal W}_n(E_e)
 N_{e, \epsilon}^{(0)}({\vct{r}}; E_e, t)]^\prime$, 
moving it to the right-hand side, 
$$
\biggl[
\frac{\partial}{\partial t} -
{\it \nabla}\cdot D({\vct{r}}; E_e){\it \nabla}  
 - \bar{\epsilon}(\vct{r}) 
\frac{\partial}{\partial E_e} {\cal W}_\epsilon(E_e)
 \biggr] {\cdot} N_{e,\epsilon}({\vct{r}}; E_e, t)
 = Q({\vct{r}}; E_e, t) +
\bar{n}(\vct{r}) \frac{\partial}{\partial E_e} {\cal W}_n(E_e)
N_{e, \epsilon}^{(0)}({\vct{r}}; E_e, t).
$$

Now, we rewrite the solution 
 $$
\vspace{-2mm}
N_{e, \epsilon}({\vct{r}}; E_e, t) = 
N_{e, \epsilon}^{(0)}({\vct{r}}; E_e, t) + 
\tilde{N}_{e, n}^{(0)}({\vct{r}}; E_e, t),
$$
leading to
$$
\vspace{2mm}
\biggl[
\frac{\partial}{\partial t} -
{\it \nabla}\cdot D({\vct{r}}; E_e){\it \nabla}  
 - \bar{\epsilon}(\vct{r}) \frac{\partial}{\partial E_e}
 {\cal W}_\epsilon(E_e) \biggr] 
{\cdot} \tilde{N}_{e, n}^{(0)}({\vct{r}}; E_e, t) = 
\bar{n}(\vct{r}) \frac{\partial}{\partial E_e} {\cal W}_n(E_e) 
N_{e, \epsilon}^{(0)}({\vct{r}}; E_e, t).
$$
Thus for the steady state ($\partial/\partial t = 0$),
we have the solution of the second order approximation
$$
\tilde{N}_{e, n}^{(0)}({\vct{r}}; E_e) = 
\int_0^{\infty} \! \!
\tilde{\it \Pi}_\epsilon^{(0)} ({\vct{r}}; y)\tilde{f}_n^{(0)}(y; E_e)dy,
$$
with
$$
\Bigl[\bar{\epsilon}(\vct{r}) c\frac{\partial}{\partial y} -
{\it \nabla}\!\cdot\!D({\vct{r}}){\it \nabla} \Bigr]{\cdot}
\tilde{\it \Pi}_\epsilon^{(0)}({\vct{r}}; y)
= \tilde{Q}^{(0)}(\vct{r}) \delta(y),
$$
$$
\blankline
\Bigl[ c E_e^{\alpha} \frac{\partial}{\partial y} - 
\frac{\partial}{\partial E_e}{\cal W}_\epsilon (E_e)
 \Bigr] {\cdot} \tilde{f}_n^{(0)}(y; E_e) = 0,
$$
where $\tilde{Q}^{(0)}(\vct{r})$ and $\tilde{f}_n^{(0)}(0; E_e)$ 
are given by replacing 
$Q^{(0)}(\vct{r})$ $[\equiv Q(\vct{r})]$ and $f_\epsilon^{(0)}(0; E_e)$ 
(see eqs.\,[24b] and [28b]) with
$$
Q^{(0)}(\vct{r}) = Q_0\mbox{e}^{-r/\rQ-|z|/\zQ}
\ \Rightarrow \ 
\tilde{Q}^{(0)}(\vct{r}) = \bar{n}(\vct{r})\mbox{e}^{-|z|/\zD}
\frac{2Q_0}{\bar{\epsilon}_0}
\frac{{\cal J}_{\kappa}({\omega}_\kappa,{U}_\kappa)}{J_{\kappa}({U}_\kappa)},
$$ 
$$
 f_\epsilon^{(0)}(0; E_e) = E_e^{-\gamma-\alpha}\ \Rightarrow \
 \tilde{f}_n^{(0)}(0; E_e) = E_e^{-\alpha} \frac{1}{c}
 \frac{\partial}{\partial E_e}
\Bigl[{\cal W}_n(E_e)F_{r,\epsilon}^{(0)}(E_e)\Bigr],
$$
with
$$
\vspace{2mm}
{\cal J}_{\kappa}({\omega},{U}) = 
\int_0^1  t^{\omega} 
J_{\kappa}({U}t)dt,
$$
see Table~4 for $\kappa, \omega_\kappa$, and $U_\kappa$, and 
$J_{\kappa}(U)$ is the Bessel function of the index with $\kappa$ (Paper~I).

Corresponding to the replacement of $Q^{(0)}(\vct{r}) \Rightarrow 
\tilde{Q}^{(0)}(\vct{r})$, the scale heights in the source, $\rQ$ and $\zQ$,
must be replaced as
$$
\frac{1}{\rQ}\ \Rightarrow\ \frac{1}{r_n},
\ \ \ 
\frac{1}{\zQ}\ \Rightarrow\ \frac{1}{z_n}+\frac{1}{\zD} = \frac{2}{\bar{z}_n},
$$
leading to the following replacements,
$$
\omega_\kappa = \biggl(\frac{1}{\zQ} - \frac{1}{2z_\epsilon}\biggr) 
\ \Rightarrow \  \biggl(\frac{2}{\bar{z}_n}
 - \frac{1}{2z_\epsilon}\biggr) \equiv \tilde{\omega}_\kappa,
$$
while the radial scale height in the source, $\rQ$, doesn't appear explicitly
in this procedure.

Now the Laplace transform of $\tilde{f}_n^{(0)}(y; E_e)$ is 
immediately given (see eq.\,[31]) by
$$
\frac{\tilde{F}_{r,n}^{(0)}(E_e)}{F_{r,\epsilon}^{(0)}(E_e)} = 
\int_{E_e}^{\infty}\! dE_0\,
\frac{[{\cal W}_n(E_0) F_{r,\epsilon}^{(0)}(E_0)]^\prime}
{{\cal W}_\epsilon(E_e)F_{r,\epsilon}^{(0)}(E_e)}\,
\mbox{e}^{-Y_{r,\epsilon}(E_e, E_0)},
\eqno{\rm (B1)}
$$
where $[\,\cdots\,]^\prime$ denotes the differential with respect to $E_0$, 
and see \S\,4.2 for $Y_{r,\epsilon}(E_e, E_0)$, and we obtain
$$
\frac{\tilde{N}_{e,n}^{(0)}(\vct{r}; E_e)}{N_{e,\epsilon}^{(0)}(\vct{r}; E_e)}
=  
 \frac{2}{\bar{\eta}_0}
\frac{{\cal J}_{\kappa}(\tilde{\omega}_\kappa,{U}_\kappa)}
{J_{\kappa}({U}_\kappa)} 
\frac{\tilde{F}_{r,n}^{(0)}(E_e)}{F_{r,\epsilon}^{(0)}(E_e)},
\eqno{\rm (B2)}
$$
with
$$
\bar{\eta}_r = {\bar{\epsilon}_r}/{\bar{n}_r}= 
\bar{\eta}_0 \mbox{e}^{-2r(1/\bar{r}_\epsilon-1/\bar{r}_n)};
 \ \ \bar{\eta}_0 = \bar{\epsilon}_0/\bar{n}_0.  
$$

In Figure~24a, we show 
${\tilde{N}_{e,n}^{(0)}(\vct{r}; E_e)}/{N_{e,\epsilon}^{(0)}(\vct{r}; E_e)}$
against $E_e$ at the SS with
$[\bar{\eta}_\odot,\ \bar{r}_\epsilon]$\,=\,[2\,eV, 8\,kpc],  
corresponding to 
[$\bar{\epsilon}_\odot, \bar{n}_\odot$]\,=\,[2\,eVcm$^{-3}$, 1\,cm$^{-3}$].
Then one finds that
the perturbative contribution due to the energy change in proportion to
the gas density, $\bar{n}(\vct{r})$, is less than 10\% in the energy
region $E_e\ \gsim \ E_c^+$.

We finally obtain 
$$
N_{e,\epsilon}(\vct{r}; E_e) = 
\frac{2Q_0}{\bar{\epsilon}_0}
\frac{{\cal J}_{\kappa}({\omega}_\kappa,{U}_\kappa)}{J_{\kappa}({U}_\kappa)}
F_{r,\epsilon}(E_e) \mbox{e}^{-|z|/\zD},
\eqno{\rm (B3)}
$$
with 
$$
F_{r,\epsilon}(E_e) = F_{r,\epsilon}^{(0)}(E_e) + 
\frac{2}{\bar{\eta}_0}
\frac{{\cal J}_{\kappa}(\tilde{\omega}_\kappa,{U}_\kappa)}
{J_{\kappa}({U}_\kappa)}
\tilde{F}_{r,n}^{(0)}(E_e).
\eqno{\rm (B4)}
$$

\begin{figure}[!b]
\vspace{-2mm}
    \includegraphics[width=7.8cm]{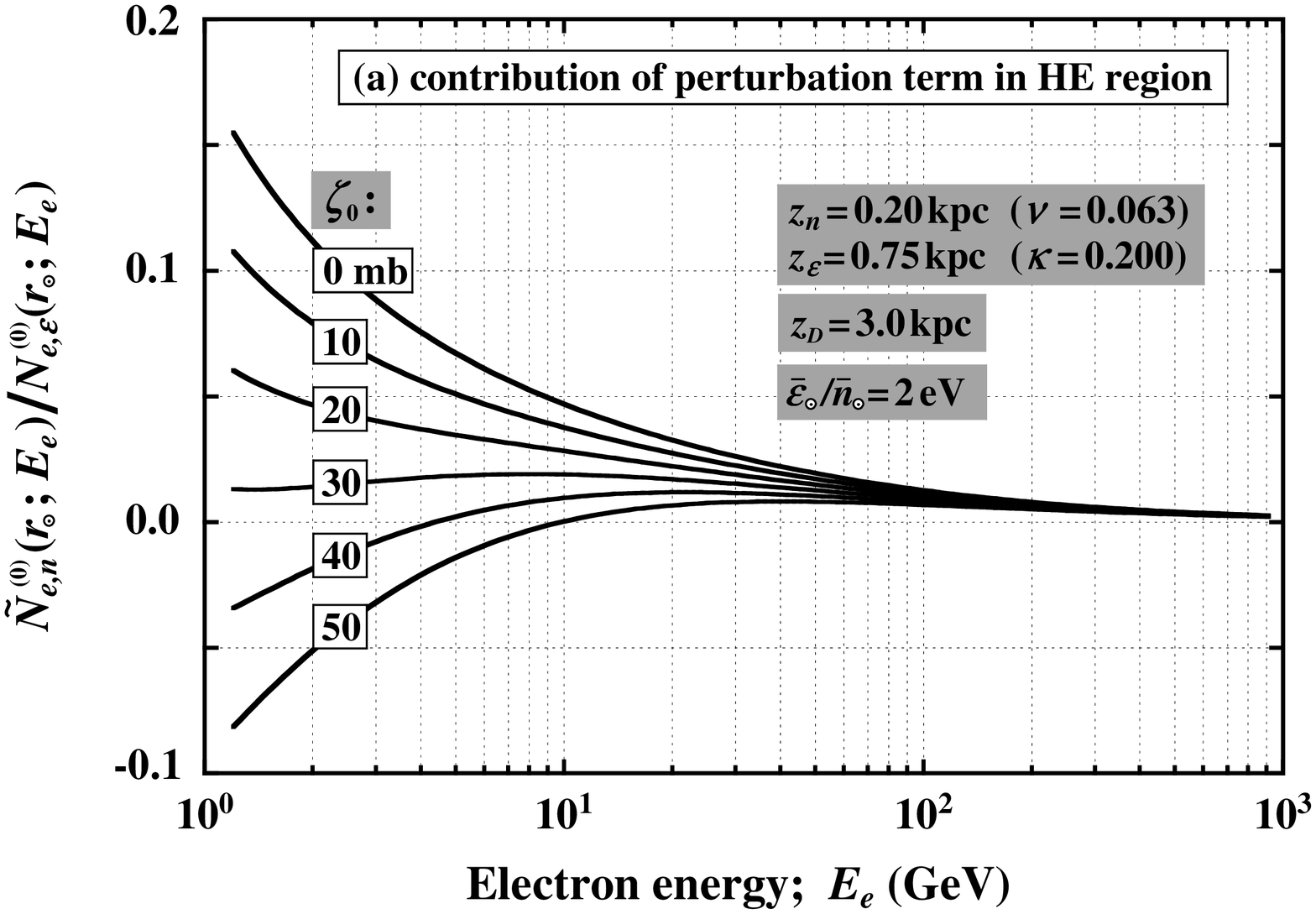}
    \includegraphics[width=7.8cm]{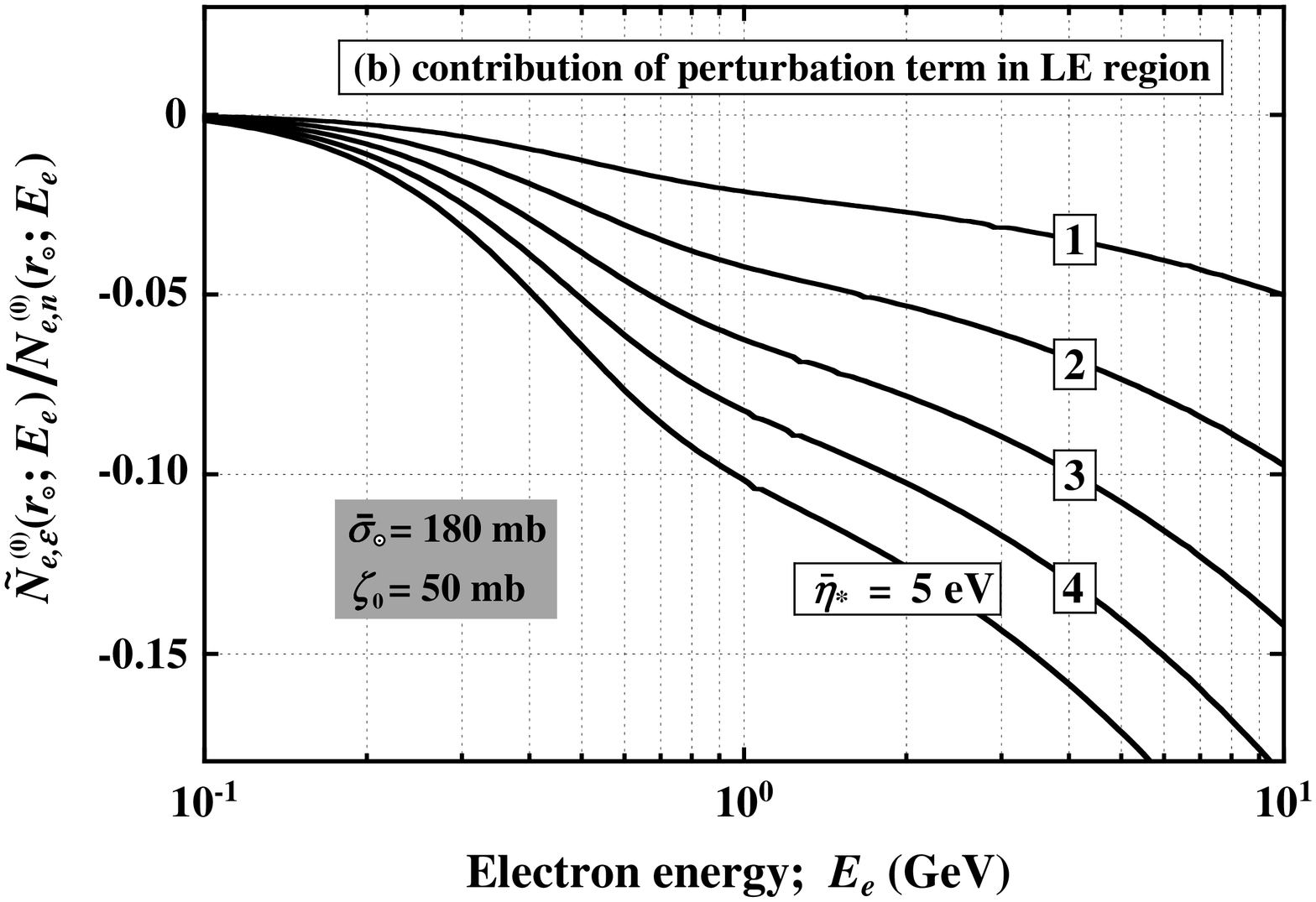}
  \caption{
Numerical results of the contribution of the
  perturbative terms in the (a) HE region and (b) LE region.
}
\end{figure}

\begin{center}
B2. LOW ENERGY REGION $E_e \ \lsim \ E_c^{+}$
\end{center}

In the LE region,  the fluctuation term due to the reacceleration , 
${\LARGE \langle} {\it \Delta}E_e^{\,2}/{{\it \Delta} t} 
{\LARGE \rangle}_{\hspace{-0.2mm}\mbox{\scriptsize rea}}$, 
becomes now effective 
as compared to the {\it average} energy-loss term due 
to the synchrotron-IC effect,
${\LARGE \langle}{\it \Delta}E_e/{{\it \Delta} t} 
{\LARGE \rangle}_{\hspace{-0.2mm}\mbox{\scriptsize sic}}$,
in proportion to the energy density, $\bar{\epsilon}(\vct{r})$.
So equation (25b) is approximately written as
$$
\bar{n}(\vct{r}) {\cal W}_n(E_e) +
{\LARGE O}[\bar{\epsilon}(\vct{r}){\cal W}_\epsilon(E_e)] 
 \simeq 
 \bar{n}(\vct{r}) [{\cal W}_n(E_e) + \bar{\eta}_*
{\cal W}_\epsilon(E_e)], 
\eqno{\rm (B5)}
$$
with
$$
\vspace{2mm}
{{\bar{\epsilon}}(\vct{r})}/{\bar{n}({\vct{r}})} \approx
 {\large \mbox{$\langle$}} {{\bar{\epsilon}}(\vct{r})}/{\bar{n}(\vct{r})}
 {\large \mbox{$\rangle$}}_{\scriptsize \mbox{eff}} \equiv
  \bar{\eta}_*,
\eqno{\rm (B6)}
$$
where the {\it effective} value of $\bar{\eta}_*$ is of the 
magnitude of [1--5]\,eV, and for instance 
$\bar{\eta}_* = \bar{\eta}_\odot \approx 2$\,eV at the SS.

Neglecting the second term in equation (B5),
 we have the solution for the principal term,
 corresponding to equation (32), (see \S\,3.3 and Table 4 for $\nu$, $U_\nu$)
$$
{N}_{e,n}^{(0)}(\vct{r}; E_e) = \frac{2 Q_0}{\bar{n}_0 c} 
\frac{{\cal J}_{\nu}({\omega}_\nu,{U}_\nu)} {J_{\nu}({U}_\nu)}
{F}_{r,n}^{(0)}(E_e) 
 \mbox{e}^{-|z|/\zD}, 
\eqno{\rm (B7)}
$$
with
$$
{F}_{r,n}^{(0)}(E_e) = \frac{c}{|{\cal W}_n(E_e)|}
\int_{E_{\scriptsize \mbox{min}}}^{E_{\scriptsize \mbox{max}}}
dE_0 E_0^{-\gamma} \mbox{e}^{-{Y}_{r,n}(E_e, E_0)},
\eqno{\rm (B8)}
$$
and
$$
{Y}_{r,n}(E_e, E_0)
 = c \bar{\sigma}_r \int_{E_e}^{E_0}\! 
\frac{E^\alpha}{{\cal W}_n(E)} dE.
\eqno{\rm (B9)}
$$

Now putting
$$
{\cal W}_n^*(E_e) = {\cal W}_n(E_e) + \bar{\eta}_*
{\cal W}_\epsilon(E_e), 
$$
the electron density with the synctrotron-IC effect (perturbative term here)
 is immediately
$$
{N}_{e,n}^{*(0)}(\vct{r}; E_e) = \frac{2 Q_0}{\bar{n}_0 c} 
\frac{{\cal J}_{\nu}({\omega}_\nu,{U}_\nu)} {J_{\nu}({U}_\nu)}
{F}_{r,n}^{*(0)}(E_e) 
 \mbox{e}^{-|z|/\zD}, 
\eqno{\rm (B10)}
$$
where ${F}_{r,n}^{*(0)}(E_e)$ is given by replacing 
${\cal W}_n$ with ${\cal W}_n^*$ in equations (B8) and (B9), and it is
 related to the perturbvative term, $\tilde{N}_{e,\epsilon}^{(0)}$,
 discussed in \S4.1 as
$$
N_{e,n}^{*(0)}({\vct{r}}; E_e) = 
N_{e,n}^{(0)}({\vct{r}}; E_e) +
\tilde{N}_{e,\epsilon}^{(0)}({\vct{r}}; E_e).
\eqno{\rm (B11)}
$$

Here one should be careful of the integral range of $E_0$ in
equation (B8), 
$[E_{\scriptsize \mbox{min}}, E_{\scriptsize \mbox{max}}]$,
since we have two zero points
in ${\cal W}_n^*(E_e)$ at two energies, 
${E}_{c*}^{-}$ and ${E}_{c*}^{+}$, 
for instance, $[{E}_{c*}^{-}, {E}_{c*}^{+}] \approx
[0.1, 1]$\,GeV for $\zeta_0$\,=\,50\,mbarn and 
$\bar{\eta}_* = 2$\,eV, and \vspace{1mm}
\begin{displaymath}
\hspace{1cm}
\mbox{[}E_{\scriptsize \mbox{min}},
 E_{\scriptsize \mbox{max}}\mbox{]}     
= \left\{ \begin{array}{ll} \vspace{2mm}
\mbox{[} {E}_e,\ {E}_{c*}^{-}{]}
 \ \ \ \ \  \mbox{for \ \hspace{1.4cm}$E_e < {E}_{c*}^{-}$,} \\ \vspace{2mm}
\mbox{[} {E}_{c*}^{-},\ E_e\mbox{]}
 \ \ \ \ \  \mbox{for \ \ \ ${E}_{c*}^{-} < E_e < {E}_{c*}^{+}$,} 
\\ 
\mbox{[}E_e,\,+\infty \mbox{]}
\ \ \ \ \,\mbox{for \ \hspace{1.45cm}$E_e > {E}_{c*}^{+}$.}
                       \end{array} 
                       \right. 
\end{displaymath}

We present 
$\tilde{N}_{e,\epsilon}^{(0)}/{N}_{e,n}^{(0)}$ against $E_e$
for several sets of $[\bar{\eta}_*; z_n, \zD]$ at the SS
in Figure~24b, and one finds
it is much less than 10\% in the low energy region $E_e\,\lsim\,E_c^{+}$.

\begin{center}
B3. CONTRIBUTION FROM THE FLUCTUATION IN THE REACCELERATION
\end{center}

Once we confirm that the contribution of 
${\LARGE O}[\bar{\epsilon}(\vct{r}){\cal W}_\epsilon(E_e)]$ is
approximately given by equation (B5) with the energy loss
proportional to $\bar{n}(\vct{r})$, 
it is possible
to use the path length distribution, ${\it \Pi}_n(\vct{r}; x)$, 
as presented in Paper~I,
$$
{\it \Pi}_n (\vct{r}; x) \simeq 
 \frac{2 Q_0}{\bar{n}_0 c} 
\frac{{\cal J}_{\nu}(\omega_\nu,{U}_\nu)}
  {J_{\nu}({U}_\nu)}
 \mbox{e}^{-\bar{\sigma}_r x-|z|/\zD}, 
\eqno{\rm (B12)}
$$
but the slab equation is now slightly cumbersome,
$$
\Bigl[c \bar{\sigma}_r E_e^{\alpha} - 
\frac{\partial}{\partial E_e}{\it {\cal W}}_n^*(E_e)
 - c \zeta_0\frac{1}{4}
\frac{\partial^2}{\partial E_e^2} E_e^{2-\alpha}
 \Bigr] {\cdot} F_{r,n}(E_e) = c E_e^{-\gamma}.
$$

Remembering that ${F}_{r,n}^{*(0)}(E_e)$ is a solution of the
equation
$$
\Bigl[c \bar{\sigma}_r E_e^{\alpha} - 
\frac{\partial}{\partial E_e}{\it {\cal W}}_n^*(E_e)\Bigr] {\cdot}
 {F}_{r,n}^{*(0)}(E_e) = c E_e^{-\gamma}, 
$$
we can write $F_{r,n}(E_e)$ as
$$
F_{r,n}(E_e) = {F}_{r,n}^{*(0)}(E_e) + {F}_{r,n}^{(1)}(E_e),
\eqno{\rm (B13)}
$$
with
$$
\Bigl[c \bar{\sigma}_r E_e^{\alpha} - 
\frac{\partial}{\partial E_e}{\it {\cal W}}_n^*(E_e)\Bigr] {\cdot}
 {F}_{r,n}^{(1)}(E_e) = c \zeta_0\frac{1}{4}
\frac{\partial^2}{\partial E_e^2} E_e^{2-\alpha}
{F}_{r,n}^{*(0)}(E_e).
\eqno{\rm (B14)}
$$

The solution for equation (B14) is immediately obtained after 
the following replacement in equation (28b) 
$$
c E_e^{-\gamma} \Rightarrow 
c \zeta_0\frac{1}{4}\frac{\partial^2}{\partial E_e^2} E_e^{2-\alpha}
{F}_{r,n}^{*(0)}(E_e),
$$
and the explicit form is given by
$$
\blankline
\frac{{F}_{r, n}^{(1)}(E_e)}{{F}_{r, n}^{*(0)}(E_e)} = c\zeta_0
\frac{1}{4} 
\int_{E_e}^{\infty}\! dE_0\,
\frac{[E_0^{2-\alpha} {F}_{r, n}^{*(0)}(E_0)]^{\prime \prime}}
{|{\cal W}_n^*(E_e)|{F}_{r, n}^{*(0)}(E_e)}\,
\mbox{e}^{-{Y}_{r, n}^*(E_e, E_0)},
$$
where ${Y}_{r, n}^*(E_e, E_0)$ is given by replacing
 ${\cal W}_n(E_e)$ in equation (B9) with ${\cal W}_n^*(E_e)$.

Now, from equations (B12) and (B13),
 the electron density in the LE region is given by
$$
N_{e, n}(\vct{r}; E_e) = 
\frac{2Q_0}{\bar{n}_0}
\frac{{\cal J}_{\nu}({\omega}_\nu,{U}_\nu)}{J_{\nu}({U}_\nu)}
F_{r, n}(E_e) \mbox{e}^{-|z|/\zD},
\eqno{\rm (B15)}
$$
where $F_{r, n}(E_e)$ is given by equation (B13).

Corresponding to equation (26b), we rewrite equation (B15),
 dividing into three terms as (see also eq.\ [B11]) 
$$
N_{e, n}(\vct{r}; E_e) = 
N_{e,n}^{(0)}({\vct{r}}; E_e) +
\tilde{N}_{e,\epsilon}^{(0)}({\vct{r}}; E_e) +
N_{e,n}^{(1)}({\vct{r}}; E_e), 
$$
 and in Figure 7 in the text, we present 
$N_{e,n}^{(1)}/[N_{e,n}^{(0)} +
\tilde{N}_{e,\epsilon}^{(0)}]$
against $E_e$. One finds that the contribution of the fluctuation is
 effective around GeV region,  boosting the electron density without
 the fluctuation by approximately 25\%.

\appendix 
\blankline
\begin{center}
APPENDIX C\\ EMISSIVITY of $\gamma$'s COMING FROM 
 INVERSE COMPTON PROCESS
\end{center}

In this appendix, we omit the suffix $i$ for simplicity.
The production rate of $\gamma$'s per unit time due to the bremsstrahlung, 
$P_{\mbox{\tiny EB}}(\vct{r}; E_e, E_\gamma)$ $\equiv$ $n(\vct{r}) c
\sigma_{\mbox{\tiny EB}}(E_e, E_\gamma)$ in equation (34), must be
replaced by
$$
\vspace{2mm}
P_{\mbox{\tiny IC}}(\vct{r}; E_e, E_\gamma)
 \equiv \!
 \int_{E_m}^{\infty} \! \! n_{\mbox{\scriptsize ph}}(\vct{r};
 E_{\mbox{\scriptsize ph}}) c 
\sigma_{\mbox{\tiny IC}}(E_e, E_\gamma;
 E_{\mbox{\scriptsize ph}})dE_{\mbox{\scriptsize ph}},
\eqno{\rm (C1)}
$$
where $E_{\mbox{\scriptsize ph}}$
 is the energy of the target photon before electron
scattering, and 
$E_m$ is given by solving equation (A3) 
with $E_{\mbox{\tiny M}} \equiv E_{\gamma}$ with respect to 
$E_{\mbox{\scriptsize ph}}$ $(\equiv E_m)$,
$$
E_m \equiv E_m(E_e, E_\gamma) = 
k_{\mbox{\tiny B}}T_{\mbox{\scriptsize ph}}{\it \Theta}_x^2,
$$
with
$$
\vspace{2mm}
{\it \Theta}_x \equiv
 {\it \Theta}_x(E_\gamma, T_{\mbox{\scriptsize ph}}; x) =
\frac{m_e c^2/2}
{\sqrt{(k_{\mbox{\tiny B}}T_{\mbox{\scriptsize ph}}) E_\gamma}}
 \frac{x}{\sqrt{1-x}};\ \ 
 x = \frac{E_\gamma}{E_e}, 
$$
and $\sigma_{\mbox{\tiny IC}}
(E_e, E_\gamma; E_{\mbox{\scriptsize ph}})$
 is the production cross-section of $\gamma$'s 
due to IC scattering, which is summarized in the right-hand side of
Table~5.

Remarking that the integral range,
 $E_m$\,$\le$\,$E_{\mbox{\scriptsize ph}}$\,$\le$\,$\infty$,
in equation (C1) corresponds to 
$0 \approx (m_e c^2/2E_e)^2 \le q \le 1$, 
using a parameter $q$  (Blumenthal \& Gould 1970) 
appearing in the fourth line of the
right-hand side of Table~5, equation (C1) is rewritten as
$$
P_{\mbox{\tiny IC}}(\vct{r}; E_e, E_\gamma)
= \epsilon_{\mbox{\scriptsize ph}}(\vct{r}) w_{\mbox{\tiny T}}
{\it \Phi}_{\mbox{\tiny IC}}(x, E_\gamma),
$$
with $\epsilon_{\mbox{\scriptsize ph}}$ in units of eVcm$^{-3}$,
and see equation (A5) for $w_{\mbox{\tiny T}}$, and 
$$
\frac{{\it \Phi}_{\mbox{\tiny IC}}(x, E_\gamma)}
{[E_\gamma/\mbox{GeV}]} = 
9\biggl(1-\frac{1}{x}\biggr)^2 \! \!
\int_0^1\! W_{\mbox{\scriptsize ph}}({\it \Theta}_x^2/q)
\phi_{\mbox{\tiny IC}}(x, q)q dq,
\eqno{\rm (C2)}
$$ 
where 
$W_{\mbox{\scriptsize ph}}(k)$ with $k \equiv {\it \Theta}_x^2/q$ 
corresponds to the two types of energy spectra 
of the photon gas, the Planck function for the CMB, and the
gaussian function for the stellar radiation and the re-emission
from the dust grains,  and  
$\phi_{\mbox{\tiny IC}}(x, q)$ is given in the right-hand side of 
Table~5.





\begin{flushleft}
{\bf REFERENCES}\\

   Abdo, A.\ A., et al.\ 2009, Phys.\ Rev.\ Lett., 102, 181101\\

   Abdo, A.\ A., et al.\ 2010a, arXiv:1003.0895v1 [astro-ph.CO] 3 Mar 2010\\
 
   Abdo, A.\ A., et al.\ 2010b, submitted to PRL\\

   Adriani, O., et al.\ 2009, Phys.\ Rev.\ Lett., 102, 051101\\

   Adriani, O., et al.\ 2010,  http://arxiv.org/abs/1007.0821\\

   Aguilar, M., et al.\ 2002, Phys.\ Rep., 366, 331\\

   Ahn, H.\ S., et al.\ 2005,
   Proc.\ 29th Int.\ Cosmic Ray Conf., (Pune), 3, 57\\

   Ahn, E., Bertone, G., Merritt, D., \& Zhang, P.\ 2007,
   Phys.\ Rev.\ D76, 023517\\

   Ahn, H.\ S., et al.\ 2008, Astropart. Phys.  30, 133\\

   Aharonian, F.\ A., Atoyan, A.\ M., \& V\"{o}lk, H.\ J.\ 1995, 
A\&A, 294, L41\\

   Aharonian, F.\ A., et al.\ 2008, 
arXiv:0811.3894v1 [astro-ph.HE] 24 Nov 2008\\

   Aharonian, F.\ A., et al.\ 2009, 
arXiv:0905.0105v1 [astro-ph.HE] 1 May 2009\\

   Berezhko, E., \& Ksenofontov, L.\ 1999, J.\ Exp.\ Theor.\ Phys., 89, 391\\

   Berezinskii, V.\ S., et al.\ 1990, 
   Astrophysics of Cosmic Rays (Amsterdam: North Holland)\\

   Bergstr\"{o}m, L.\ 2000, Rep. Prog. Phys., 63, 793\\

   Bertsch, et al.\ 1993, ApJ, 416, 587\\

   Bindi, V.\ 2009, Nucl.\ Instr.\ and Meth.\ A,
 doi:10.1016/j.nima.2009.10.090\\

   Blumenthal, G.\ R., \& Gould, R.\ J.\ 1970, Rev.\ Mod.\ Phys., 42, 237\\

   Boezio, M., et al.\ 2000, ApJ, 532, 653\\

   Caballero, R., et al.\ 2008, 
   Proc.\ 30th Int.\ Cosmic Ray Conf., (Merida), 3, 1313\\
 
   Case, G., \& Bhattacharya, D.\ 1996, A\&AS, 120C, 437\\

   Chang, J.\ et al.\ 2008, Nature, 456, 362\\

   Cheng, H.\ C., Feng, J.\ L.,\& Matchev, K.\ T.\ 2002, 
   Phys.\ Rev.\ Lett., 89, 211301\\
 
   Chiang, J., \& Mukherjee, R.\ 1998, ApJ, 496, 752\\

   Clemens, D.\ P., Sanders, D.\ B., \& Scoville, N.\ Z.\ 1988, ApJ, 327, 139\\

   Cordes, J.\ M., et al.\ 1991, Nature, 354, 121\\

   Cowsik, R., \& Lee, M.\ A.\ 1979, ApJ, 228, 297\\

 Cutler, D.\ J.\ \& Groom, D.\ E.\ 1991, 
   ApJ, 376, 322\\

   Davis, A.\ J., et al.\ 2000, AIP Conf.\ Proc., 528, 421\\

   Delahaye, T.\ et al.\ 2010, arXiv:1002.1920v1 [astro-ph.HE] 9 Feb 2010\\

   Derbina, V.\ A., et al.\ 2005, ApJ, 628, L41\\

   Diplas, A., \& Savage, B.\ D.\ 1991, ApJ, 377, 126\\

   Dogiel, V.\ A., \& Urysson, A.\ V.\ 1988, A\&A, 197, 335\\

   DuVernois, M.\ A., et al.\ 2001, ApJ, 559, 296\\

   Ellison, D.\ C., et al.\ 2000, ApJ, 540, 292\\

   Ferriere, K.\ M.\ 2001, Rev.\ Mod.\ Phys., 73, 1031\\

   Freudenreich, H.\ T.\ 1998, ApJ, 492, 495\\

   Gabici, S., \& Blasi, P.\ 
   2003, Astropart.\ Phys., 19, 679\\

   Gaisser, T.\ K.\ 1990, Cosmic Rays and Particle Physics
   (Cambridge: Cambridge Univ.\ Press)\\

   Ginzburg, V.\ L.\ 1979, 
   Theoretical Physics and Astrophysics (Oxford: Pergamon Press)\\

   Ginzburg, V.\ L., Khazan, Y.\ A., \& Ptuskin, V.\ S.\ 1980, 
   Ap\&SS, 68, 295\\

   Golden, R.\ L., et al. 1994, ApJ, 436, 769\\

   Gould, R.\ J.\ 1969, Phys.\ Rev., 185, 72\\

   Heinbach, U., \& Simon, M.\ 1995, ApJ, 441, 209\\

   Henderson, A.\ P., Jackson, P.\ D., \& Kerr, F.\ J.\ 1982, ApJ, 263, 116\\

   Hendrick, S.\ P., \& Reynolds, S.\ P.\ 2001, ApJ, 559, 903\\

   Hotta, N.\ et al.\ 1980, Phys.\ Rev., D22, 1\\

   Hunter, S.\ D., et al.\ 1997, ApJ, 481, 205\\

   Ishikawa, T.\ 2010, Master Thesis for Aoyama-Gakuin University\\

   Jones, F.\ C.\ 1965, Phys.\ Rev., 137, B1306\\

   Jones, F.\ C.\ 1968, Phys.\ Rev., 167, 1159\\

   Kasahara, K.\ 1985, Phys.\ Rev., D31, 2737\\

   Kappadath, S.\ C., 1996, A\&AS, 120, 619\\

   Kerr, F.\ J., \&  Lynden-Bell, D.\ 1986, MNRAS, 221, 1023\\

   Keshet, U., Waxman, E., \& Loeb, A.\ 2004, J.\ Cosmol.\ Astropart.\ Phys.,
   04, 006\\

   Kinzer, R.\ L., Purcell, W.\ R., \& Kurfess, J.\ D.\ 1999, ApJ, 515, 215\\

   Kobayashi, T.\ et al.\  1999, 
   Proc.\ 26th Int.\ Cosmic Ray Conf., (Salt Lake City), 3, 61\\

   Kobayashi, T., Komori, Y., Yoshida, K.\ \& Nishimura, J.\ 2004, 
   ApJ, 601, 340\\

   Koch, H.\ W., \& Motz, J.\ W.\ 1959, Rev.\ Mod.\ Phys., 31, 920\\

   Kodaira, K.\ 1974, Pub.\ Astr.\ Soc.\ Japan, 26, 255\\

   Kounine, A.\ 2010, Invited talk at 16th ISVHECRI (Fermilab)

   Kulkarni, S.\ R., Blitz, L., \& Heiles, C.\ 1982, ApJ, 259, L63\\

   Loeb, A., \& Waxman, E.\ 2000, Nature, 405, 156\\

   Mathis, J.\ S., Mezger, P.\ G., \& Panagia, N.\ 1983, A\&A, 128, 212\\

   Miniati, F.\ et al.\ 2000, ApJ, 542, 608\\

   Moskalenko, I.\ V.\ et al.\ 2002, ApJ, 565, 280\\

   M\"{u}cke, A., \& Pohl, M.\ 2000, MNRAS, 312, 177\\

 M\"{u}ller, D.\ 
   2009,  Proc.\ 31st Int.\ Cosmic Ray Conf., (Lodz), in press\\

Nagashima, K., et al.\ 1989, 
   Nuovo Cimento, 12C, 695\\

Nishimura, J.\ 1964, Prog.\ Theor.\ Phys.\
   Suppl.\ 6, 93\\

 Nishimura, J., Fujii, M.\ \& 
   Taira, T.\ 1979,  Proc.\ 16th Int.\ Cosmic Ray Conf., (Kyoto), 1, 488\\

   Nishimura, J.\ et al.\ 1980, ApJ, 238, 394\\

   Okamoto, M., \& Shibata, T.\ 1987, Nucl.\ Instr.\ Methods A257, 155\\

   Picozza, P., et al.\ 2007,
   Proc.\ 30th Int.\ Cosmic Ray Conf., (Merida), 2, 19\\
 
   Pohl, M., \& Esposito, J.\ A.\ 1998, ApJ, 507, 327\\

   Porter, T.\ A., et al.\ 2008, ApJ, 682, 400\\

   Porter, T.\ A.\ 2009, Talk at Fermi Symposium (2--5 November,
 Washington DC)\\

   Ptuskin, V.\ S., \& Ormes, J.\ F.\ 1995, 
   Proc.\ 24th Int.\ Cosmic Ray Conf., (Rome), 3, 56\\

   Ptuskin, V.\ S., Jones, F.\ C., \& Ormes, J.\ F.\ 1996, ApJ, 465, 972\\

   Reynolds, S.\ P., \& Keohane, J.\ W.\ 1999, ApJ, 525, 368\\

 Sakakibara, S.\ 1965, 
   J.\ Geomag.\ Geoelectr., 17, 99\\

 Sato, Y., \& Sugimoto, H.\ 1979,
     Proc.\ 16th Int.\ Cosmic Ray Conf., (Kyoto), 7, 42\\

 Seo, E.\ S.\ 
   2009,  Proc.\ 31st Int.\ Cosmic Ray Conf., (Lodz), in press\\

   Shen, C.\ S.\ 1970, ApJ, 162, L181\\
 
   Shibata, T., Hareyama, M., Nakazawa, M., \& Saito, C.\ 2004,
   ApJ, 612, 238 (Paper~I)\\

   Shibata, T., Hareyama, M., Nakazawa, M., \& Saito, C.\ 2006,
   ApJ, 642, 882 (Paper~II)\\

   Shibata, T., \& Ito, T.\ 2007, ApJ, 655, 892 (Paper~III)\\

   Shibata, T., Honda, N., \& Watanabe, J., 
   2007, Astropart.\ Phys., 27, 411 (Paper~V)\\
 
   Shibata, T., Futo, Y., \& Sekiguchi, S.\ 2008, ApJ, 678, 907 (Paper~IV)\\

   Simon, M., Heinrich, W., \& Mathis, K.\ D.\ 1986, ApJ, 300, 32\\

   Sreekumar, P., et al.\ 1998, ApJ, 494, 523\\
 
   Stecker, F.\ W., \& Jones, F.\ 1977, ApJ, 217, 843\\

   Stecker, F.\ W., \& Salamon, M.\ H.\ 1996, ApJ, 464, 600\\

   Stecker, F.\ W., Hunter, S.\ D., \& Kniffen, D.\ A.\ 
   2008, Astropart.\ Phys., 29, 25\\

   Strong, A.\ W., et al.\ 1988,  A\&A, 207, 1\\

   Strong, A.\ W., \& Moskalenko, I.\ V.\ 1998, ApJ, 509, 212;\\ 
   http://www.gamma.mpe.-garching.mpg.de./$\sim$aws/aws.html\\
 
   Strong, A.\ W., Moskalenko, I.\ V.\ \& Ptuskin, V.\ S.\ 2007, 
   Annu.\ Rev.\ Nucl.\ Part.\ Sci.\ 2007, 57, 285\\

   Strong, A.\ W., et al.\ 1999,  Astrophys.\ Lett.\ Commun., 39, 209\\

   Strong, A.\ W., Moskalenko, I., \& Reimer, O.\ 2000, ApJ, 537, 763\\
 
   Strong, A.\ W., Moskalenko, I., \& Reimer, O.\ 2004, ApJ, 613, 962\\

   Suzuki, R., Watanabe, J., \& Shibata, T.\ 
   2005, Astropart.\ Phys., 23, 510\\

   Syrovatskii, S.\ I., 1959, 36, 17\\

   Torii, S., et al.\ 2001, ApJ, 559, 973\\

   Torii, S., et al.\ 2006, Adv.\ Polar Upper Atmos.\ Res., 20, 52\\

   Totani, T., \& Kitayama, T.\ 2000, ApJ, 545, 572\\

   Ullio, P., Bergstr\"{o}m, L., Edsj\"{o}, J., \& Lacey, C.\ 2002,
   Phys.\ Rev.\ D66, 123502 \\

   Yamazaki, R., et al.\ 2006, MNRAS, 371, 1975\\

\end{flushleft}

\end{document}